\documentclass[11pt]{article}

\usepackage{amssymb}
\newcommand{\reals}{{\mathbb{R}}}

\newcommand{\pow}[1]{{\mathfrak{P}}(#1)}

\newcommand{\prob}[2]{\item[{\bf #1}] {\sl #2}}

\font\sf=cmss10
\newcommand{\Nats}{{\hbox{\sf I\kern-.13em\hbox{N}}}}   
\newcommand{\Reals}{{\hbox{\sf I\kern-.14em\hbox{R}}}}  
\newcommand{\Ints}{{\hbox{\sf Z\kern-.43emZ}}}          
\newcommand{\CC}{{\hbox{\sf C\kern -.48emC}}}           
\newcommand{\QQ}{{\hbox{\sf C\kern -.48emQ}}}           

\newtheorem{theorem}{Theorem}
\newtheorem{lemma}[theorem]{Lemma}
\newtheorem{corollary}[theorem]{Corollary}
\newtheorem{definition}{Definition}[section]

\newenvironment{proof}{\QuadSpace\par\noindent{\bf Proof}:}{\EndProof\HalfSpace}
\newenvironment{proofsketch}{\QuadSpace\par\noindent{\bf Proof sketch}:}{\EndProof\HalfSpace}

\newenvironment{proposition}{\QuadSpace\par\noindent{\bf Proposition}:}{\HalfSpace}

\newcommand{\QuadSpace}{\vspace{0.25\baselineskip}}
\newcommand{\HalfSpace}{\vspace{0.5\baselineskip}}

\newcommand{\EndProof}{ \hfill \vrule width 1ex height 1ex depth 0pt }

\usepackage{epsfig}

\def\expect{\mathbb{E}}
\def\prob{{\mbox{Pr}}}

\title{Nearly-Exponential Size Lower Bounds for  Symbolic Quantifier Elimination Algorithms and OBDD-Based Proofs of Unsatisfiability}

\author{Nathan Segerlind \\ Department of Computer Science \\ Portland State University \\ Portland, Oregon \\ nsegerli@cs.pdx.edu}

\begin{document}

\maketitle

\begin{abstract}
We demonstrate a family of propositional formulas in conjunctive normal form
so that a formula of size $N$ requires size $2^{\Omega(\sqrt[7]{N/logN})}$
to refute using the tree-like OBDD refutation system of
Atserias, Kolaitis and Vardi~\cite{akv2004}
with respect to  all variable orderings. 
All known  symbolic quantifier elimination algorithms for satisfiability
generate tree-like proofs when run on unsatisfiable CNFs, so this lower bound
applies to the run-times of these algorithms.  
Furthermore, the lower bound generalizes earlier
results on OBDD-based
proofs of unsatisfiability in that it applies for all variable
orderings,  it applies when the clauses are  processed according to
an arbitrary schedule, and it applies when variables are eliminated via
quantification.
\end{abstract}

\section{Introduction}\label{IntroductionSect}
Ordered binary decision diagrams (OBDDs) are data structures  for representing
Boolean functions~\cite{bryant86,bryant92,meinelTheobaldOBDD} that
are widely used when solving problems in  circuit synthesis and model 
checking (cf. ~\cite{bryant86,bryant92,mcmillanSMC1992,CGPmodelChecking}).
A large number of OBDD-based algorithms have been implemented for solving
the Boolean satisfiability 
problem~\cite{bryant86,uribeStickel1994,groote1996,chatalicSimon2000a,chatalicSimon2000b,aguirre01random,motMarkBCP2002,motMarkCASSATT2002,aloulMSIWLS2002,cdasv2003,panVardi2004,huangDarwiche2004,akv2004,jusSinBiere2006}. 
Many of these algorithms are known to efficiently generate proofs of
unsatisfiability for CNFs known to require exponential running times for
other methods, such as the pigeonhole principle that states $n+1$
objects cannot be placed into $n$ holes without a collision, and it is not
immediately clear what the limitations of OBDD-based methods are.
While
it would immediately follow from the hypothesis $P \neq NP$ that
such methods cannot solve all satisfiability instances in time 
polynomially-bounded by the input size, that sort of thinking
strikes us as begging the question, and  here we present
{\em{unconditional}} limitations for algorithms of this kind:
We unconditionally show that a wide class of 
OBDD-based satisfiability algorithms cannot 
solve all satisfiability instances in
sub-exponential time.  
Prior analyses of the runtimes of OBDD-based satisfiability methods
have been limited in their application
because of assumptions on the order of processing the input clauses~\cite{grooteZantema2004,groote2006} or an assumption on the variable ordering used when
building the OBDDs~\cite{akv2004}, so this is the first unconditional
lower bound that applies even to a system that explicitly constructs
the OBDD for a CNF by  selecting a variable ordering and then
conjoining the clauses according to a heuristically
chosen order.

More formally, we present  superpolynomial  size lower bounds for
the {\em{tree-like OBDD refutation system}} and satisfiability algorithms
based on  {\em{explicit OBDD construction}} and
{\em{symbolic quantifier elimination}}.  
  We give two motivations for
studying minimum refutation sizes for  proof systems and satisfiability
algorithms.  The first is that it
is a necessary and tractable step towards understanding larger
questions:  Whether or not there is a polynomial-time algorithm
for satisfiability,  and whether or not propositional proof systems 
manipulating
Boolean circuits can prove every tautology in size bounded by a polynomial
in the size of the tautology (formalized as whether or the 
{\em{extended-Frege proof systems}} are polynomially bounded,
cf.~\cite{krajicekBook}).
Both of these problems seem well beyond our current understanding.  Rather
than try to understand {\em{all}} polynomial-time computations or {\em{all}}
extended-Frege proofs,  we study the sizes of proofs of unsatisfiability
for a particular class of satisfiability algorithms and extended-Frege proofs:
In this case,  tree-like OBDD refutations. Under this interpretation,  the main
result of this paper can be interpreted as saying ``As far as symbolic
quantifier elimination algorithms are concerned, $P$ is different from $NP$.''  The second motivation is to develop
 taxonomy of satisfiability methods and identify the  kinds of
reasoning best suited to each method. Under this interpretation, the main 
result of this paper can be interpreted as saying ``While symbolic quantifier
elimination methods can perform efficiently on some structured formulas
such as the  $n+1$ to $n$ pigeonhole principle,  such methods inherently
face an exponential blow-up when reasoning about the behavior of a system 
acted upon by a permutation.''

\subsection{Using OBBDs for Satisfiability and Propositional Proofs}

One motivation for developing satisfiability algorithms based on
OBDDs
is the hope to escape the limitations of the 
resolution proof system.  Most current
satisfiability engines, in particular, the DLL with clause learning 
approach~\cite{grasp1996,chaff2001,berkmin2002,minisat2003},   implement the resolution proof
system~\cite{Robinson-1965} and therefore require exponential running times
on the many CNFs known to require exponential size resolution
refutations~\cite{Haken:1985:IR,Urq:harer,JACM::ChvatalS1988,Ben-Sasson:2001:SPN,Raz01,Beame:2002:ERD}.
The hope is that by developing algorithms that implement proof systems 
other than resolution,
new satisfiability algorithms will be able to  efficiently
solve  satisfiability instances not yet efficiently solvable.

An OBDD is a read-once branching program in which the variables appear
according to a fixed order along every path (ie. the nodes are arranged
in levels,  all nodes at a level query the same variable, 
and each variable corresponds to at most one level).  
The choice
of variable ordering can affect the size of the OBDD by an exponential factor
and choosing a suitable variable ordering for a task is of utmost importance.
The primary utility of the ordering restriction  is that with respect to
each fixed ordering,  the OBDD computing a Boolean function is unique, up
to a linear-time reduction to normal form (cf.~\cite{meinelTheobaldOBDD}).
Because of this canonicity property,  the equality test for two Boolean 
functions represented as OBDDs is simply a check that their OBDDs
are identical.
  Many simple but useful functions have small OBDDs with respect to
some variable ordering, 
and many set operations, such as union and intersection, can be computed in
polynomial time from two OBDDs.  These properties make OBDDs well-suited
for reasoning about symbolically encoded sets of states, and their use
revolutionized the field of model 
checking~\cite{mcmillanSMC1992,CGPmodelChecking}.
In light of this success,  a
number of attempts have been made to utilize OBDDs for more efficient
satisfiability algorithms.  This results of this paper apply to two such
methods,  explicit construction and symbolic quantifier elimination,
but do not clearly apply to a third,  compressed resolution.

{\bf{Explicit construction.}} In the literature, this is sometimes called the
 ``OBDD apply'' method.  
In this method,
a variable ordering is selected,    the OBDD  for the CNF with
respect to that ordering is constructed, and it is checked whether
this OBDD is the constant false~\cite{bryant86}. 
Proofs in this system are straightforward: We begin with the OBDDs
representing each clause, 
and we repeatedly conjoin them together until we obtain an OBDD for the 
conjunction of all the clauses. 
There are two opportunities for cleverness - the variable ordering used to
construct the OBDDs,  and the order in 
which the clauses are joined together, 
cf. \cite{uribeStickel1994,aguirre01random,huangDarwiche2004}.
Empirical studies~\cite{uribeStickel1994,cdasv2003} and a mathematical
analysis of the implementation in which the clauses are conjoined in the same
order as the input presentation~\cite{grooteZantema2004} have
suggested that this method
is incomparable with resolution-based methods.

{\bf{Symbolic quantifier elimination.}}  
This method extends the explicit construction
method by strategically eliminating variables via the application of
existential quantifiers~\cite{groote1996,aguirre01random,panVardi2004,huangDarwiche2004,sinzBiereCSR2006}.
In particular, to determine if a CNF $\bigwedge_{i=1}^m C_i(\vec{x})$ is 
satisfiable,  rather than
build an OBDD for $\bigwedge_{i=1}^m C_i(\vec{x})$,  it suffices to build one
for $\exists \vec{x} \bigwedge_{i=1}^m C_i(\vec{x})$. This is can be
more efficient because it is often the case that the OBDD for
$\exists \vec{x} F(\vec{x},\vec{y})$ are significantly smaller than the OBDD
for $F(\vec{x},\vec{y})$.  One example of this approach is to first
use heuristic methods  to partition the
variables into sets $X_1, \ldots
X_k$ and the clauses into sets $A_1, \ldots A_k$ so that for each
$i=1, \ldots k$, the variables of $X_i$ do not appear in the clauses belonging
to sets $A_{i+1}, \ldots A_k$,
then  construct
the OBDD for the quantified Boolean formula:
\[ \exists X_k \left(  \ldots \left( \exists X_2 \left(\exists X_1 \bigwedge_{C \in A_1} C(X_1,\ldots X_k) \right) \wedge \bigwedge_{ C \in A_2} C(X_2, \ldots X_k) \right)  \ldots \right) \wedge \bigwedge_{C \in A_k} C(X_k) \]
It has been observed that symbolic quantifier elimination leads to significant
speed-ups over explicit OBDD construction on random $3$-CNFs~\cite{groote1996,aguirre01random},
and that, on a certain mix of structured benchmarks, symbolic quantifier
elimination solves more instances before time-out
than  solvers based on resolution  or compressed 
resolution~\cite{huangDarwiche2004,panVardi2004}.

When formalized as proof systems,  these algorithms can be viewed as
treelike versions of the  OBDD propositional 
proof system described by Atserias, Kolaitis
and Vardi~\cite{akv2004}.  
This proof system is highly non-trivial:  OBDDs are  circuits
not formulas, so this proof system is a kind of weak extended-Frege 
system\footnote{For uninitiated,  {\em{Frege systems}} are basically
the standard textbook style systems of propositional logic manipulating Boolean
formulas whereas {\em{extended Frege systems}} manipulate Boolean
circuits.  From a computational complexity perspective,
Frege systems can be thought of as manipulating concepts definable in
$NC^1$ and extended Frege systems can be thought of as manipulating
concepts definable in $P$.}.
Because it is not believed possible to convert OBDDs into formulas without
an exponential blow-up,  the OBDD proof system is not expected to be
$p$-simulatable by Frege systems 
(in the sense of Cook and Reckhow~\cite{cookReckhow}).  The tree-like OBDD system
possesses polynomial-size refutations of the $n+1$ to $n$ pigeonhole
principle,  and it
can $p$-simulate several interesting proof
systems, such as tree-like resolution,  Gaussian refutations over
a finite field, and tree-like cutting planes refutations with unary 
coefficients~\cite{akv2004}.

{\bf{Compressed resolution and compressed search.}}  The analysis of this
paper does not apply to these systems in a clear way,  and we take a few
paragraphs to to discuss  why not.
Compressed resolution and search methods use OBDDs (or sometimes,  a 
variant known as {\em{ZDDs}} or
{\em{zero-suppressed binary decision diagrams, cf.~\cite{meinelTheobaldOBDD}}})
to encode exponentially large resolution refutations. 
A well-known example of this method is {\em{multiresolution}}, developed by
Chatalic and Simon~\cite{chatalicSimon2000a,chatalicSimon2000b}.  In 
multiresolution,   the set of clauses in the refutation is represented 
symbolically with a ZDD, and the Davis-Putnam variable elimination step
is performed using ZDD operations, so that many resolution steps
are handled simultaneously.   
In addition to the DP procedure,  clause learning and breadth-first
search algorithms have been implemented in the compressed
 setting~\cite{motMarkCASSATT2002,motMarkBCP2002,motRoyMark2005}.

The reason that the lower bound
of this paper does not seem to apply to  ``compressed proof systems''
is that in these systems,
 the OBDDs are not over the same variables as the input
CNF. The OBDDs  symbolically encode a large resolution proof, so
they work over new variables that encode clauses 
over the original variables.
A typical encoding  has for each literal $l$ over original input CNF 
variables, a new variable $y_l$ 
that corresponds to whether or not the literal $l$ is present in a clause.
In this  way,  compressed methods are akin to the ``implicit proofs''
described by Kraj{\'{i}}{\v{c}}ek~\cite{krajicekImp2004}.

\subsection{The Result and Comparisons with Earlier Work}

The main result of this paper is that
for infinitely many values of $N$,  there is an unsatisfiable CNF
$\Phi$ of size $N$ so that every tree-like OBDD refutation of  $\Phi$
has size at least $2^{\Omega(\sqrt[7]{N/\log N})}$ (Theorem~\ref{payoff}).
This lower bound generalizes earlier work on proving size lowerbounds
for OBDD-based proofs of unsatisfiability in three ways:  
The proofs can use variable elimination via existential quantifiers,
the clauses of the input CNF can be processed in any order (so long as they
are recombined according to a tree-structure), and
the variable ordering of the OBDDs can be arbitrary.
The two previously published results regarding 
size lower bounds for OBDD-proofs of unsatisfiability either made use of 
a restriction on the order in which the clauses are processed,  or held only
for a fixed ordering on the variables.

In~\cite{grooteZantema2004}, Groote and Zantema prove a size lower bound for
refutations in the OBDD-apply system that
 conjoins the clauses of the CNF in the order of the input listing
(ie. to process  $C_1 \wedge (C_2 \wedge C_3)$, an OBDD for $C_2 \wedge C_3$ is
built and then one for $C_1 \wedge (C_2 \wedge C_3)$ is built).  In fact,
in that paper they give a size lower bound for refutations of a formula of the 
form $\neg x \wedge (x \wedge \psi)$, which is trivial to refute if the formula
is processed as $(\neg x \wedge x) \wedge \psi$.
Qualitatively, Theorem~\ref{payoff} generalizes their bound by applying to 
systems that eliminate variables by quantification, and by applying to systems
that allow the clauses to be processed in an arbitrary manner. However,
their bound is 
quantitatively stronger: Where $N$ is the size of the difficult CNF,  
their bound on refutation size is $2^{\Omega(\sqrt{N})}$ whereas ours is
$2^{\Omega(\sqrt[7]{N})}$.

In~\cite{akv2004},  Atserias, Kolaitis, and Vardi 
formalized the OBDD-based propositional proof system  incorporating
symbolic quantifier elimination, and
proved that
for each fixed variable ordering,
there is a CNF of size $N$ that requires size $2^{N^{\Omega(1)}}$
to refute in the OBDD proof system using that particular variable ordering.
The two results are incomparable.
The bound of~\cite{akv2004} 
applies to the general (DAG-like) system, whereas
Theorem~\ref{payoff} only applies to the tree-like system. On the other hand,
Theorem~\ref{payoff} shows that there is a CNF for which every refutation
with respect to every variable ordering has nearly-exponential size.
The result of~\cite{akv2004} says that for each variable ordering, there is a
CNF for which that ordering is a poor choice, and does not elminate the 
possibility that for each CNF there is a variable ordering for which the
CNF will be refuted in (say) time linear in the size of the CNF.
Theorem~\ref{payoff} eliminates this possibility for the tree-like case,
which includes all known implementations of these algorithms.

The analysis of Theorem~\ref{payoff} is the first that applies 
to all symbolic quantifier elimination algorithms so far
developed~\cite{groote1996,aguirre01random,panVardi2004,huangDarwiche2004,sinzBiereCSR2006}.  It is not hard to see upon inspection that these algorithms 
generate proofs of unsatisfiability in the tree-like OBDD system.  Moreover,
the results of~\cite{grooteZantema2004} do not apply to these methods
as the methods typically perform a preprocessing analysis that chooses the
order in which clauses are combined, and the methods eliminate variables
via existential quantification.
 The results of~\cite{akv2004} do not apply
to these methods because the variable ordering is typically selected by some
static analysis of the input CNF.

\subsection{The Technique and its Comparison with Earlier Work}

The argument is a reduction: We produce a CNF so that
if there is a small refutation of the CNF in the tree-like OBDD
proof system,  then there is a low-communication randomized two-player
protocol for the set-disjointness function.  The set-disjointness function
 is known to require high communication~\cite{kalyaSchnit92,razborovDisj},
so all refutations of this CNF must be large.
The reduction is obtained by the interpolation by a 
communication game technique  that has been well-used in
the propositional 
proof complexity community for some time now~\cite{ipu1994,akv2004}.
However,  there is a wrinkle that complicates our return to
this well-trodden path. 
Accounting for all possible variable orderings for the OBDDs
corresponds to proving communication lower bounds that
hold under all ways of partitioning the inputs, the so-called {\em{best-case}}
partition model in communication complexity.

The analysis takes a turn from the beaten path 
at how the reduction fares under this best-case partitioning of variables.
Indeed, the reduction can be thought of a variant of the reduction
given by Raz and Wigderson~\cite{razWigMatch92} in which an adversarial
partitioning of the variables has taken place.  
The reductions 
in~\cite{razWigMatch92,ipu1994,akv2004} show that there is a search problem
in variables $\vec{U}$ and $\vec{V}$, $Search(\vec{U},\vec{V})$,
and a  randomized one-sided-error reduction from set-disjointness (in variables
$\vec{X}$ and $\vec{Y}$) to $Search(\vec{U},\vec{V})$ in which player I
 creates an assignment to
$\vec{U}$ using $\vec{X}$ and player II  creates an assignment to $\vec{V}$
using $\vec{Y}$.  These reductions make heavy use of the structure inherent
in the fixed partition of the variables of the search problem.
In the best-case partition scenario that our reduction handles,  we provide a 
search problem $Search(\vec{W})$ and show that no matter how the variables
of $\vec{W}$ are partitioned into two equal-sized sets $\vec{U}$ and $\vec{V}$,
there is a reduction from set-disjointness to the search problem
in which player I to creates an assignment to
$\vec{U}$ using $\vec{X}$ and player II to creates an assignment to $\vec{V}$
using $\vec{Y}$. 

Over the course of the analyzing the randomized reduction, in particular, its
 distribution on placing gadgets, we develop a framework for
passing local density results that hold for the uniform distribution to
hold for distributions that we say are ``generated by dependent domains with
blocking processes''.  While these techniques are quite simple,
they may be of interest 
for analyzing other random processes and reductions that exploit structure
in dense graphs or set systems.

\subsection{Outline of this Article}

Sections~\ref{notationSect} and~\ref{proofSystems} are notation and 
background.
The CNF that we prove difficult for OBDD refutations is introduced in
Section~\ref{cnfSect}.  Because of the central role of handling the
partition of the variables,  Section~\ref{partitionSect} is dedicated to 
the bookkeeping involved with handling  partitions and defining the
density of a partition, which is the parameter governing the quality of
the reduction from set-disjointness.

We present the reduction and its analysis in an order that  emphasizes
the similarities with the reductions of~\cite{ipu1994} 
and~\cite{razWigMatch92},
while encapsulating the differences in some lemmas that are proved in later
sections. Section~\ref{refToSearch} includes the standard argument that a 
small treelike refutation yields a low-communication search protocol,
although some work is needed to guarantee that the search protocol works
for a partition of density $\Omega(1)$.  Section~\ref{reductionSect}
details the reduction proves the lower bound, modulo a lemma about
the distribution on the gadgets used to build the reduction,
Lemma~\ref{key-lemma}.
The marquee lower bound is presented in Subsection~\ref{lowerBoundSubSect},
Theorem~\ref{payoff}.

In Section~\ref{layoutSect}, we construct the objects claimed in
Lemma~\ref{key-lemma}.  The distribution is very far from uniform, 
and this makes 
the analysis quite different from that of~\cite{razWigMatch92}.
However, to make the reduction work, we need only two properties to
hold. The first is that the probabilities assigned to 
objects at Hamming distance $\Omega(1)$ differ by at most a constant factor
(encapsulated as Lemma~\ref{continuity-lemma}, the ``continuity lemma''),
  and the second is that
events ensuring correctness of the reduction
 occur with  probability not-too-much-less than they
would under the uniform distribution 
(encapsulated as Lemma~\ref{switchable-density}, the ``completeness lemma'').
Because the reduction is based on randomly 
flinging gadgets into the dense corners of a graph,
the distributions get messy and  it seems wise to pass to a cleaner
framework as soon as possible.  We call this framework 
{\em{distributions from dependent domains  with blocking processes}},
or {\em{DDWB distributions}}.  Section~\ref{sect-probbackground} lays out
the notation used for the probability calculations and states some simple
calculations that are needed, while
Section~\ref{DDWB-section} is devoted to
DDWB distributions and their properties.
In Section~\ref{analyzeDSect}, we show that the distribution of 
Lemma~\ref{key-lemma} is a DDWB distribution and use this to prove 
the continuity lemma and the completeness lemma, which
 guarantee the correctness of the reduction.

\subsection{Open Questions}\label{concSect}

The main question left open by this paper is to increase the constants 
for Theorem~\ref{payoff}. 
The constant hidden in the $\Omega()$ of the
$2^{\Omega(\sqrt[7]{N/  \log N})}$ lowerbound of Theorem~\ref{payoff}
is extremely small. Not logician small, but somewhere above Ramsey theorist
small and way below computer scientist small. It is well below $2^{-500}$.
It is doubtful that
this is strongest refutation-size lower bound that holds for the system,
even for these particular CNFs.

The next question is whether or not we can go from the tree-like to
the DAG-like case,  ie. can a superpolynomial size lower bound be
 proved for DAG-like OBDD refutations of some family of CNFs?
This would fully resolve the question posed in~\cite{akv2004}.

What can be said about the expected size of a (tree-like) OBDD refutation of a
random $3$-CNF?
This is  open even for the  explicit OBDD construction method.
It would be especially interesting if such an analysis could explain some of the
threshold behavior observed in~\cite{cdasv2003,aguirre01random}.

It is common for OBDD packages to include a feature that
dynamically recomputes the variable ordering when the OBDDs grow too large.
The analysis  of Theorem~\ref{payoff}
does not cover this as the conversion from
refutation to search (Lemma~\ref{searchLemma}) 
seems to depends on every OBDD in a derivation using the same variable
ordering. 
Current work with symbolic quantifier elimination algorithms for satisfiability
has suggested  that, given  current technology,  static variable
orderings generally
lead to better performance than dynamic variable  
orderings~\cite{aguirre01random,huangDarwiche2004}. This may be 
because these studies compare a default dynamic reordering heuristic 
against a static order that is customized for the satisfiability problem. 
A dynamic variable reordering method that consistently outperforms
static methods remains unseen. On the other hand,  there is no 
 explanation of why static orderings should perform just as well as dynamic
orderings.
An interesting extension of this work  would be
to find a proof system that formalizes OBDD-proofs that include
dynamic variable reordering and to use this to formally compare methods that
use dynamic reordering with those that use static variable orderings. And
of course, proving unconditional proof size lower bounds for algorithms
that incorporate dynamic variable reordering would  be interesting.

To the best of our knowledge, no non-trivial size lower bounds are known
for any of the compressed
methods~\cite{chatalicSimon2000a,chatalicSimon2000b,motMarkCASSATT2002,motMarkBCP2002,motRoyMark2005}.  Because these systems work with OBDDs, there is a
similar flavor with the systems studied in this article. However, 
the fact the systems
build OBDDs in different variables than those of
the input CNF prevents an immediate
application of Theorem~\ref{payoff} to these systems.

\subsection{Acknowledgments}

This work was inspired by conversations with
Albert Atserias and Moshe Vardi about the OBDD refutation system
at the Workshop on New Directions in Proof Complexity held at the 
Isaac Newton  Institute for Mathematics,
where Moshe discussed the findings  of~\cite{akv2004}.  
The author would like to thank Moshe and Albert for
their enthusiasm and encouragement,  as well as workshop co-organizer
Jan Kraj{\'{i}}{\v{c}}ek, who was helpful 
securing the author's attendance.  The author also thanks Jan Friso Groote
for answering some questions about~\cite{grooteZantema2004}.
Paul Beame provided useful comments on an early draft of the paper.
Special thanks  go to 
Cindy Brown and Barton Massey of
Portland State University for their generous hospitality.

\section{Notation and Communication Complexity Background}\label{notationSect}

\begin{definition}
The real numbers are denoted by ${\mathbb{R}}$ and $[0,1]$ denotes the
closed unit interval. Let $n$ be an integer. 
The set of integers $\{1, \ldots n \}$ is
denoted by $[n]$.  
 For a set $S$ and a non-negative integer $k$,
the set of all $k$-tuples over $S$ is denoted by $S^k$ and the of all
size $k$ subsets of $S$ is denoted by ${S \choose k}$.
For a set $S$ we let $\chi_S$ denote the indicator function for $S$ 
with $\chi_S(a)=1$ if $a \in S$, $\chi_S(a)=0$ is $a \not\in S$. The domain
of $\chi_S$ will  always clear from context.  For a product space 
$\prod_{i \in I} X_i$ where $I$ is a finite set, we will sometimes
say that the product space is ``$|I|$ dimensional'' even though is no
algebraic structure defined on $\prod_{i \in I} X_i$.
\end{definition}
Note that
${[n] \choose k }$ is a set with
$\left|{[n] \choose k }\right| = {n \choose k}$.

\begin{definition}
We use the word ``graph'' to mean a  simple, loopless undirected graph.
We use $\subseteq$ to denote the (not necessarily induced)
subgraph relation, ie. $G \subseteq H$ if $G=(V,E)$ and $H=(W,F)$ with
$V \subseteq W$ and $E \subseteq F$ (as sets).  For any two disjoint
nonempty sets $A$ and $B$,  we write $K(A,B)$ to denote the complete
bipartite graph  with partition $\{A, B\}$.
Let $G= (V,E)$ be a graph.  Let $V_0 \subseteq V$ and let
$E_0 \subseteq E$. The set of edges {\em{$E_0$ restricted to $V_0$}}, written
{\em{$E_0 \left[ V_0 \right]$,}} is defined as
$E_0 \left[ V_0  \right]  = \{ e \in E_0 \mid e \subseteq V_0 \}$.
\end{definition}

We use standard results on the randomized two-party communication complexity
of the set-disjointness function.  For a more thorough introduction to this
subject, consult~\cite{kncc97}.
\begin{definition}
Let $f(\vec{X},\vec{Y})$ be a function.
A {\em{randomized two-player protocol for $f$}} is a two-party communication
protocol in which Player I has private access to $\vec{X}$,
Player II has private access to $\vec{Y}$, and the players share access to a 
source of random bits, so  that for all inputs $\vec{X}$ and $\vec{Y}$,
with probability at least $2/3$, the players agree upon the correct value of
$f(\vec{X},\vec{Y})$.  A {\em{deterministic protocol}} is one in which
the answer arrived at by the players is independent of any randomness and is
uniquely determined by the input $\vec{X},\vec{Y}$.
The {\em{cost of a protocol}} is the maximum number
of bits communicated between the two players taken over settings of the input 
and the random bits. The {\em{randomized communication complexity of $f$}}
is the minimum cost of a randomized two-player protocol that computes $f$.
The {\em{set-disjointness function on $n$ bits}} is a Boolean function
$setdisj_n: \{0,1\}^n \times \{0,1\}^n \rightarrow \{0,1\}$ with
\[setdisj(\vec{X},\vec{Y}) = \left\{ \begin{array}{cc}  1 & {\mbox{ if $\exists i \in [n], \ X_i=Y_i =1$}} \\ 0 & {\mbox{ otherwise}}
\end{array} \right.
\]
\end{definition}

\begin{theorem}\label{sdBound}(\cite{kalyaSchnit92,razborovDisj}, cf.~\cite{kncc97})  The two-party randomized communication complexity of $setdisj_n$ 
 is  $\Omega(n)$.   
\end{theorem}

\section{The Ordered-Binary Decision Diagrams Refutation System}{\label{proofSystems}}

\begin{definition}(cf.~\cite{bryant92,meinelTheobaldOBDD})
A {\em{binary decision diagram}} (also known as a {\em{branching program}})
is a rooted, directed acyclic graph in which every
nonterminal node $u$ labeled by a variable $x_u$ and has two out-arcs,
one two a node $t_u$ and the other to a node $f_u$.  Sinks are labeled by
Boolean values.   The function represented by a branching program is calculated
by starting at the root and following a path to the sink as follows: If the
current node $u$ is labeled by the variable $x_u$, and $x_u$ is assigned the
value true,  then follow the arc $t_u$, otherwise follow the arc labeled $f_u$.
The value that the function takes is the value labeled on the sink.
The {\em{size}} of a binary decision diagram is its number of nodes as a DAG.
An {\em{ordered binary decision diagram (OBDD)}}
is a binary decision diagram in 
which:   Along every path from the source to a sink,  every variable is
queried at most once, and,  there is fixed ordering of the variables
$\preceq$ so that along
all paths from the source to a sink,  the order in which
variables are queried is consistent with $\preceq$.
\end{definition}

For the purposes of our argument,
we do not care if the OBDDs are reduced to canonical normal form. 
Indeed, all that is actually used about OBDDs is a 
simple connection between OBDDs and communication complexity
that is the starting point for our reduction.  We do not use it explicitly
in this article,  however,  it is an ingredient  for the proof of 
Lemma~\ref{fcs-lemma}.
\begin{proposition}
If there is size $S$ OBDD for a function $f(x_1, \ldots x_n)$ with respect
to some variable order $x_{i_1}, \ldots x_{i_n}$,  then for each $k \in [n]$,
there is a  two-party communication protocol computing  $f$ with
respect to the variable partition $\{x_{i_1}, \ldots x_{i_k}\}, \{x_{i_k+1}, \ldots x_{i_n}\}$ that uses $\lceil \log S \rceil$
 many bits of communication.
\end{proposition}
\begin{proofsketch}
The first player broadcasts the index of the node that is reached
in the OBDD after following the path given by the assignment to 
$\{x_{i_1}, \ldots x_{i_k}\}$. The second player continues computation
from this node, using the values $\{x_{i_k+1},\ldots x_{i_n}\}$. No further
communication is necessary because of the ordering on queries.
\end{proofsketch}

It is easy to see that the size of the OBDD representing a clause is no more
than the size of the clause, plus the two sink nodes for
 ``true'' and ``false''.  For this reason, we do not distinguish between
a clause and its OBDD with respect to some order.
\begin{proposition}
Let $C$ be a clause containing $l$ literals.  For every variable ordering,
$C$ can be represented by an OBDD of size at most $l+2$.
\end{proposition}

\begin{definition}\label{obddSystem-defn}
Let ${\cal{C}}$ be a set of clauses in variables from a set $V$.
A {\em{OBDD derivation from ${\cal{C}}$
with respect to a variable ordering $\preceq$ on $V$}}
is a sequence of OBDDs $F_1, \ \ldots, \ F_m$ so that each OBDD is built
from the variables of $V$ with respect to the order $\preceq$, 
and each $F_i$ either is a clause in ${\cal{C}}$,
or follows from the preceding $F_1, \ldots F_{i-1}$
by an application of one of the following inference rules:
($A$, $A_0$, and $B$ are OBDDs in the variables $V$
 with ordering $\preceq$, where
$A \Rightarrow A_0$ as Boolean functions, and
$\vec{x}$, $\vec{y}$, $\vec{z}$ are tuples of variables from $V$):
\begin{center}
\begin{tabular}{rlrlrl}
Subsumption: &
$\displaystyle \frac{A}{A_0}$
&
~~~Conjunction: &
$\displaystyle
 \frac{A(\vec{x},\vec{y}) \ \ B(\vec{y},\vec{z})}{ A(\vec{x},\vec{y}) \wedge B(\vec{y},\vec{z})} $
  &Projection: &
$\displaystyle
 \frac{ A(x,\vec{y})}{ \exists x A(x,\vec{y})}$
\end{tabular}
\end{center}

For a set of clauses ${\cal{C}}$, an {\em{OBDD refutation of ${\cal{C}}$ }} is
a derivation from ${\cal{C}}$ whose final line is the OBDD ``false''.
The size of an $OBDD$ refutation is the sum of the sizes of its OBDDs.
An OBDD derivation $F_1, \ldots F_m$ 
is said to be {\em{treelike}} if each $F_i$ 
is used at most once as an antecedent to an inference.
\end{definition}
It is easily checked that the symbolic quantifier elimination algorithms for
satisfiability all generate treelike OBDD refutations in the above system when run on
unsatisfiable CNFs~\cite{groote1996,aguirre01random,huangDarwiche2004,panVardi2004} (so long as a dynamic variable reordering package is not in use).

The lower bound of Theorem~\ref{payoff} actually pertains to many different
formulations of the tree-like OBDD refutation system. In particular, most
sensible inference rules and axioms can be added and the lower bound will
still apply.  This is
because the conversion from refutation to  search protocols 
(cf.~\cite{ipu1994,akv2004}) requires only that
(1) the refutation structure is tree-like (2) the OBDDs are in the same
variables as the input CNF (3) the OBDDs are each built according to the
same variable ordering, and (4)  the inference
rules are sound and of fan-in at most two. Lemma~\ref{permutation-lemma}
of the current work requires that the proof structure is preserved under
under simultaneous permutations of the variables (such a substitution
does change the variable ordering $\preceq$, however).

\section{The Difficult CNF: Indirect Matching Principles}\label{cnfSect}

The CNF  $IndMatch_m$ is a propositional encoding of the fact that in a graph
on $3m$ vertices, it is impossible to simultaneously  have a perfect
matching on $2m$ vertices and an independent set of size $2m+1$.
It is similar to CNF $Match_m$ used by Impagliazzo,
Pitassi, and Urquhart to prove  size lower bounds for
the tree-like cutting planes system~\cite{ipu1994}.  However,  in order to
prove the CNFs difficult for tree-like OBDD refutations
with respect to any variable ordering,   we introduce
a level of indirection via  permutations.

\subsection{The CNF $Match_m$}
There are two distinct kinds of variable used in the CNF $Match_m$:

\begin{enumerate}
\item {\em{The edge variables.}}
There are are $m \cdot {3m \choose 2}$ many variables used to
specify the  matching: One variable 
$x^i_e$ for each $i=1, \ldots m$ and each $e \in [3m]^2$.  The intended
semantics is that the variable $x^i_e$ is equal to one if and only if the 
edge $e$ is the $i$'th edge of the matching.

\item {\em{The vertex variables.}} There
are $(2m+1)3m = 6m^2+3m$ many variables used to specify the
independent set: One variable $y^j_k$ for each $j =1, \ldots 2m+1$ and
each $k= 1, \ldots 3m$. The intended semantics is that the variable $y^j_k$
is equal to one if and only if the element $k$ is the $j$'th element of
the independent set. 
\end{enumerate}

The set of all these variables is  $MVars_m$.
The  following clauses form the CNF $Match_m$:

\begin{enumerate}
\item\label{mclauses-edgeCount} (At least $m$  edges in the matching.)
 For each $i \in [m]$:  $\bigvee_{e \in [3m]^2} x^i_e$

\item\label{mclauses-matching} (Edges form a matching.) For each $i, j \in [2m]$ with $i \neq j$ and each $e,f \in [3m]^2$ with
$e \cap f \neq \emptyset$:   $\neg x^i_e \vee \neg x^j_f$

\item\label{mclauses-vertexUB} (At least $2m+1$ vertices in the independent set.) For each $j \in [2m+1]$:  $\bigvee_{u \in [3m]} y^j_u$

\item\label{mclauses-vertexLB} (Vertices in the independent set are distinct.) For each $i,j \in [2m+1]$ with $i \neq j$ and each $u \in [3m]$:
$\neg y^i_u \vee \neg y^j_u$

\item\label{mclauses-independence} (The vertices are independent.)
For each $e \in [3m]^2$
with $e=\{u,v\}$, each $k \in [m]$ and each $i,j \in [2m+1]$:
$\neg y^i_{u} \vee \neg y^j_{v} \vee \neg x^k_e$
\end{enumerate}
Notice that the CNF $Match_m$ has size $O(m^5)$.

\subsection{ The CNF $IndMatch_m$}

The difference between the CNF $IndMatch_m$ and the CNF $Match_m$
is that we add variables specifying a permutation $\pi$, 
and for an assignment $A$ to $MVars_m$,  we
interpret the independent set not as 
$\{u \mid \exists j \in [2m+1], \ A(y^j_u)= 1 \}$ but instead as
$ \{\pi(u) \mid \exists j \in [2m+1], \ A(y^j_u)= 1 \}$.

\begin{definition}
Let $N$ be given.  A set $\Pi$ of permutations of $N$
is said to be {\em{pairwise}} independent if for all $a, b, c, d \in [N]$
with $a \neq b$ and $c \neq d$:
\[\prob_{\pi \in \Pi} \left[ \pi(a)=c \ \wedge \ \pi(b)=d \right] = \frac{1}{N(N-1)}\]
\end{definition}

It is well-known that for any finite field, the
set of mappings 
$\{ x \mapsto ax+b \mid a \in {\mathbb{F}}^*, \ b \in {\mathbb{F}} \}$ is a pairwise independent family of permutations of size
$|{\mathbb{F}}|(|{\mathbb{F}}|-1)$.

\begin{proposition}\label{pwip-prop}
Whenever $m$ is a power of $3$, there is  a pairwise-independent family of
permutations of $[3m]$, $\Pi_m$, with $|\Pi_m|=9m^2-3m$.
\end{proposition}

The variables used in the CNF $IndMatch_m$ are the variables used in
$Match_m$, along with new variables for encoding a permutation:
There are $l=\lceil \log(|\Pi|) \rceil $ many variables that encode 
a permutation from $\Pi$:  $z_1, \ldots z_l$.
The intended semantics
is that the variables $z_1, \ldots z_l$ encode the permutations of $\Pi$
in some surjective fashion.  This set of permutation variables is denoted
$PVars_m$.
The set of variables $IMVars_m$ is 
$MVars_m \cup PVars_m$.
The CNF $IndMatch_m$ has the same clauses of type~\ref{mclauses-edgeCount}, type~\ref{mclauses-matching}, type~\ref{mclauses-vertexUB} and 
type~\ref{mclauses-vertexLB} that $Match_m$ has,  whereas the clauses enforcing  independence are as follows:
\begin{description}
\item\label{clauses-independence} (Independence between vertices after
application of the permutation.)
For each  $\alpha_1, \ldots \alpha_l \in \{0,1\}$, 
each $e \in [3m]^2$
with $e=\{u,v\}$, each $k \in [m]$ and each $i,j \in [2m+1]$,  with
$\pi$ denoting the element of $\Pi$ encoded by $\vec{\alpha}$:
$\bigvee_{i=1}^L z_i^{1-\alpha_i} \vee \neg y^i_{\pi(u)} \vee \neg y^j_{\pi(v)} \vee \neg x^k_e$
\end{description}
Notice that the CNF $IndMatch_m$ has  $O(m^7)$ many clauses, and size
$O(m^7 \log m)$.

\begin{definition}
Let $\pi$ be a permutation of $[3m]$.  For each variable $v \in MVars_m$
we define
\[\pi(v)  = \left\{\begin{array}{cc}
y^j_{\pi(u)} & {\mbox{if $v= y^j_u$ for some $j \in [2m+1]$, $u \in [3m]$}} \\
x^i_e  & {\mbox{if $v = x^i_e$ for some $i \in [m]$, $e \in {[3m] \choose 2}$}}
 \end{array}  \right. \]
\end{definition}

\begin{lemma}\label{permutation-lemma}
Let $\pi \in \Pi$ be fixed. 
If $\Gamma$ is  a size $S$ refutation of $IndMatch_m$
with variable ordering $v_1, \ldots v_N$,  
then there is a size $S$ refutation of $Match_m$ that
uses the variable ordering $\pi(v_1), \ldots \pi(v_N)$.
\end{lemma}
\begin{proof}
Let $\alpha$ be the assignment to $\vec{z}$ that selects the
permutation $\pi^{-1}$.   We  apply the restriction $\alpha$ to
$\Gamma$,  and we see that
the clauses of $IndMatch_m$ that that are not satisfied
are the non-independence clauses 
that do not use any $\vec{z}$ variables (ie. all clauses 
of type~\ref{mclauses-edgeCount}, type~\ref{mclauses-matching}, 
type~\ref{mclauses-vertexUB}, and type~\ref{mclauses-vertexLB}),
and the independence clauses of the form
$ \neg y^i_{\pi^{-1}(u)} \vee \neg y^j_{\pi^{-1}(v)} \vee \neg x^k_e$,
for $i,j \in [2m+1]$, $u,v \in [3m]$,  $k \in [m]$, and 
 $e \in {[3m] \choose 2}$.
We now  replace every occurrence of the
variable $y^i_u$ by $y^i_{\pi(u)}$.  For the variable ordering,  this
means that $y^i_u$ takes the place of $y^i_{\pi(u)}$ in the ordering.
In each OBDD,  each query to $y^i_u$ is replaced by a query to
$y^i_{\pi(u)}$.    
Every OBDD is now constructed according to the query order 
$\pi(v_1), \ldots \pi(v_N)$.  It is easily checked that 
the proof structure is preserved under this substitution so that
the new derivation is a derivation with respect to the order
$\pi(v_1),\ldots \pi(v_N)$ in the sense of Definition~\ref{obddSystem-defn}.
Moreover, each clause 
$\neg y^i_{\pi^{-1}(u)} \vee \neg y^j_{\pi^{-1}(v)} \vee \neg x^k_e$,
becomes $ \neg y^i_u \vee \neg y^j_v \vee \neg x^k_e$, so that
the new refutation is a refutation of $Match_m$.
\end{proof}

\section{Variable Partitions and Their Densities}\label{partitionSect}

The central task in the proof of
Theorem~\ref{payoff} is to generate reductions
from set-disjointness to the false-clause-search of $IndMatch_m$,
given  an arbitrary partitioning of the variables $IMVars_m$.    In this
brief subsection we present the machinery for analyzing these partitions.
We view the partition as splitting the players into an {\em{edge player}},
with access to variables in ${\cal{V}}_I$,  and a {\em{vertex player}},
with access to variables in ${\cal{V}}_{II}$.  
In the reduction,  the edge player will place his set
disjointness variables $X_l$ on edge variables $x^i_e$ and the vertex player
will place his set-disjointness variables $Y_l$ on vertex variables $y^j_u$.

\begin{definition}
Let $m$ be a positive integer,  and let $\left({\cal{V}}_I, {\cal{V}}_{II}
\right)$ be a partition of $MVars_m$.
For each $i=1, \ldots m$,
define $E_i({\cal{V}}_I)$ to be $\{ e \in [3m]^2 \mid x_e^i \in {\cal{V}}_I \}$.
For each $j=1, \ldots 2m+1$,
define $V_j({\cal{V}}_{II})$ to be $\{ u \in [3m] \mid y^j_u \in {\cal{V}}_{II} \}$. Except for in the proof of Lemma~\ref{part-lemma},
we do not discuss more than one variable partition at a time, so
we usually write $E_i$ instead of $E_i({\cal{V}}_I)$ and $V_j$ instead
of $V_j({\cal{V}}_{II})$.
\end{definition}

It is helpful to think of the variables of $MVars_m$ as being organized into
$m$ rows of edge variables and $2m+1$ rows of vertex variables, with $E_i$
being the set of 
edge variables in row $i$ available to Player I, and $V_j$ being
the set of vertex variables in row $j$ available to Player II.
A very important complication  is that for distinct $i_1,i_2 \in [m]$,   it is
 possible that $E_{i_1} \neq E_{i_2}$.  This means that not only does the edge
used in assignment matter,  but the identity of the variable specifying the
edge matters as well.  The same complication is in play regarding the sets
$V_{j_1}$ and $V_{j_2}$.
Because the identity of the variables matters, in contrast with the reduction
of~\cite{razWigMatch92},  we must treat the objects seen by the players as
assignments to the variables,  not  merely sets of vertices and edges. 

\begin{definition}\label{densityDefn} Let $({\cal{V}}_I,{\cal{V}}_{II})$
be a partition of $MVars_m$.
The {\em{density of $\left({\cal{V}}_I, {\cal{V}}_{II}\right)$,
$\delta \left({\cal{V}}_I, {\cal{V}}_{II} \right)$,}} is defined as follows: 

\[\delta\left({\cal{V}}_{I}, {\cal{V}}_{II}\right) := \frac{1}{m^2(2m+1)^5}\sum_{\vec{\imath} \in [m]^2} \sum_{\vec{\jmath} \in [2m+1]^5} \frac{| \bigcap_{k=1}^5  E_{i_1} \left[ V_{j_k} \right]  \cap   E_{i_2} \left[ V_{j_k} \right] |}{ {3m \choose 2}} \]
\end{definition}

\section{From Refutation to Search}\label{refToSearch}
We  transform small refutations of the $IndMatch_m$ principles into
a low-communication protocol for a search problem in the variables $Mvars_m$.

\begin{definition}
Let $A$ be an assignment to $MVars_m$. 
We say that $A$ is {\em{non-degenerate}} if
it satisfies all of the clauses from $Match_m$ of
type~\ref{mclauses-edgeCount},
type~\ref{mclauses-matching},
type~\ref{mclauses-vertexUB},
and type~\ref{mclauses-vertexLB}.
(Informally, this means that the assignment selects $m$ distinct edges
and $2m+1$ distinct vertices.)
An edge $e \in {[3m] \choose 2}$ is said to be  {\em{bad  for $A$}}
if $e = \{u,v\}$ and there exist $i,j \in [2m+1], k \in [m]$ with
$A(y^i_u) =1$, $A(y^j_v) =1$, and $A(x^k_e) =1$.
\end{definition}

\begin{proposition}
If $A$ is a non-degenerate assignment to $MVars_m$, then there exists
an edge that is bad for $A$.
\end{proposition}

\begin{definition}\label{matchSearchDefn}
Let $m$ be a positive integer,  and let $\left({\cal{V}}_I, {\cal{V}}_{II}\right)$ be
a partition of $MVars_m$.
The search problem {\em{$FindBadEdge_m\left({\cal{V}}_I, {\cal{V}}_{II}\right)$}} is defined as follows:

\begin{enumerate}
\item Player I has private access to the variables of ${\cal{V}}_I$.
\item Player II has private access to the variables of ${\cal{V}}_{II}$.
\item Given a non-degenerate assignment $A$ to $MVars_m$, the players must find
a bad edge of $A$.
\end{enumerate}
\end{definition}

The partition $({\cal{V}}_I,{\cal{V}}_{II})$ 
of $MVars_m$ will play an important role in the quality of the
reduction from set-disjointness. We will see that
the larger the density of the partition, the larger the instances of
set-disjointness that can be reduced to 
$FindBadEdge_m({\cal{V}}_I,{\cal{V}}_{II})$. In particular,
when  $\delta\left( {\cal{V}}_I, {\cal{V}}_{II}\right) = \Omega(1)$,
$FindBadEdge_m({\cal{V}}_I, {\cal{V}}_{II})$ requires communication
$\Omega(m)$.

\begin{lemma}\label{searchLemma}
There a exists a constant $c>0$ so that for  all $m \ge 84651$,
if there is a size $S$ tree-like OBDD refutation of $IndMatch_m$ then there 
is a partition $({\cal{V}}_I,{\cal{V}}_{II})$ 
of $MVars_m$ so that
$\delta \left({\cal{V}}_I,{\cal{V}}_{II}\right) \ge 2^{-13}$ and
there exists a deterministic two-player protocol for the search problem
$FindBadEdge_m \left( {\cal{V}}_I,{\cal{V}}_{II} \right)$ that uses
at most $c \log S$ many bits of communication.
\end{lemma}

\subsection{The Proof of Lemma~\ref{searchLemma}}

The following lemma follows from standard arguments.
\begin{lemma}\label{fcs-lemma}(cf. \cite{ipu1994,akv2004})
There exists a constant $c>0$ so that for all $m$, 
and every partition $\left( {\cal{V}}_I,  {\cal{V}}_{II} \right)$  of 
$MVars_m$,
if there is treelike OBDD refutation of $Match_m$ of size $S$ that uses
a variable order in which either every variable of ${\cal{V}}_I$ precedes
every variable of ${\cal{V}}_{II}$,  or vice-versa,
then for each $i \in [n]$,  then there is a deterministic
two-player protocol for
$FindBadEdge_m \left( {\cal{V}}_{I}, {\cal{V}}_{II} \right)$
that uses at most $c\log S $ many bit of communication.
\end{lemma}

\begin{lemma}\label{part-lemma}
For $m \ge 84651$,
if there exists size $S$ refutation of $IndMatch_m$, then there
exists
a partition of $MVars_m$,  $\left({\cal{V}}_I,{\cal{V}}_{II}\right)$,
with $\delta({\cal{V}}_I,{\cal{V}}_{II}) \ge 2^{-13}$,
and a size $S$ refutation of $Match_m$  in which every variable of
${\cal{V}}_I$ precedes every variable of ${\cal{V}}_{II}$, or vice-versa.
\end{lemma}

\begin{proof}
Let $v_1, \ldots v_N$ be the variable ordering of $IMVars_m$ used by the
refutation of $IndMatch_m$.
Let $i_0$ be the first
position to split either the set of vertex variables or the set of
edge variables in half.  More formally,  
for each $i=1, \ldots N$,  let $vvars(i)$ be the number of vertex variables
in $\{v_1, \ldots v_i\}$, let $evars(i)$ be the number of edge
variables in $\{v_1, \ldots v_i\}$,
and let $i_0$ least integer with
either $evars(i_0) \ge \frac{m}{2} \cdot {3m \choose 2}$ or 
$vvars(i_0) \ge \frac{2m+1}{2} \cdot 3m$.  
Notice that there are two possible cases:  The first is that
$evars(i_0) \ge \frac{m}{2} \cdot {3m \choose 2}$ so that
 $\{v_1, \ldots v_{i_0}\}$ contains  exactly 
$\frac{m}{2} \cdot {3m \choose 2}$ many edge variables and 
$\{v_{i_0+1}, \ldots v_N\}$ contains at least $ \frac{1}{2} \cdot (6m^2 + 3m)$
many vertex variables. The second is that
$vvars(i_0) \ge \frac{2m+1}{2} \cdot 3m$ so that
 $\{v_1, \ldots v_{i_0}\}$ contains  exactly 
$ \frac{1}{2} \cdot (6m^2 + 3m)$ many vertex variables and 
$\{v_{i_0+1}, \ldots v_N\}$ contains at least
$\frac{m}{2} \cdot {3m \choose 2}$ many edge variables.
In the first case, we set ${\cal{V}}_I = \{v_1, \ldots v_{i_0} \}$ and
${\cal{V}}_{II} = \{ v_{i_0+1}, \ldots v_N\}$.  In the second case,
we set ${\cal{V}}_{II} = \{v_1, \ldots v_{i_0} \}$ and
${\cal{V}}_{I} = \{ v_{i_0+1}, \ldots v_N\}$.
In either case, 
$\frac{1}{m} \sum_{i=1}^m |E_i| \ge  \frac{1}{2} {3m \choose 2}$ and
 $\frac{1}{2m+1} \sum_{i=1}^{2m+1} |V_j| \ge \frac{3m}{2} $.
Therefore,  by Lemma~\ref{convexity-lemma}:
 $\frac{1}{m^2} \sum_{\vec{\imath} \in [m]^2} |E_{i_1} \cap E_{i_2}| \ge  \frac{1}{4} {3m \choose 2}$,  and 
$\frac{1}{(2m+1)^5} \sum_{\vec{\jmath} \in [2m+1]^5} |V_{j_1} \cap V_{j_2} \cap V_{j_3} \cap V_{j_4} \cap V_{j_5} | \ge \frac{3m}{32} $.

We now  calculate the expected value of
$\delta(\pi({\cal{V}}_I),\pi({\cal{V}}_{II}))$ over $\pi \in \Pi$.
We begin by noting that for all $i \in [m]$,
$E_i(\pi({\cal{V}}_I)) = E_i({\cal{V}}_I) = E_i$
and for all $j \in [2m+1]$,
$V_j(\pi({\cal{V}}_{II})) = \pi(V_j({\cal{V}}_{II})) = \pi(V_j)$.
For each $\vec{\imath} \in [3m]^2$,  let 
$E_{\vec{\imath}} = E_{i_1} \cap E_{i_2}$
and for each $\vec{\jmath} \in [2m+1]^5$,  let
$V_{\vec{\jmath}} = V_{j_1} \cap V_{j_2} \cap V_{j_3} \cap V_{j_4} \cap V_{j_5}$.
For each $\{u,v\} \in {[3m] \choose 2}$,  by the pairwise independence of the
permutations,  we have that:
\begin{eqnarray*}
 \prob_{\pi \in \Pi} \left[ \{\pi(u),\pi(v)\} \in E_{\vec{\imath}}  \right] & =& \sum_{\{a,b\} \in E_{\vec{\imath}}} \left(\prob_{\pi \in \Pi} \left[ \pi(u)=a, \ \pi(v)=b  \right] +\prob_{\pi \in \Pi} \left[ \pi(u) = b, \ \pi(v) =a \}\right] \right)\\ &=& \frac{2|E_{\vec{\imath}}|}{3m(3m-1)} =   \frac{|E_{\vec{\imath}}|}{{3m \choose 2 }}
\end{eqnarray*}

Therefore,  by linearity of expectation, we have that:
\[ \expect_{\pi \in \Pi} \left[ |E_{\vec{\imath}} \left[\pi \left( V_{\vec{\jmath}} \right) \right] | \right] =  \sum_{\{u,v\} \in { V_{\vec{\jmath}} \choose 2}} \prob_\pi \left[\{\pi(u),\pi(v)\} \in E_{\vec{\imath}}    \right]  =   \frac{|E_{\vec{\imath}}|}{{3m \choose 2}}  {|V_{\vec{\jmath}}| \choose 2}  \]

And thus we bound $\expect_{\pi \in \Pi} \left[ \delta(\pi({\cal{V}}_I,{\cal{V}}_{II}))  | \right] $ from below as follows:
\begin{eqnarray*}
\lefteqn{\expect_{\pi \in \Pi} \left[ \frac{1}{m^2(2m+1)^5}\sum_{\vec{\imath} \in [m]^2} \sum_{\vec{\jmath} \in [2m+1]^5} \frac{| \bigcap_{k=1}^5  E_{i_1}(\pi({\cal{V}}_I)) \left[ V_{j_k}(\pi({\cal{V}}_{II})) \right]  \cap   E_{i_2}(\pi({\cal{V}}_I)) \left[ V_{j_k}(\pi({\cal{V}}_{II})) \right] |}{ {3m \choose 2}} \right]} \\
 & = & \expect_{\pi \in \Pi} \left[ \frac{1}{m^2(2m+1)^5}\sum_{\vec{\imath} \in [m]^2} \sum_{\vec{\jmath} \in [2m+1]^5} \frac{| \bigcap_{k=1}^5  E_{i_1} \left[ \pi(V_{j_k}) \right]  \cap   E_{i_2} \left[ \pi(V_{j_k}) \right] |}{ {3m \choose 2}} \right] \\
 &= & \expect_{\pi \in \Pi} \left[ \sum_{\vec{\imath} \in [m]^2} \sum_{\vec{\jmath} \in [2m+1]^5} |E_{\vec{\imath}} \left[\pi\left( V_{\vec{\jmath}}\right) \right] | \right]  = 
 \sum_{\vec{\imath} \in [m]^2} \sum_{\vec{\jmath} \in [2m+1]^5}  \expect_{\pi \in \Pi} \left[ |E_{\vec{\imath}} \left[ \pi\left( V_{\vec{\jmath}}\right) \right] | \right] \\
 &=&  \sum_{\vec{\imath} \in [m]^2}   \frac{|E_{\vec{\imath}}|}{{3m \choose 2}} \sum_{\vec{\jmath} \in [2m+1]^5} {|V_{\vec{\jmath}}|  \choose 2}
  \ge  \sum_{\vec{\imath} \in [m]^2}   \frac{|E_{\vec{\imath}}|}{{3m \choose 2}}  (2m+1)^5  {3m/32 \choose 2}  \\
 &=&  (2m+1)^5 {3m/32 \choose 2} \sum_{\vec{\imath} \in [m]^2}     \frac{|E_{\vec{\imath}}|}{{3m \choose 2}} 
 \ge  (2m+1)^5{3m/32 \choose 2} m^2  \left( \frac{1}{4} \right)     \\
& = &  \frac{m^2(2m+1)^5}{4} \frac{(3m/32)(3m/32 -1)}{2}   =  \frac{m^2(2m+1)^5}{4\cdot (32)^2} \frac{(3m)(3m -32)}{2} \\
& = &  \frac{m^2(2m+1)^5}{4\cdot (32)^2} \left( {3m \choose 2} - \frac{(3m)(31)}{2}\right) =  \frac{m^2(2m+1)^5}{2^{12}} {3m \choose 2}\left(1  - \frac{31}{3m-1}\right) \\
& = & \left(2^{-12} - \frac{31}{3m-1}\right) m^2(2m+1)^5 {3m \choose 2} \end{eqnarray*}

Choose a permutation $\pi$  with
$ \sum_{\vec{\imath} \in [m]^2} \sum_{\vec{\jmath} \in [2m+1]^5} |E_{\vec{\imath}} \left[ \pi\left(V_{\vec{\jmath}}\right) \right] |  \ge \left(2^{-12} - \frac{31}{3m-1}\right) m^2(2m+1)^5 {3m \choose 2}  $.
By Lemma~\ref{permutation-lemma}, 
there is a size $S$ refutation of $Match_m$ that uses the variable
ordering $\pi(v_1), \ldots \pi(v_N)$. 
Notice that in this order,  either every variable of 
$\pi({\cal{V}}_I)$ precedes every variable of $\pi({\cal{V}}_{II})$, or
every variable of $\pi({\cal{V}}_{II})$ precedes every variable of
$\pi({\cal{V}}_I)$.  
By the above calculation,
$\delta(\pi({\cal{V}}_I), \pi({\cal{V}}_{II})) \ge 2^{-12} -\frac{31}{3m-1}$.
Because $m \ge 84651$, we have 
$\frac{31}{3m-1} \le 2^{-13}$, so
$\delta(\pi({\cal{V}}_I),\pi({\cal{V}}_{II})) \ge 2^{-12}- 2^{-13} = 2^{-13}$.

\end{proof}

To prove Lemma~\ref{searchLemma}, simply take the partition of $MVars_m$ and
the size $S$ refutation of $Match_m$ 
guaranteed by Lemma~\ref{part-lemma} and feed them
 into Lemma~\ref{fcs-lemma}.

\section{Reduction and Lower Bound}\label{reductionSect}

The correctness of the reduction from $setdisj_n$ to
$FindBadEdge_m({\cal{V}}_I,{\cal{V}}_{II})$
depends on the following lemma:

\begin{lemma}\label{key-lemma}(proof in Section~\ref{layoutSect})
For every $\delta > 0$,  there exist $c_0, c_1 > 0$ so that
for  all  $m \ge 31 (2/\delta)^8$,
and all partitions of $MVars_m$,  
$\left({\cal{V}}_I,{\cal{V}}_{II}\right)$ with $\delta({\cal{V}}_I,{\cal{V}}_{II})\ge \delta$,
for all $n$ with
$n \le c_0 m$, 
there exists
a set ${\cal{L}}$, a distribution ${\cal{D}}$ on ${\cal{L}}$ with measure
function $\mu$, a function 
$A : {\cal{L}} \times  \{0,1\}^n \times \{0,1\}^n \rightarrow \{0,1\}^{MVars_m}$, and
 a function $pe: {\cal{L}}  \rightarrow {[3m] \choose 2}$
so that:
\begin{enumerate}

\item\label{cond-compute} For all $L \in {\cal{L}}$, $(\vec{X},\vec{Y}) \in \{0,1\}^n \times \{0,1\}^n$, all $v \in {\cal{V}}_I$, $A_{L,\vec{X},\vec{Y}}(v)$ is determined 
by $L$ and $\vec{X}$, and for all $v \in {\cal{V}}_{II}$,
$A_{L,\vec{X},\vec{Y}}(v)$ is determined by $L$ and $\vec{Y}$.

\item\label{cond-nondeg} For all $L \in {\cal{L}}$, all 
$(\vec{X},\vec{Y}) \in \{0,1\}^n \times \{0,1\}^n$,
  the assignment $A_{L,{\vec{X}},\vec{Y}}$
is non-degenerate.

\item\label{cond-plant} For all 
$(\vec{X},\vec{Y}) \in \{0,1\}^n \times \{0,1\}^n$, and all 
$e \in {[3m] \choose 2}$,  if $e$ is bad for 
$A_{L,\vec{X},\vec{Y}}$,  then $e= pe(L)$ or
$setdisj_n(\vec{X},\vec{Y})=1$.

\item\label{cond-weird} For all
$(\vec{X},\vec{Y}) \in \{0,1\}^n \times \{0,1\}^n$ with 
$setdisj_n(\vec{X},\vec{Y})=1$, there exists ${\cal{S}} \subseteq {\cal{L}}$
with $\mu ({\cal{S}}) \ge \delta^8/2^9$ so that
for all
$A \in \{A_{{L},\vec{X},\vec{Y}} \mid {L} \in {\cal{S}} \}$:
\[ \max_{e \in {[3m] \choose 2} } \mu ( pe(L) =e \mid A_{{L},\vec{X},\vec{Y}}= A, \ {L} \in {\cal{S}} ) \le 1 - c_1 \]
\end{enumerate}

\end{lemma}
It is helpful to think of $L \in {\cal{L}}$ as a ``layout'' guiding the
construction of an $MVars_m$ assignment from $\vec{X}$, $\vec{Y}$. 
$A_{L,\vec{X},\vec{Y}}$ is simply the assignment constructed using layout $L$
with set-disjointness instance $(\vec{X},\vec{Y})$.  
Condition~\ref{cond-compute} 
is the requirement that the Player I can compute the value of $A_{L,\vec{X},\vec{Y}}(v)$ for $v \in {\cal{V}}_I$ without communicating with Player II,
and that player II can compute $A_{L,\vec{X},\vec{Y}}(v)$ for
$v \in {\cal{V}}_{II}$ without communication.
  Condition~\ref{cond-nondeg} guarantees that the assignment
created is a valid instance
of the $FindBadEdge_m({\cal{V}}_I,{\cal{V}}_{II})$ problem.
The function $pe$ can be 
thought of as a ``planted bad edge'': The reduction is based on the idea of
having positions with $X_k=Y_k=1$ create bad edges.  However,  because
the assignment is nondegenerate, there must always be some bad edge,
even when $setdisj_n(\vec{X},\vec{Y})=0$. The players knowingly
create one such edge and we call this edge the
planted edge for the layout,  $pe(L)$. Condition~\ref{cond-plant} states that
 when $setdisj_n(\vec{X},\vec{Y})=0$, the only bad edge is the planted edge. Condition~\ref{cond-weird} states that when 
$setdisj_n(\vec{X},\vec{Y})=1$,
conditioned on the layout coming from
the set ${\cal{S}}$, no assignment is overly-correlated with
a particular planted edge.

\begin{lemma}\label{reduction-lemma}
For all $\delta > 0$, there exist $C_0, C_1>0$ so that
for all $m \ge 31(2/\delta)^8$,
for all partitions of $MVars_m$, $({\cal{V}}_I,{\cal{V}}_{II})$,
with $\delta({\cal{V}}_I,{\cal{V}}_{II}) \ge \delta$,
 for all $n \le C_0 m$,
if there is a two-player deterministic protocol $SEARCH$ that solves
$FindBadEdge_m({\cal{V}}_I,{\cal{V}}_{II})$ using $r$ bits of communication,
then the randomized communication complexity of $setdisj_n$  
is $\le C_1 r$.
\end{lemma}
\begin{proof} 
Let $C_0$ be the  $c_0$ as in the statement of Lemma~\ref{key-lemma}.
We give a one-sided
reduction that never gives a wrong answer when
$setdisj_n({\vec{X}},{\vec{Y}})=0$, and 
when $setdisj_n({\vec{X}},{\vec{Y}})=1$, it gives the correct answer with
probability $\ge c_1 \delta^8/2^9$, where $c_1$ is the second constant
guaranteed by Lemma~\ref{key-lemma}.  
Repeating the protocol a constant number of times
and returning a $0$ only if all
runs produce  a $0$ gives a protocol with correctness $\ge 2/3$.

\begin{enumerate}

\item Using public randomness, the players
select a reduction layout $L$ according to the distribution ${\cal{D}}$
guaranteed by Lemma~\ref{key-lemma}.

\item  The players run the protocol $SEARCH$ using the
assignment $A_{{L},\vec{X},\vec{Y}}$ and
let $e$ be the edge returned by the protocol $SEARCH$.

\begin{enumerate}
\item If $pe(L) = e$ then return $0$.
\item If $pe(L) \neq e$ then return $1$.
\end{enumerate}
\end{enumerate}

By Lemma~\ref{key-lemma}, Condition~\ref{cond-compute}, 
the players can compute the
needed values of $A_{L,\vec{X},\vec{Y}}$
with no communication.  By Lemma~\ref{key-lemma}, Condition~\ref{cond-nondeg},
the assignment $A_{L,\vec{X},\vec{Y}}$ is non-degenerate, and is therefore
a legal input for the problem $FindBadEdge_m({\cal{V}}_I,{\cal{V}}_{II})$.
Consider the case when $\vec{X}$ and $\vec{Y}$ are disjoint. 
By Lemma~\ref{key-lemma}, Condition~\ref{cond-plant},  the only bad edge in
$A_{{L},\vec{X},\vec{Y}}$ is $pe(L)$, 
so the protocol returns  $0$.
Consider the case when $\vec{X}$ and $\vec{Y}$ are intersecting. 
Apply Lemma~\ref{key-lemma}, Condition~\ref{cond-weird},
and let ${\cal{S}}$ be the set guaranteed
for the pair $\vec{X}$, $\vec{Y}$.
Define the event ${\cal{B}}$ as 
${\cal{B}} = \{ {L} \in {\cal{S}} \mid SEARCH(A_{{L},\vec{X},\vec{Y}}) = pe(L)\}$.
This is the
event that the layout belongs to ${\cal{S}}$ and  the protocol gives an
erroneous answer.
Let $A_{\cal{S}} = \{ A_{{L},\vec{X},\vec{Y}} \mid {L} \in {\cal{S}} \}$.  For each $A \in A_{\cal{S}}$,  let 
${\cal{S}}_A = \{ {L} \in {\cal{S}} \mid A_{{L},\vec{X},\vec{Y}} = A\}$ and let
${\cal{B}}_A = \{ {L} \in {\cal{B}} \mid A_{{L},\vec{X},\vec{Y}} = A\}$.   Because the protocol $SEARCH$ is deterministic, for 
each $A$  on the set ${\cal{B}}_A$, the function 
${L} \mapsto pe(L)$ is the constant function
taking the value
returned by $SEARCH(A)$.
Therefore,  by Lemma~\ref{key-lemma}, Condition~\ref{cond-weird},
for each $A \in A_{\cal{S}}$,  
$\mu({\cal{B}}_A) \le (1-c_1)\mu({\cal{S}}_A)$, and so:
\[\mu({\cal{B}})  =  \sum_{A \in A_{\cal{S}}} \mu({\cal{B}}_A)
  \le  \sum_{A \in A_{\cal{S}}} \mu({\cal{S}}_A) (1-c_1)  =  (1-c_1) \mu({\cal{S}})  \]

Therefore $ \mu({\cal{S}} \setminus {\cal{B}}) \ge c_1 \mu({\cal{S}}) \ge 
c_1 \delta^8/2^9$.
Of course,   ${\cal{S}} \setminus {\cal{B}}$ is  the event that
$L \in {\cal{S}}$ and the protocol gives the answer $1$. 

\end{proof}

\subsection{The Lower Bound}\label{lowerBoundSubSect}

\begin{theorem}\label{payoff} There exists a constant $C > 0$ so that for 
sufficiently large $m$,  every tree-like OBDD refutation of 
$IndMatch_m$ has size at least $2^{C m }$.
\end{theorem}
\begin{proof}
Apply Theorem~\ref{sdBound} and 
choose $N \ge 0$ and $c^* > 0$ so that for every $n \ge N$,
randomized two-player protocols for solving $setdisj_n$
require $\ge c^*n$ bits of communication.
Let $C_0$ and $C_1$ be the constants of Lemma~\ref{reduction-lemma},
and let $m$ be so large that
$m \ge 31 (2/ (2^{-13}))^8 = 31 \cdot 2^{112}$ (so that we
can apply Lemma~\ref{reduction-lemma} with $\delta \ge 2^{-13}$),
and
$N \le \lfloor C_0 m \rfloor $ (so that we can apply Theorem~\ref{sdBound}).
Set $n=  \lfloor C_0 m \rfloor $.
Let $c>0$ be the constant from Lemma~\ref{searchLemma}.
Let $\Gamma$ be a tree-like OBDD 
refutation of $IndMatch_m$ of size $S$.
Because $m  > 84651$,
we may apply Lemma~\ref{searchLemma} and choose a partition
$({\cal{V}}_I,{\cal{V}}_{II})$  so that 
$\delta({\cal{V}}_I,{\cal{V}}_{II})  \ge 2^{-13}$ and a two-player 
deterministic communication protocol
$FindBadEdge_m({\cal{V}}_I,{\cal{V}}_{II})$ that uses at most 
$c\log S$ bits of
communication.  
By Lemma~\ref{reduction-lemma},  there
is a two-party randomized communication
protocol for $setdisj_n$ on inputs from ${\cal{P}}_n$
that exchanges at most $C_1 \log S$ bits of 
communication.
Therefore, applying the communication bound for set-disjointness,
$ C_1 \log S \le  c^*n = c^* \lfloor C_0 m \rfloor $, and thus
$S \le 2^{\frac{c^* \lfloor C_0 m \rfloor}{C_1}}$
\end{proof}

\section{Reduction Layouts}\label{layoutSect}

The reduction from set-disjointness by randomly
generates ``reduction layouts''. A reduction layout is a framework
for generating instances of the search problem from instances of 
set-disjointness, a collection of gadgets.
We now take a moment to discuss the gadgets underlying the reduction
from set-disjointness to the problem of finding a bad edge.

\begin{figure}
\begin{center}
\epsfig{file=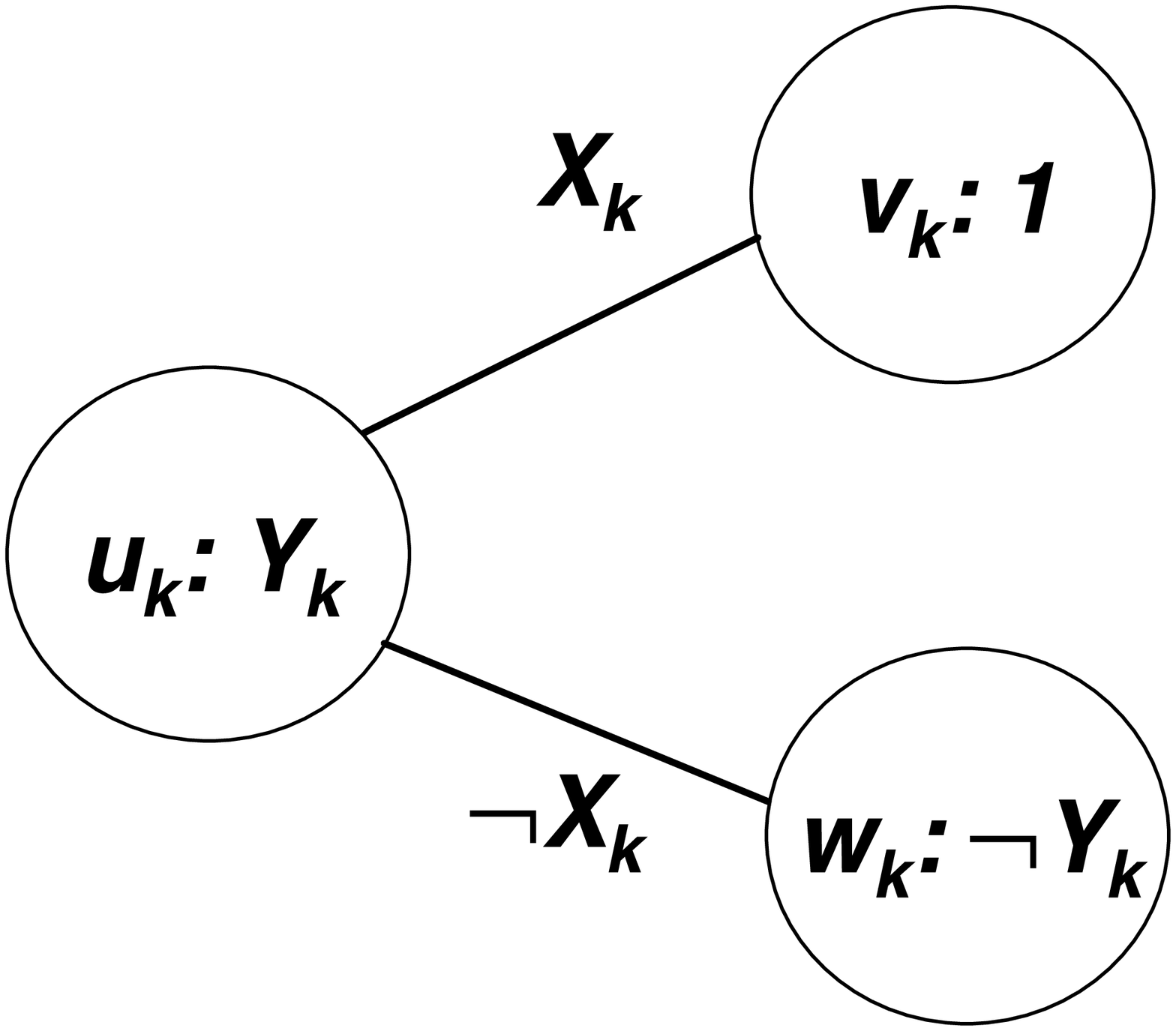,height=3cm}
\caption{\label{gadget-fig}The basic set-disjointness gadget.
A bad edge corresponds to the situation when an edge and both of
its endpoints receive the label $1$.
The assignment uses: $x^{i_k}_{\{u_k,v_k\}} = X_k$, 
$x^{i_k}_{\{u_k,w_k\}} = \neg X_k$, 
$y^{j_{k,1}}_v = 1$, $y^{j_{k,2}}_{u_k} = Y_k$, and $y^{j_{k,2}}_{w_k} = \neg Y_k$.
Notice that $\{u_k,w_k\}$ is never
a bad edge, and that  $\{u_k,v_k\}$ is a bad edge if and only if
$X_k=Y_k=1$. } \end{center} \end{figure}

The basic idea is to create a bad edge for each $k$ with
$X_k=Y_k=1$. To do this without communicating,
 the players use the public randomness to choose
$u_k,v_k,w_k \in [3m]$ with the intent to place $\{u_k,v_k\}$
in the matching if
$X_k=1$ and $\{u_k,w_k\}$ in the matching if $X_k=0$,  and to place
$v_k$ in the independent no matter what, but to include $u_k$ if $Y_k=1$
and to include $w_k$ if $Y_k=0$. 
Of course, we must specify which variables are used to place the
gadget, and those variables must be available to the players
under the partition.
The players  use the public randomness
to choose $i_k \in [m]$  with 
$x^{i_k}_{\{u_k,v_k\}}$, $x^{i_k}_{\{u_k,w_k\}} \in {\cal{V}}_I$ 
(equivalently,  $\{u_k,v_k\}, \{u_k,w_k\} \in E_{i_k}$)  and
$j_{k,1}, j_{k,2} \in [m]$ with 
$y^{j_{k,1}}_{v_k},  y^{j_{k,2}}_{u_k}, y^{j_{k,2}}_{w_k} \in {\cal{V}}_{II}$, 
(equivalently,  $v_k \in V_{j_{k,1}}$ and $u_k,w_k \in V_{j_{k,2}}$). 
The situation  resembles that in
Figure~\ref{gadget-fig}, with a bad edge occurring only if $X_k=Y_k=1$
and only then only at $\{u_k,v_k\}$. The reduction plants one of these
gadgets for each $k=1, \ldots n$.

Because there are $m$ edges in the matching
and $2m+1$ vertices in the set,  one more vertex must be placed
in addition to the two associated with each set-disjointness gadget.
A final gadget (thought of as being at position $n+1$)
will contain the ``planted bad edge'', in which three vertices $u_{n+1}$,
$v_{n+1}$, and $w_{n+1}$ are all placed in the set, and the edge
$\{u_{n+1},w_{n+1}\}$ is included. Because all three vertices are placed
in the set,  three variables $y^{j_{n+1,1}}_{u_{n+1}}$, $y^{j_{n+1,2}}_{v_{n+1}}$ and
$y^{j_{n+1,3}}_{w_{n+1}}$ are needed with $u_{n+1} \in V_{j_{n+1,1}}$,
$v_{n+1} \in V_{j_{n+1,2}}$, and $w_{n+1} \in V_{j_{n+1,3}}$.

\begin{figure}
\begin{center}
\epsfig{file=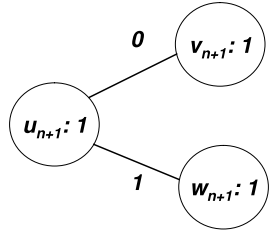,height=3cm}
\caption{\label{gadgetPlant-fig} The set-disjointness gadget at the
position with a planted bad edge. All three vertices $u_{n+1},v_{n+1},w_{n+1}$
are placed in the set of vertices and the edge $\{u_{n+1},w_{n+1}\}$ is placed
in the set of edges.  The edge $\{u_{n+1},w_{n+1}\}$ is
a bad edge.
The assignment uses: $x^{i_{n+1}}_{\{u_{n+1},v_{n+1}\}} = 0$, 
$x^{i_{n+1}}_{\{u_{n+1},w_{n+1}\}} = 1$, 
$y^{j_{n+1,1}}_{u_{n+1}} = 1$, $y^{j_{n+1,2}}_{v_{n+1}} = 1$,  $y^{j_{n+1,3}}_{w_{n+1}} = 1$.
} \end{center} \end{figure}

The basic idea of the reduction is to
randomly plant  these $n+1$ gadgets on disjoint variables. However, to 
ensure that the probabilities work out as claimed in Lemma~\ref{key-lemma},
we make use of the density of the partition.

\begin{definition}\label{gdefn}
Fix a partition of $MVars_m$,  $({\cal{V}}_I,{\cal{V}}_{II})$.  Set
$\delta = \delta({\cal{V}}_I,{\cal{V}}_{II})$.
For each $i \in [m]$ let $E_i = E_i({\cal{V}}_I)$ and for each
$j \in [2m+1]$ let $V_j = V_j({\cal{V}}_{II})$.
For each $i \in [m]$,  let $N_3(i) = \{ (j_1,j_2,j_3) \in [2m+1]^3 \mid j_1\neq j_2, \ j_2 \neq j_3, \ j_3 \neq j_1, \ |E_i [V_{j_1} \cap V_{j_2} \cap V_{j_3}]| \ge (\delta/3){3m \choose 2} \}$,  and let
$N_2(i) = \{ (j_1,j_2) \in [2m+1]^2 \mid \exists j_3 \in [2m+1], \ (j_1,j_2,j_3) \in N_3(i)\}$.
Set $G = \{i \in [m] \mid |N_3(i)| \ge (\delta/12) (2m+1)^3 \}$.
Of course, each of $G$, $N_3(\cdot)$, and $N_2(\cdot)$ depend upon 
the partition $({\cal{V}}_I,{\cal{V}}_{II})$, but we drop that from
notation as we will never discuss more than one partition at a time.
\end{definition}

\begin{lemma}\label{lemma-gdense}(Proof in Appendix, Section~\ref{APP-layoutSect})
Let $\delta \in [0,1]$ and let $m$ be an integer
$ \ge 3/\delta$.
Let $({\cal{V}}_I,{\cal{V}}_{II})$ be a  partition of $MVars_m$ with
$\delta({\cal{V}}_I,{\cal{V}}_{II}) \ge \delta$.
$|G| \ge \left(\delta/12\right)  m$
\end{lemma}

\begin{definition}\label{defn-reduction-layout}
Fix an integer $m$, a partition $\left( {\cal{V}}_I, {\cal{V}}_{II}\right)$
of $MVars_m$..  A {\em{reduction layout 
 (with respect to  $({\cal{V}}_I, {\cal{V}}_{II})$, of length $n$)}}
is a tuple 
$(i_1, \ldots i_{n+1}, (j_{1,1},j_{1,2}), \ldots (j_{n,1},j_{n,2}),(j_{n+1,1},j_{n+1,2},j_{n+1,3}),$ $(u_1,v_1,w_1), \ldots (u_{n+1},v_{n+1},w_{n+1}))$  
 from the set
$[m]^{n+1} \times ([2m+1]^2)^n \times ([2m+1]^3) \times \left([3m]^3\right)^n$
with the following properties:
\begin{enumerate}

\item\label{ld-idist}The indices $i_1, \ldots i_{n+1}$ are distinct.

\item\label{ld-jdist} The indices $ j_{1,1}, j_{1,2}, \ldots j_{n,1}, j_{n,2},j_{n+1,1}, j_{n+1,2},j_{n+1,3}$ are 
distinct.

\item\label{ld-uvwdist} The integers $u_1, \ldots u_{n+1}, v_1, \ldots v_{n+1}, w_1, \ldots w_{n+1}$ are distinct.

\item\label{ld-edges} For each $k=1, \ldots n+1$,
$\{u_k, v_k\} \in E_{i_k}$ and  $\{u_k, w_k\} \in E_{i_k}$.

\item\label{ld-uvwV1V2} For each $k=1, \ldots n+1$,
 $u_k, v_k, w_k \in V_{j_{k,1}} \cap V_{j_{k,2}}$.

\item\label{ld-plant-j0} $u_{n+1},v_{n+1},w_{n+1} \in V_{j_{{n+1},1}} \cap V_{j_{{n+1},2}} \cap V_{j_{{n+1},3}}$.

\item\label{ld-iG} For all $k \in [n+1]$, 
$i_k \in G$.

\item\label{ld-pnbrs} $(j_{{n+1},1},j_{{n+1},2},j_{{n+1},3}) \in N_3(i_{n+1})$

\item\label{ld-jnbrs}For $k \in [n]$,  each $(j_{k,1},j_{k,2}) \in N_2(i_k)$.

\end{enumerate}

The set of all reduction layouts of length $n$ 
with respect to $({\cal{V}}_I,{\cal{V}}_{II})$ 
is denoted  ${\cal{L}}_{m,n} ({\cal{V}}_I,{\cal{V}}_{II})$.  When $m$, $n$,
and $({\cal{V}}_I,{\cal{V}}_{II})$ are clear from context, we simply
write ${\cal{L}}$ and call $L \in {\cal{L}}$ a {\em{reduction layout}}.
\end{definition}
When listing the elements of a reduction layout, we will abuse notataion write
$(\vec{\imath},\vec{\jmath},\vec{u},\vec{v},\vec{w})$ despite the fact that
a reduction layout is emphatically {\em{not}} a member of the  set
$[m]^{n+1} \times [2m+1]^{2n+3} \times [3m]^{n+1} \times [3m]^{n+1} \times [3m]^{n+1}$. 
This matters for the purpose of computing Hamming distances.
The Hamming distance between two reduction layouts in ${\cal{L}}$
is their Hamming distance as elements of the $3n+3$ ``dimensional''
product set
$[m]^{n+1} \times ([2m+1]^2)^n \times ([2m+1]^3) \times \left([3m]^3\right)^{n+1}$.
In particular,  if two reduction layouts 
${L}=(\vec{\imath},\vec{\jmath},\vec{u},\vec{v},\vec{w})$ and
${L}^* = (\vec{\imath}^*,\vec{\jmath}^*,\vec{u}^*,\vec{v}^*,\vec{w}^*)$ differ in only that $(u_k,v_k,w_k)\neq (u_k^*,v_k^*,w_k^*)$ then they
are at Hamming distance $1$.

\begin{definition}\label{defn-assignment}
Fix $m,n$, a partition $({\cal{V}}_I,{\cal{V}}_{II})$ of $MVars_m$.
Let ${L}=\left(\vec{\imath},\vec{\jmath},\vec{u},\vec{v},\vec{w}\right)$ be a reduction layout from ${\cal{L}}$,
and let $X_1, \ldots X_n, Y_1, \ldots Y_n$ be a
set-disjointness instance.   We define an assignment 
$A_{{L}, \vec{X}, \vec{Y}}$
to the variables of $MVars_m$ as follows:
Set $I = \{i_1, \ldots i_{n+1} \}$. 
Set $J = \{j_{1,1}, j_{1,2}, \ldots j_{n,1}, j_{n,2}, j_{n+1,1}, j_{n+1,2}, j_{n+1,3} \}$.
Set $V = \{u_1, \ldots u_{n+1}, v_1, \ldots v_{n+1}, w_1, \ldots w_{n+1} \}$.
Let $\beta$, $\beta\left(  {L} \right)$,
be the lexicographically first assignment to the variables
$\{ x^i_e \mid i \in [m] - I, \ e \in [3m-V]^2 \}$ $ \cup \{y^j_u \mid j \in [2m+1] - J, \ u \in [3m]-V \}$
so that $\beta$ defines a matching of size $m-n-1$ and an independent set of
size $2(m-n-1)$. Define $A_{L,\vec{X},\vec{Y}}$ as follows:

\begin{eqnarray*}
A_{{L}, \vec{X}, \vec{Y}}(x^i_e) & = & \left\{ \begin{array}{cc}
\beta(x^i_e) & {\mbox{ if $i \in [m]-I$ and $e \in \left([3m] -V\right)^2$}} \\
X_k     & {\mbox{ if $i=i_k$ and $e = \{u_k,v_k\}$ for some $k \in [n]$ }}\\
\neg X_k & {\mbox{ if $i=i_k$ and $e = \{u_k,w_k\}$ for some $k \in [n]$ }}\\
1 & {\mbox{ if $i= i_{n+1}$ and $e = \{u_{n+1},w_{n+1}\}$}} \\
0  &  {\mbox{ otherwise}}
\end{array}\right.
\end{eqnarray*}

\begin{eqnarray*}
A_{{L}, \vec{X}, \vec{Y}}(y^j_x) & = & \left\{ \begin{array}{cc}
\beta(y^j_x) & {\mbox{ if $j \in [2m+1]-j$ and $u \in [3m] -V$}} \\
1    & {\mbox{if  $j = j_{k,1}$ and $x = v_k$ for some $k \in [n] $}} \\
 Y_k  & {\mbox{if  $j = j_{k,2}$ and $x= u_k$ for some $k \in [n]$}} \\
\neg Y_k  & {\mbox{if  $j = j_{k,2}$ and $x= w_k$ for some $k \in [n]$}} \\
1  & {\mbox{if  $j = j_{n+1,1}$ and $x= u_{n+1}$}} \\
1  & {\mbox{if  $j = j_{n+1,2}$ and $x= v_{n+1}$}} \\
1  & {\mbox{if  $j = j_{n+1,3}$ and $x= w_{n+1}$}} \\
0  &  {\mbox{ otherwise}}
\end{array}\right.
\end{eqnarray*}

\end{definition}

Notice that when both players have access to the layout ${L}$,  condition~\ref{ld-edges} of Definition~\ref{defn-reduction-layout} ensures that
Player I can compute the assignment to all variables in ${\cal{V}}_I$ by only consulting his private
set-disjointness variables,  and conditions~\ref{ld-uvwV1V2} and~\ref{ld-plant-j0} similarly guarantee
that Player can compute the assignment to all variables in ${\cal{V}}_{II}$ by only consulting his
private set-disjointness variables. This accounts for 
Condition~\ref{cond-compute} of Lemma~\ref{key-lemma}.
The conditions~\ref{ld-idist}, \ref{ld-jdist} and~\ref{ld-uvwdist} 
of Definition~\ref{defn-reduction-layout} ensure
that $A_{{L},\vec{X},\vec{Y}}$ is well-defined and non-degenerate.  This
accounts for Condition~\ref{cond-nondeg} of Lemma~\ref{key-lemma}.

\begin{definition}\label{defn-pe}
Let $m$ and $n$ be given.
Let $\left({\cal{V}}_I,{\cal{V}}_{II}\right)$ be a variable partition for 
$MVars_m$.
Let $\vec{X}$, $\vec{Y}$ be a set-disjointness instance, and let
${L}= \left(\vec{\imath},\vec{\jmath},\vec{u},\vec{v},\vec{w} \right)$
be a reduction layout from ${\cal{L}}_{m,n}$.
The {\em{planted edge for $\vec{X}, \vec{Y}, {L}$,}}  $pe(L)$,
is defined to be $\{u_{n+1},w_{n+1}\}$.
\end{definition}

Condition~\ref{cond-plant} of Lemma~\ref{key-lemma} is the content of the
following lemma.
\begin{lemma}\label{layout-plant-lemma}(Proof in Appendix Section~\ref{APP-layoutSect})
Let ${L}=(\vec{\imath},\vec{\jmath},\vec{u},\vec{v},\vec{w})$
be a reduction layout.
If $e$ is a bad edge of $A_{{L},\vec{X},\vec{Y}}$ then 
$e = pe(L)$, or,  $e = \{u_l, v_l\}$ with $X_l= Y_l=1$.
\end{lemma}

\section{The Distribution on Reduction Layouts}\label{distributionSect}

There is a technical point that we defer
until after we describe the distribution:  
Why the experiment cannot ``get stuck'' and find itself in a position of
attempting to choose an item from an empty set. For $n$ a sufficiently
small constant fraction of $m$,
this is ruled out by some
calculations that follow the description of the experiment.
In the process that generates the distribution, we use the following
auxiliary definitions:
\begin{definition}
Let $E$ be a set of edges over $[3m]$, and define
${\cal{K}}_{1,2}(E) := \{ (u,v,w) \in [3m]^3 \mid v \neq w, \ \{u,v\} \in E, \ \{u,w\} \in E \}$.  Let $X$ be a set. For $U \subseteq X$ define
$pm_X(U) : \{ (u,v) \in X^2 \mid \{u,v\} \cap U \neq \emptyset \}$ and
$tm_X(U) := \{ (u,v,w) \in X^3 \mid \{u,v,w\} \cap U \neq \emptyset \}$.
(The mnemonic for this notation is ``pairs over $X$ that meet $U$'' and
``triples over $X$ that meet $U$''.)
\end{definition}

\begin{definition}\label{Ddefn}
Let $({\cal{V}}_I,{\cal{V}}_{II})$ be a variable partition for
$MVars_m$. Let $G$, $N_3(\cdot)$, and $N_2(\cdot)$ be as in 
Definition~\ref{gdefn}.
The distribution ${\cal{D}}$ on
${\cal{L}}$
is given by the following experiment:

\begin{enumerate}
\item For each $k=1, \ldots n+1$:  Choose $i_k$ from $G \setminus \{i_1, \ldots i_{k-1}\}$.

\item Set $J = \emptyset$.

\item For each $k=1, \ldots n$:

\begin{enumerate}
\item  Uniformly choose $(j_{k,1}, j_{k,2})$ from
$N_2(i_k) \setminus pm_{[2m+1]}(J)$
\item Set $J := J \cup \{ j_{k,1}, j_{k,2} \}$
\end{enumerate}

\item  Uniformly choose $(j_{n+1,1}, j_{n+1,2},j_{n+1,3})$ from
$N_3(i_{n+1}) \setminus tm_{[2m+1]}(J)$

\item Set $J := J \cup \{ j_{n+1,1}, j_{n+1,2}, j_{n+1,3} \}$

\item Set $V^* = \emptyset$.

\item\label{exp-getuvws}  For each $k=1, \ldots n$:

\begin{enumerate}

\item   Uniformly choose $(u_k,v_k,w_k)$ 
from ${\cal{K}}_{1,2}(E_{i_k}\left[\left( V_{j_{k,1}} \cap V_{j_{k,2}}\right)\right]) \setminus tm_{[3m]}(V^*)$.
\item Set $V^* = V^* \cup \{u_k,v_k,w_k\}$.

\end{enumerate}

\item   Uniformly choose $(u_{n+1},v_{n+1},w_{n+1})$ 
from ${\cal{K}}_{1,2}(E_{i_{n+1}}\left[\left( V_{j_{{n+1},1}} \cap V_{j_{{n+1},2}} \cap V_{j_{{n+1},3}} \right)\right]) \setminus tm_{[3m]}(V^*)$.

\item Return the layout $(\vec{\imath},\vec{\jmath},\vec{u},\vec{v},\vec{w})$.

\end{enumerate}
\end{definition}

\begin{proposition}\label{distFact}
For all ${L} \in {\cal{L}}$,
 $\mu({L})>0$.
\end{proposition}

The above proposition can  be checked by iteratively noting that
when we condition on the experiment producing a prefix of ${L}$,
the probability that it selects the next coordinate of ${L}$ is
non-zero.

The results of the following lemma guarantee that when $\gamma$ is sufficiently
small with respect to $\delta$, the experiment does not ``get stuck''.
The proof is in the Appendix.

\begin{lemma}\label{bountiful-lemma}
Let $\delta \in [0,1]$ and let $m$ be an integer  $\ge 450/\delta^2$.
Let $({\cal{V}}_I,{\cal{V}}_{II})$ be a partition of $MVars_m$ with
$\delta({\cal{V}}_I,{\cal{V}}_{II}) \ge \delta$.  Let $n$  given with
 $\gamma = \frac{n+1}{m}$.
For all runs of the experiment in Definition~\ref{Ddefn}, and for each $k=1, \ldots n$:
\begin{enumerate}
\item $|G \setminus \{i_1, \ldots i_{k-1} \} | > ((\delta/12)  - \gamma )m$.

\item\label{bountiful-neighbors} $|N_2(i_k) \setminus pm_{[2m+1]}(J) \ge ((\delta/3) - 2\gamma) (2m+1)^2$

\item $|N_3(i_{n+1}) \setminus tm_{[2m+1]}(J)| \ge ((\delta/3)  - 3\gamma) (2m+1)^3$

\item\label{bountiful-triples} $|{\cal{K}}_{1,2}\left( E_{i_k} \left[V_{j_{k,1}} \cap V_{j_{k,2}} \right] \right) \setminus tm_{[3m]}(V^*)| \ge (\delta^2/10- 3\gamma)(3m)^3$

\item $|{\cal{K}}_{1,2}\left( E_{i_{n+1}} \left[V_{j_{{n+1},1}} \cap V_{j_{{n+1},2}} \cap V_{j_{{n+1},3}} \right] \right) \setminus tm_{[3m]}(V^*)| \ge (\delta^2/10- 3\gamma)(3m)^3$

\end{enumerate}
\end{lemma}

The following two statements are used to prove Lemma~\ref{key-lemma}.
Their proofs depend upon calculations regarding the distribution 
${\cal{D}}$ , and seem to be best put in the framework of
 ``distributions from dependent
domains processes with blocking''.

\begin{definition}\label{defn-lswitch}
A reduction layout
 ${L} = (\vec{\imath},\vec{\jmath},\vec{u},\vec{v},\vec{w})$ 
is said to be {\em{$l$-switchable}} if 
$(j_{{n+1},2},j_{l,1},j_{l,2}) \in N_3(i_l)$ and
$K(\{u_{n+1},u_l\},\{v_{n+1},v_l,w_{n+1},w_l\}) \subseteq  E_{i_{n+1}}[V_{j_{{n+1},1}} \cap V_{j_{{n+1},2}} \cap V_{j_{{n+1},3}} ] \cap E_{i_l} [ V_{j_{l,1}} \cap V_{j_{l,2}} ]$.
Let ${\cal{S}}^{l}$
denote the set of $l$-switchable reduction layouts from ${\cal{L}}$.
\end{definition}

\begin{lemma}\label{switchable-density} (``Completeness lemma'', proof in Section~\ref{analyzeDSect})
For all $\delta > 0$,  for all $m \ge 31 (2/\delta)^8$,
 all partitions $({\cal{V}}_I,{\cal{V}}_{II})$ of
$MVars_m$ with $\delta({\cal{V}}_I,{\cal{V}}_{II}) \ge \delta$,
for all $n \le \frac{\delta^{10}}{2^{10} \cdot 3 \cdot 5^2} m$, 
for all $l \in [n]$,  $\mu({\cal{S}}^l) \ge \delta^8/2^9$.
\end{lemma}

\begin{lemma}\label{continuity-lemma} (``Continuity lemma'', proof in Section~\ref{analyzeDSect})
For every $\delta > 0$  for  every  integer  $d \ge 1$
for all  $m \ge 450/\delta^2$,
for all partitions $({\cal{V}}_I,{\cal{V}}_{II})$  of $MVars_m$
with $\delta({\cal{V}}_I,{\cal{V}}_{II}) \ge \delta$,
for all $n \le (\delta^2/60) m$,  
for all reduction layouts ${L},{L}^* \in {\cal{L}}$
with $HD({L},{L}^*) \le k$, 
$ \mu(L^*) \ge (\delta^2/20)^{2d}e^{-3d} \cdot \mu(L)$.
\end{lemma}

\subsection{The Proof of Lemma~\ref{key-lemma}}

To prove Lemma~\ref{key-lemma} we use the following helper lemma.
\begin{lemma}\label{lemma-switcheroo}(Proof immediately follows that of
Lemma~\ref{key-lemma}.)
For all $\delta > 0$, 
all  $m \ge 450/\delta^2$,  all partitions $({\cal{V}}_I,{\cal{V}}_{II})$
of $MVars_m$ with $\delta({\cal{V}}_I,{\cal{V}}_{II}) \ge \delta$, all
$n \le (\delta^2/20) m$, and all set-disjointness instances
$(\vec{X}$,$\vec{Y})$,
there exists an involution $f : {\cal{S}}^{l} \rightarrow {\cal{S}}^{l}$
so that for all ${L} \in {\cal{S}}^{l}$,
$A_{{L},\vec{X},\vec{Y}} = A_{f({L}), \vec{X},\vec{Y}}$,
$pe(f(L)) \neq pe(L)$, and
$\mu(f({L})) \ge  \mu({L})  (\delta^2/20)^{12}e^{-18}$.
\end{lemma}

\begin{proof}(of Lemma~\ref{key-lemma} from Lemma~\ref{lemma-switcheroo})
Let $\delta >0$ be given.  
Set $c_0  = \frac{\delta^{10}}{2^{10} \cdot 3 \cdot 5^2} m$
Let $m \ge 31 (2/\delta)^8$
and $n \le c_0$ be given.
Let
$({\cal{V}}_I,{\cal{V}}_{II})$ be a partition of $MVars_m$ with
$\delta({\cal{V}}_I,{\cal{V}}_{II}) \ge \delta$. 
We take
 ${\cal{L}} = {\cal{L}}_{m,n}({\cal{V}}_I,{\cal{V}}_{II})$ 
per Definition~\ref{defn-reduction-layout},
${\cal{D}} = {\cal{D}}_{m,n}({\cal{V}}_I,{\cal{V}}_{II})$  per
Definition~\ref{Ddefn}, 
$A : (L,\vec{X},\vec{Y}) \rightarrow A_{L,\vec{X},\vec{Y}}$  per
Definition~\ref{defn-assignment}, and
$pe$  per Definition~\ref{defn-pe}.

Condition~\ref{cond-compute} and Condition~\ref{cond-nondeg}
 follow immediately from
Definition~\ref{defn-reduction-layout},
and Condition~\ref{cond-plant} follows from Lemma~\ref{layout-plant-lemma}.
What remains to be shown is that Condition~\ref{cond-weird} holds.
Let $(\vec{X},\vec{Y}) \in \{0,1\}^n \times \{0,1\}^n$  with
$setdisj_n(\vec{X},\vec{Y})=1$ be given.  Choose
$l \in [n]$ with $X_l=Y_l=1$  and set ${\cal{S}} = {\cal{S}}^{l}$.
By Lemma~\ref{switchable-density},
$\mu({\cal{S}}^{l}) \ge \delta^8/2^9$.
Set $c= (\delta^2/20)^{12}e^{-18}$ (The constant of Lemma~\ref{lemma-switcheroo}.)
We now show that  for all assignments $A$ to 
$MVars_m$:
\[\max_e \mu ( pe(L) =e \mid A_{{L},\vec{X},\vec{Y}}= A, \ {L} \in {\cal{S}}^{l} ) \le 1/(1+c)\]

Let $A$ be an assignment to $MVars_m$ and let
$e \in {[3m] \choose 2}$ be given.
Let ${\cal{B}}^e_A = \{ {L} \in {\cal{S}}^{l} \mid A_{{L},\vec{X},\vec{Y}}= A, \ pe(L) =e\}$,  
let ${\cal{S}}^{l}_A = \{ {L} \in {\cal{S}}^{l} \mid A_{{L},\vec{X},\vec{Y}}= A\}$.  Take
take as $f$ guaranteed by Lemma~\ref{lemma-switcheroo}.
Because $f$ maps ${\cal{S}}^{l}$ to ${\cal{S}}^{l}$,
we have that $f({\cal{B}}^e_A) \subseteq {\cal{S}}^{l}$,  
because $A_{f({L}),\vec{X},\vec{Y}}=A_{{L},\vec{X},\vec{Y}}=A$,
we have that $f({\cal{B}}^e_A) \subseteq {\cal{S}}^{l}_A$,
and because
$pe(f(L)) \neq pe(L)=e$,
we have that  
$f({\cal{B}}^e_A) \subseteq {\cal{S}}^{l}_A \setminus {\cal{B}}^e_A$. 
Because $f$ is an 
involution of ${\cal{S}}^{l}$, it is injective, and
 because $\mu(f({L})) \ge c \mu({L})$ for all ${L}$,
 we have  that
$\mu({\cal{S}}^{l}_A \setminus {\cal{B}}^e_A) \ge \mu(f({\cal{B}}^e_A)) \ge c_1\mu({\cal{B}}^e_A)$ and therefore 
$\mu({\cal{S}}^{l}_A) = \mu({\cal{S}}^{l}_A \setminus {\cal{B}}^e_A) + \mu({\cal{B}}^e_A) \ge (1+c)\mu({\cal{B}}^e_A)$.
Therefore: $\mu ( pe(L) =e \mid A_{{L},\vec{X},\vec{Y}}= A, \ {L} \in {\cal{S}}^{l} ) =  \mu ({\cal{B}}^e_A \mid {\cal{S}}^{l}_A ) = \frac{\mu({\cal{B}}^e_A)}{{\mu({\cal{S}}^{l}_A)}}  \le  \frac{1}{1+c}$.
Noting that $1/(1+c)=1 - c/(1+c)$,  we set $c_1 = c/(1+c)$ and we 
conclude the proof of  Lemma~\ref{key-lemma}.
\end{proof}

\begin{proof}(of Lemma~\ref{lemma-switcheroo})
Let ${L}= (\vec{\imath},\vec{\jmath},\vec{u},\vec{v},\vec{w})$.  We 
define $f({L}) = \left(\vec{\imath},\vec{\jmath}^*,\vec{u}^*,\vec{v}^*,\vec{w}^* \right)$ below.
The basic the idea is to modify the reduction layout $L$
by swapping some vertices between the gadgets at positions $n+1$ and
$l$ so that the planted edge changes but
the  assignment remains the same.
This is graphically illustrated in Figure~\ref{switcheroo-fig}. Because of the
partitioning of the variables, it is not immediately the case that
$L^*$ will be a reduction layout. Among other things,
we need to ensure that
$\{u^*_l,w^*_l\} \in E_{i_l^*}$ and
$\{ j_{n+1,1}^*, j_{n+1,2}^*, j^*_{n+1,3} \} \in N_3(i_{n+1}^*)$,
which is where we  make use of the hypothesis that $L$ is 
$l$-switchable\footnote{A reader carefully
checking the case analysis below will note that the definition of 
$l$-switchable is a bit stronger than we need.  See the discussion in
Section~\ref{debriefing}.}. We give the full definition
of $L^*$ below, along with the case analysis ensuring that the conclusions of
the lemma hold.

\begin{figure}
\begin{center}
\epsfig{file=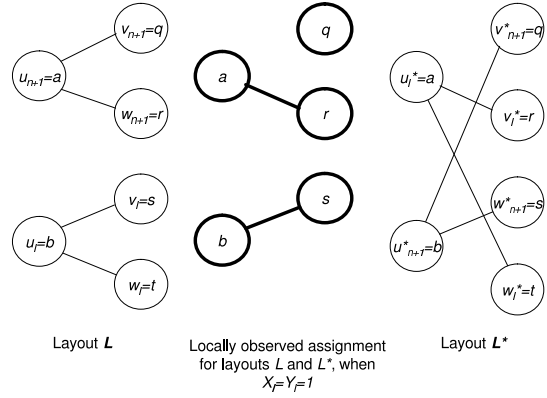,height=5cm}
\caption{\label{switcheroo-fig}
With layouts $L$ and $L^*$, when $X_l=X_l=1$,
set of vertices and edges specified by the assignments 
$A_{L,\vec{X},\vec{Y}}$ and $A_{L^*,\vec{X},\vec{Y}}$ are equal.
Notice however, that the planted edge under $L$ is $\{a,r\}$ whereas
the planted edge under $L^*$ is $\{b,s\}$.
} \end{center} \end{figure}

\[
\begin{array}{ccccccc}
i^*_k & = & \left\{ \begin{array}{cc}
i_{n+1} & {\mbox{ if $k=l$}} \\
i_l & {\mbox{ if $k={n+1}$}} \\
i_k & {\mbox{ otherwise}}
\end{array} \right.  & & u^*_i &= & \left\{ \begin{array}{cc}
u_l & {\mbox{ if $i={n+1}$}} \\
u_{n+1} & {\mbox{ if $i=l$}} \\
u_i & {\mbox{ otherwise}}
\end{array} \right. \\
j^*_{k,1} & = & \left\{ \begin{array}{cc}
j_{{n+1},3} & {\mbox{ if $k=l$}} \\
j_{l,2} & {\mbox{ if $k=n+1$}} \\
j_{k,1} & {\mbox{ otherwise}}
\end{array}  \right. & &
v^*_k &= & \left\{ \begin{array}{cc}
w_{n+1} & {\mbox{ if $k=l$}} \\
v_k & {\mbox{ otherwise}}
\end{array} \right. \\
j^*_{k,2} & = & \left\{ \begin{array}{cc}
j_{{n+1},1} & {\mbox{ if $k=l$}} \\
j_{k,2} & {\mbox{ otherwise}}
\end{array} \right. &&
w^*_k &= & \left\{ \begin{array}{cc}
v_l & {\mbox{ if $k={n+1}$}} \\
w_k & {\mbox{ otherwise}} \\
\end{array} \right. \\
j^*_{{n+1},3} & =&  j_{l,1} & && &
\end{array}
\] 

We now check each of the properties required by Lemma~\ref{lemma-switcheroo}.
This is just case analysis and rewriting. However, in order to show that
$f(L) \in {\cal{S}}^{l}$ we make use of the hypothesis that $L$ is 
$l$-switchable.

\begin{description}
\item[]The mapping $f$ is an involution.
This is  verified by iterating
the definition of $f$.  The details are carried out in the Appendix, 
Section~\ref{APP-distributionSect}.

\item[$A_{{L},\vec{X},\vec{Y}} = A_{f({L}),\vec{X},\vec{Y}}$.]
This is follows from expanding the definitions and doing a little bookkeeping,
we put the argument in the Appendix, Section~\ref{APP-distributionSect}.

\item[$pe(L) \neq pe(f(L))$.]  
Because ${L} = (\vec{\imath},\vec{\jmath},\vec{u},\vec{v},\vec{w})$
is a reduction layout,
$\{ u_{n+1},w_{n+1} \} \cap \{u_l,v_l\} = \emptyset$.
Applying Definition~\ref{defn-pe},  we see that $pe(L) =\{u_{n+1},w_{n+1}\} \neq  \{u_l,v_l \} = \{u^*_{n+1},w^*_{n+1}\} = pe(f(L))$.

\item[$\mu(f({L})) \ge \mu({L}) \cdot (\delta^2/20)^{12}e^{-18}$.]
In order to show this, we need that $\mu({L})>0$ (which holds
because ${L} \in {\cal{L}}$) and
$\mu(f({L}))>0$ (which depends on the fact
that $f({L}) \in {\cal{L}}$, which we show
below).  For now we take the non-zero mass of $f({L})$ as a 
given.
The differences between ${L}$ and $f({L})$ occur only with:
$i_{n+1} \neq i_{n+1}^*$, $i_l \neq i_l^*$, 
$(j_{{n+1},1},j_{{n+1},2},j_{{n+1},3}) \neq (j_{{n+1},1}^*,j_{{n+1},2}^*,j_{{n+1},3}^*)$,  
$(j_{l,1},j_{l,2}) \neq (j_{l,1}^*,j_{l,2}^*)$,  
$(u_l,v_l,w_l) \neq (u_l^*,v_l^*,w_l^*)$, and
$(u_{n+1},v_{n+1},w_{n+1}) \neq (u_{n+1}^*,v_{n+1}^*,w_{n+1}^*)$.
Therefore $HD({L},f({L}^*)) \le 6$.  We apply 
Lemma~\ref{continuity-lemma} to deduce that 
$\mu(f({L})) \ge \mu({L}) \cdot  (\delta^2/20)^{12}e^{-18}$.

\item[] For each ${L} \in {\cal{S}}^{l}$,  $f({L}) \in {\cal{S}}^{l}$.
First we check that
$f({L})=(\vec{\imath},\vec{\jmath},\vec{u},\vec{v},\vec{w})$
is indeed a reduction layout. 
We check each property from Definition~\ref{defn-reduction-layout}:

\begin{enumerate}

\item The indices $i^*_1, \ldots i^*_{n+1}$ are distinct: This holds because $\vec{\imath}^*$ is 
a permutation of $\vec{\imath}$.

\item The indices $j_{1,1}^*, j_{1,2}^*, \ldots j_{n,1}^*, j_{n,2}^*,j_{n+1,1}^*, j_{n+1,2}^*, j_{n+1,3}^*$ are 
distinct:
This holds because $\vec{\jmath}^*$ is a permutation of $\vec{\jmath}$.

\item The integers $u_1^*, \ldots u_{n+1}^*, v_1^*, \ldots v_{n+1}^*, w_1^*, \ldots w_{n+1}^*$ are
distinct:  This is true because 
$u_1^*, \ldots u_{n+1}^*,$ $ v_1^*, \ldots v_{n+1}^*,$ $w_1^*, \ldots w_{n+1}^*$ 
is a permutation of 
$u_1, \ldots u_{n+1}, v_1, \ldots v_{n+1}, w_1, \ldots w_{n+1}$.

\item For each $k=1, \ldots n+1$,
$\{u_k^*, v_k^*\} \in E_{i_k^*}$ and  $\{u_k^*, w_k^*\} \in E_{i_k^*}$:
Because
\[K(\{u_l,u_{n+1}\},\{v_l,v_{n+1},w_l,w_{n+1}\}) \subseteq E_{i_{n+1}} [V_{j_{{n+1},1}} \cap V_{j_{{n+1},2}} \cap V_{j_{{n+1},3}}  ] \cap E_{i_l}[  V_{j_{l,1}} \cap V_{j_{l,2}} ]\]
we have that 
$\{u_l^*, v_l^*\}  = \{u_{n+1}, w_{n+1}\} \in  E_{i_{n+1}} = E_{i_l^*}$,
$\{u_l^*, w_l^*\} =\{u_{n+1}, w_l \} \in E_{i_{n+1}} =  E_{i_l^*}$,
$\{u_{n+1}^*, v_{n+1}^*\}  = \{ u_l, v_{n+1}\} \in  E_{i_l} = E_{i_{n+1}^*}$,  and $\{u_{n+1}^*, w_l^* \}= \{u_l,w_l \} \in E_{i_l} =  E_{i_{n+1}^*}$.
For $k \in [n] \setminus \{l\}$, we have that
$\{u_k^*, v_k^*\} = \{u_k, v_k\} \in E_{i_k} =  E_{i_k^*}$ and  $\{u_k^*, w_k^*\} =\{u_k, w_k\} = E_{i_k}  \in E_{i_k^*}$.

\item For each $k=1, \ldots n+1$, 
$\{u^*_k,v^*_k,w^*_k\} \subseteq V_{j_{k,1}^*} \cap V_{j_{k,2}^*}$:
Because 
\[K(\{u_l,u_{n+1}\},\{v_l,v_{n+1},w_l,w_{n+1}\}) \subseteq E_{i_{n+1}} [V_{j_{{n+1},1}} \cap V_{j_{{n+1},2}} \cap V_{j_{{n+1},3}}] \cap E_{i_l}[ V_{j_{l,1}} \cap V_{j_{l,2}} ]\]
we have that 
$\{u_l^*,  v_l^*, w_l^*\} = \{u_{n+1}, w_{n+1}, w_l  \} \subseteq V_{j_{{n+1},3}} \cap V_{j_{{n+1},1}} = V_{j_{l,1}^*} \cap V_{j^*_{l,2}}$.
For the same reason, 
$\{u_{n+1}^*,  v^*_{n+1},w_{n+1}^*\} = \{u_l, v_{n+1}, v_l  \} \subseteq V_{j_{l,2}} \cap V_{j_{n+1,2}} =  V_{j_{{n+1},1}^*} \cap V_{j^*_{{n+1},2}}$.
For $k \in [n] \setminus \{l\}$, we have that 
$\{u_k^*, v_k^*, w_k^*\} = \{u_k, v_k, w_k\} \subseteq V_{j_{k,1}} \cap V_{j_{k,2}} = V_{j^*_{k,1}} \cap V_{j^*_{k,2}}$.

\item 
We have that
$\{u^*_{n+1},v^*_{n+1},w^*_{n+1}\} = \{u_l,v_{n+1},v_l\} \subseteq V_{j_{l,1}}= V_{j_{{n+1},3}^*}$,
because
\[K(\{u_l,u_{n+1}\},\{v_l,v_{n+1},w_l,w_{n+1}\}) \subseteq E_{i_{n+1}} [V_{j_{{n+1},1}} \cap V_{j_{{n+1},2}} \cap V_{j_{{n+1},3}}] \cap E_{i_l}[ V_{j_{l,1}} \cap V_{j_{l,2}} ]\]

\item For each $k \in [n+1]$,  $i^*_k \in G$: 
 This holds because $\vec{\imath}^*$ is a permutation
of $\vec{\imath}$ and for each $k \in [n+1]$, $i_k \in G$.

\item $(j_{{n+1},1}^*,j_{{n+1},2}^*,j_{{n+1},3}^*) \in N_3(i_{n+1}^*)$:
Because ${L}$ is $l$-switchable, 
$(j_{{n+1},1},j_{l,1},j_{l,2}) \in N_3(i_l)$,  therefore,
$(j^*_{{n+1},1},j^*_{{n+1},2},j^*_{{n+1},3}) = (j_{l,2},j_{n+1,2},j_{l,1}) \in N_3(i_l) = N_3(i_{n+1}^*)$.

\item For each $k=1, \ldots n$:
$(j_{k,1}^*, j_{k,2}^*) \in N_2(i_k^*)$.
For $k \in [n] \setminus \{l\}$,  we have that $(j_{k,1}^*,j_{k,2}^*)= (j_{k,1},j_{k,2}) \in N_2(i_k)=N_2(i_k^*)$. When $k=l$,
because ${L}$ is a reduction layout, we have that
$(j_{n+1,1},j_{n+1,2},j_{n+1,3}) \in N_3(i_{n+1})$, and therefore 
$(j_{{n+1},3},j_{{n+1},1}) \in N_2(i_{n+1})$.  Thus:
$(j_{l,1}^*,j_{l,2}^*)= (j_{{n+1},3},j_{{n+1},1}) \in N_2(i_{n+1}) = N_2(i_l^*)$.

\end{enumerate}

This establishes that $f({L}) \in {\cal{L}}$. 
That $f({L}) \in {\cal{S}}^{l}$ follows immediately from
the hypothesis that $L \in {\cal{S}}^{l}$ and the definitions:
$(j_{n+1,2}^*,j_{l,1}^*,j_{l,2}^*) = (j_{n+1,2},j_{{n+1},3},j_{{n+1},1}) \in N_3(i_{n+1}) = N_3(i_l^*)$ and
\begin{eqnarray*}
K(\{u_l^*,u_{n+1}^*\},\{v_l^*,v_{n+1}^*,w_l^*,w_{n+1}^*\}) & = &  K(\{u_l,u_{n+1}\},\{v_l,v_{n+1},w_l,w_{n+1}\}) \\
& \subseteq    &   E_{i_{n+1}}[ V_{j_{{n+1},1}} \cap V_{j_{{n+1},2}} \cap V_{j_{{n+1},3}} ] \cap  E_{i_l}[V_{j_{l,1}} \cap V_{j_{l,2}} ] \\
& = & E_{i_l} [V_{j_{l,2}}  \cap V_{j_{{n+1},2}} \cap V_{j_{l,1}}  ] \cap E_{i_{n+1}}[ V_{j_{{n+1},3}} \cap V_{j_{{n+1},1}} ] \\
& = & E_{i_{n+1}^*} [V_{j_{{n+1},1}^*}  \cap V_{j_{{n+1},2}^*} \cap V_{j_{{n+1},3}^*}  ] \cap E_{i_l^*}[ V_{j_{l,1}^*} \cap V_{j_{l,2}^*} ]  
\end{eqnarray*}

\end{description}

\end{proof}

\section{Probability  Notation and Background}\label{sect-probbackground}

\begin{definition} Let $X_i$, $i \in I$, be a family of sets indexed by a
set $I$; we write $X_I$ as an abbreviation for the product 
$\prod_{i \in I} X_i$.  
Let $\prod_{i \in I} X_i$ and $\prod_{j \in J} X_j$ be product spaces with
$I \cap J = \emptyset$.  For $\vec{x} \in \prod_{i \in I} X_i$ and
$\vec{y} \in \prod_{i \in J} X_i$ we write $\vec{x}\vec{y}$ to denote the
concatenation of $\vec{x}$ and $\vec{y}$ (an element of 
$\prod_{i \in I \cup J} X_i$).   We use the same indices for elements in tuples
as we do for the factors of the product, ie. for
$\vec{u} \in \prod_{i=j}^t X_i$,  we write  $\vec{u} = (u_j, \ldots u_t)$, we
{\em{do not}} write $\vec{u}=(u_1, \ldots u_{t-j+1})$.
Let $f$ be a function whose domain is a product space $\prod_{i=1}^t X_i$.
For each $j \in [t]$,  for each $\vec{x} \in \prod_{i=1}^j X_i$,  we
write $f^{\vec{x}}$ to denote the curried function with domain
$\prod_{i=j+1}^t X_i$, that is,
$f^{\vec{x}}( \vec{y})= f(\vec{x}\vec{y})$.
\end{definition}

\begin{definition} Let $\eta$ be a probability distribution over a set $X$
and let $f : X \rightarrow \reals$.
We write $\expect_\eta[f]$ to denote the expectation of $f$ with respect to
$\eta$.  At times,  the uniform distribution over a set will be written as
$U$. Other times, we will write with $E \subseteq S$, we will
write $\prob_{x \in S}[E]$
to denote the probability that $x \in E$ holds when
$x$ is selected uniformly from $S$. 
\end{definition}

\begin{definition}
Let $\eta$ be a probability distribution on a product space 
$\prod_{i=1}^t X_i$. For each $I \subseteq [t]$,  let $\eta_I$ be the
marginal distribution of $\eta$ on $\prod_{i \in I}X_i$.
 For each $j \in [t]$ and each
$\vec{x} \in \prod_{i=1}^j X_i$,   let $\eta^{\vec{x}}$ be the probability
distribution on $\prod_{i=j}^t X_i$ given by the formula
$\eta^{\vec{x}}(\vec{y}) = \frac{\eta(\vec{x}\vec{y})}{\eta_{[j]}(\vec{x})}$
if $\eta_{[j]}(\vec{x}) \neq 0$ and $0$ otherwise.  
\end{definition}

Notice that $\eta^{\vec{x}}$ is the marginal  distribution 
of   $\eta$ to the coordinates $[t]\setminus[j]$
 conditioned on the event that the first
$j$ coordinates take the value $\vec{y}$.
An immediate consequence of the definitions:
\begin{lemma}
Let $f : \prod_{i=1}^t X_i \rightarrow {\mathbb{R}}$, let 
$I = \{1, \ldots i_0\}$:  $\expect_\eta[f] = \sum_{\vec{u}\in X_I} \eta_I({\vec{u}}) \expect_{\eta^{\vec{u}}} [f^{\vec{u}}]$
\end{lemma}

Unsurprisingly for a technique based on finding structure in a dense
family of sets, we beat the stuffing out 
Jensen's Inequality, its relatives, and any averaging arguments that
we find in the neighborhood.

\begin{proposition}(Jensen's Inequality)
Let $f : D \rightarrow {\mathbb{R}}$,   let 
$g: {\mathbb{R}} \rightarrow {\mathbb{R}}$  be a convex function, and let 
$\eta$ be a probability distribution on $D$. 
$\expect_\eta [g \circ f]  \ge g\left(\expect_\eta [f] \right)$.
\end{proposition}

\begin{lemma}\label{convexity-lemma} (Proof in the Appendix, section~\ref{APP-notationSect}.)
Let $X$ be a finite set,
and let $Y_1, \ldots Y_n$ be a family of subsets of $X$.
Set $\alpha = \frac{1}{n} \sum_{i=1}^n |Y_i|/|X|$, and let
$k$ be a non-negative integer:
$\frac{1}{n^k} \sum_{\vec{\imath} \in [n]^k} | \bigcap_{l=1}^k Y_{i_l}| \ge \alpha^k |X|$.
\end{lemma}

\begin{lemma}\label{supersaturation} (Proof in the Appendix, section~\ref{APP-notationSect}.)
There exists a constant $c>0$ so that for every undirected graph
$G = (V,E)$ with $|V| = N$ and $|E| \ge \alpha {N \choose 2}$.
We have that:
\[
\begin{array}{lcl}
\prob_{\vec{u} \in V^3} [ K(\{u_1\},\{u_2,u_3\}) \subseteq G]  &\ge &  \alpha^2 - (5/N) \\
\prob_{\vec{u} \in V^6} \left[ K\left(\{u_1,u_2\},\{u_3,u_4,u_5,u_6\}\right) \subseteq G \right]&  \ge &  \alpha^8 - (23/N) 
\end{array} \]
\end{lemma}

\begin{proposition}
Let $\eta$ be a probability measure on a space $X$, 
and let $f : X \rightarrow [0,1]$ be measurable.  For all $\epsilon \in [0,1]$
and all $c>0$
$\eta(\{ x \mid f(x) \ge \frac{1}{c} \expect_\eta[f]  \}) \ge (1-1/c) \expect_\eta[f]$.
\end{proposition}

\section{Distributions from DDWB Processes}\label{DDWB-section}

To prove the completeness lemma (Lemma~\ref{switchable-density}) and the
continuity lemma (Lemma~\ref{continuity-lemma}),
we make some detailed calculations about
the distribution ${\cal{D}}$.  
It seems that by moving to slightly
more general framework,  some of the calculations and case analyses are
simplified. In Lemma~\ref{lemma-DIsNBP} in 
Section~\ref{analyzeDSect} we show that the distribution
${\cal{D}}$ falls into this framework and use the machinery of DDWB processes
developed in
this section to finish the proofs of Lemma~\ref{switchable-density}
and Lemma~\ref{continuity-lemma}.

\begin{definition}\label{defn-nonblockingproc}
Let $t$ be an integer, $X_1, \ldots X_t$ be sets, and let
$S_i : \prod_{k=1}^{i-1} X_k \rightarrow \pow{X_i}$,  and
$F_i : \prod_{k=1}^{i-1}X_k \rightarrow \pow{X_i}$ be families of
maps with $i \in [t]$.
Assume that for all $i=1, \ldots t$, and all 
$(u_1, \ldots u_{i-1}) \in \prod_{k=1}^{i-1}X_i$,
$S_i(u_1, \ldots u_{i-1}) \setminus F(u_1, \ldots u_{i-1}) \neq \emptyset$ and
$S_i(u_1, \ldots u_{i-1}) \setminus F_i(u_1, \ldots u_{i-1}) \neq \emptyset$.

The {\em{distribution given by the dependent domains with
blocking process of $\vec{S}$ and $\vec{F}$}}
is the distribution $\pi (=\pi_{\vec{S},\vec{F}})$ 
on  $\prod_{i=1}^t X_i$  given by the
random process that generates a sequence
$(u_1, \ldots u_t)$ as follows:  For $i=1, \ldots t$, 
choose $u_i$ uniformly from 
$S_i(u_1, \ldots u_{i-1}) \setminus F_i(u_1, \ldots u_{i-1})$.
The {\em{blockage bound}} of a DDWB process $\vec{S}$, $\vec{F}$
is the smallest
$\beta \ge 0$ so that for all $i=1, \ldots t$ and all
$\vec{u} \in \prod_{k=1}^{i-1} X_k$,
$|F_i(\vec{u})| \le \beta |S_i(\vec{u})|$.
The {\em{covering bound}} for $\vec{S}$, $\vec{F}$ is the 
largest $\kappa \in [0,1]$ so that for all
$i=1, \ldots t$ and all $\vec{u} \in \prod_{k=1}^{i-1} X_k$,
$|S_i(\vec{u}) \setminus F_i(\vec{u})| \ge \kappa |X_i|$.
\end{definition}

The following easy fact is the crux of an induction argument.

\begin{proposition}\label{lemma-restrictedNBP}
Let $\pi$ be the distribution on $\prod_{i=1}^t X_i$ given
by the DDWB process  $\vec{S}$, $\vec{F}$.  
For each $a \in X_1$,
The distribution $\pi_a$ is generated by the DDWB process on 
$\prod_{i=2}^t X_i$ given by $S_2^a, \ldots S^a_t$, $F^a_2, \ldots F^a_t$.
If the process $\vec{S}$, $\vec{F}$ has  a blockage bound $\le \beta$,
then the process $\vec{S}^a$, $\vec{F}^a$ has a blockage bound $\le \beta$.
\end{proposition}

\subsection{Loss of Expectation Lemma for  DDWB Distributions}\label{density-subsubsec}

The following lemma is used to 
pass density results 
for the uniform distribution, such as Lemma~\ref{supersaturation}, to
certain DDWB distributions.  This is how 
Lemma~\ref{switchable-density} will be proved.  It is a simple but careful
combination of two observations:
If the domains $S_i$ contain the support of a $[0,1]$ valued function, then
uniformly selecting over the $S_i$'s (instead of all of $X_i$) will only
increase the expectation.  Of course the blocking of the $F_i$'s could reduce
the expectation,  but for a DDWB with blockage bound $\beta$,  
each coordinate that the event
depends upon can reduce the expectation  by at most  $\beta$.

\begin{lemma}\label{lemma-unif2nonblock}
Let $\prod_{i=1}^t X_i$ be  a product space, and
let $f : \prod_{i=1}^t X^i \rightarrow [0,1]$
be a function that depends upon at most $k$
coordinates, $i_1, \ldots i_k$.
Let $U$ be the uniform distribution on $\prod_{i=1}^t X_i$,  and let
$\pi$ be a DDWB distribution on $\prod_{i=1}^t X_i$ given by
some $\vec{S}$ and $\vec{F}$.
If the following two conditions are satisfied:
\begin{enumerate}
\item The DDWB process $\vec{S}$, $\vec{F}$ has blockage bound $\le \beta$.
\item For all $\vec{a} \in \prod_{i=1}^t X_i$,  if
$f(\vec{a}) > 0$ then  for all $j=1, \ldots k$, 
$a_{i_j} \in S_{i_j}(a_1, \ldots a_{i_j-1})$.
\end{enumerate}
Then    $\expect_\pi[f]  \ge  \expect_U[f] - k\beta$.
\end{lemma}

\begin{proof}
We prove the claim by induction on $k$.  The lemma clearly holds for $k=0$,
as in that case $f$ is constant over $\prod_{i=1}^t X_i$, and therefore
 $\expect_\pi[f]=\expect_U[f]$.
We now assume that the lemma holds for functions that
depend on only $k$ coordinates, and demonstrate that it holds for functions
that depend on only $k+1$ coordinates.

Let $t$,  $\prod_{i=1}^t X_i$,  $\pi$, $\vec{S}$, $\vec{F}$, and
be given as in the statement of the lemma- with $f$ dependent only upon 
$k+1$  coordinates,   $i_1, \ldots i_{k+1}$.  Let $i=i_1$ be the first 
coordinate upon which the function $f$ depends.  
Set $I = [i -1]$ and $J = [t] \setminus [i]$.
Let $X_I = \prod_{k \in I} X_k$ and $X_J = \prod_{k \in J} X_k$.

We reduce to the induction hypothesis by showing that for each
$\vec{u} \in X_I$,  $a \in X_i$, the conditions of the induction
hypothesis are met for 
the function $f^{\vec{u}a}$,
with process $\vec{S}^{\vec{u}a}$, $\vec{F}^{\vec{u}a}$, and
distribution $\pi^{\vec{u}a}$.
Observe that the distribution $\pi^{\vec{u}a}$ is given by the DDWB
process $S_{i+1}^{\vec{u}a}, \ldots S_t^{\vec{u}a}$ and $F_{i+1}^{\vec{u}a}, \ldots F_t^{\vec{u}a}$,  a process with blockage bound $\le \beta$ because
$\vec{S}$, $\vec{F}$ has blockage bound $\le \beta$.
Moreover,  the function $f^{\vec{u}a} : \prod_{j=i+1}^t X_i \rightarrow [0,1]$
depends on at most $k$ coordinates. 
By specializing the hypothesis
``for all $\vec{a}$, if $f(\vec{a})>0$ then  for all $j=1, \ldots k$, 
$a_{i_j} \in S_{i_j}(a_1, \ldots a_{i_j-1})$''
to inputs with prefix $\vec{u}a$ and weakening its conclusion
to cover only $j=2, \ldots k$,
we have that ``for all $\vec{b} \in X_J$ so that
$f(\vec{u}a\vec{b})>0$, for all $j=2, \ldots k$, 
$b_{i_j} \in S_{i_j}(\vec{u},a,b_{i+1}, \ldots b_{i_j-1})$''. This
is equivalent to 
``for all $\vec{b} \in X_J$ so that
$f^{\vec{u}a}(\vec{b})>0$, for all $j=2, \ldots k$, 
$b_{i_j} \in S^{\vec{u}a}_{i_j}(b_{i+1}, \ldots b_{i_j-1})$''. 
Therefore by the induction hypothesis
 we have that
$\expect_{\pi^{\vec{u}a}}[f^{\vec{u}a}] \ge \expect_{U^{\vec{u}a}}[f^{\vec{u}a}]  - k  \beta$.

Furthermore, from the hypothesis
``for all $\vec{u} \in \prod_{i=1}^t X_i$,
if $f(\vec{u})>0$ then $\forall j \in [k+1],
\ u_{i_j} \in S_{i_j}(u_1,\ldots u_{i_j-1})$'' we conclude
that  for all $\vec{v} \in \prod_{j=1}^i X_j$
 with $\expect_{U^{\vec{v}}}[f^{\vec{v}}]>0$,
 $v_i \in S_i(v_1, \ldots v_{i-1})$.
Therefore,  for all $\vec{u}=(u_1, \ldots u_{i-1}) \in X_I$
\[ \expect_{U^{\vec{u}}}[f^{\vec{u}}] =\sum_{a \in X_i} \frac{1}{|X_i|} \expect_{U^{\vec{u}a}}[f^{\vec{u}a}] = \sum_{a \in S_i(\vec{u})} \frac{1}{|X_i|} \expect_{U^{\vec{u}a}}[f^{\vec{u}a}] \le \sum_{a \in S_i(\vec{u})} \frac{1}{|S_i(\vec{u})|} \expect_{U^{\vec{u}a}}[f^{\vec{u}a}]\]

We now bound the expectation of $f$ with respect to $\pi$ from below:
\begin{eqnarray*}
\expect_\pi[f] & = & \sum_{\vec{u} \in X_I} \pi_I({\vec{u}}) \sum_{a \in X_i} \sum_{\vec{b} \in X_J} \pi^{\vec{u}}(a \vec{b}) f(\vec{u}a\vec{b}) \\
 & = & \sum_{\vec{u} \in X_I} \pi_I({\vec{u}}) \sum_{a \in X_i} \sum_{\vec{b} \in X_J} \frac{\chi_{S_i(\vec{u})\setminus F_i(\vec{u})}(a)}{|S_i(\vec{u})\setminus F_i(\vec{u})|}\pi^{\vec{u}a}(\vec{b}) f(\vec{u}a\vec{b}) \\
 & = &     \sum_{\vec{u} \in X_I} \pi_I({\vec{u}}) \sum_{a \in S_i(\vec{u})} \sum_{\vec{b} \in X_J } \frac{1-\chi_{F_i(\vec{u})}(a)}{|S_i(\vec{u}) \setminus F_i(\vec{u})|}  \pi^{\vec{u}a}({\vec{b}}) f(\vec{u}a{\vec{b}})\\
 & = &     \sum_{\vec{u} \in X_I} \pi_I({\vec{u}}) \sum_{a \in S_i(\vec{u})}  \frac{1-\chi_{F_i(\vec{u})}(a)}{|S_i(\vec{u}) \setminus F_i(\vec{u})|}   \sum_{\vec{b} \in X_J } \pi^{\vec{u}a}({\vec{b}})f(\vec{u}a{\vec{b}})\\
 & = &     \sum_{\vec{u} \in X_I} \pi_I({\vec{u}}) \sum_{a \in S_i(\vec{u})} \frac{1-\chi_{F_i(\vec{u})}(a)}{|S_i(\vec{u}) \setminus F_i(\vec{u})|}   \expect_{  \pi^{\vec{u}a}}[f^{\vec{u}a}] \\
 & \ge &     \sum_{\vec{u} \in X_I} \pi_I({\vec{u}}) \sum_{a \in S_i(\vec{u})} \frac{1-\chi_{F_i(\vec{u})}(a)}{|S_i(\vec{u}) \setminus F_i(\vec{u})|}     \left( \expect_{U^{\vec{u}a}}[f^{\vec{u}a}] - k \beta\right) \\
 & = &  -k\beta+    \sum_{\vec{u} \in X_I} \pi_I({\vec{u}}) \sum_{a \in S_i(\vec{u})} \frac{1-\chi_{F_i(\vec{u})}(a)}{|S_i(\vec{u}) \setminus F_i(\vec{u})|}      \expect_{U^{\vec{u}a}}[f^{\vec{u}a}] \\
 & \ge &  -k\beta+    \sum_{\vec{u} \in X_I} \pi_I({\vec{u}}) \sum_{a \in S_i(\vec{u})}  \frac{1-\chi_{F_i(\vec{u})}(a)}{|S_i(\vec{u}) |}      \expect_{U^{\vec{u}a}}[f^{\vec{u}a}]  \\
 &\ge &  -k\beta+    \sum_{\vec{u} \in X_I} \pi_I({\vec{u}}) \left( \sum_{a \in S_i(\vec{u})}  \frac{1}{|S_i(\vec{u}) |} \expect_{U^{\vec{u}a}}[f^{\vec{u}a}]  - \sum_{a \in S_i(\vec{u})}  \frac{\chi_{F_i(\vec{u})}(a)}{|S_i(\vec{u}) |}      \right) \\
 & \ge &  -k\beta+    \sum_{\vec{u} \in X_I} \pi_I({\vec{u}}) \left( \sum_{a \in S_i(\vec{u})}  \frac{1}{|S_i(\vec{u}) |}  \expect_{U^{\vec{u}a}}[f^{\vec{u}a}] -    \frac{|F_i(\vec{u}) |}{|S_i(\vec{u}) |}      \right) \\
 & \ge &  -k\beta+    \sum_{\vec{u} \in X_I} \pi_I({\vec{u}}) \left( \sum_{a \in S_i(\vec{u})}  \frac{1}{|S_i(\vec{u}) |}  \expect_{U^{\vec{u}a}}[f^{\vec{u}a}] -  \beta      \right) \\
 & = &  -(k+1)\beta+    \sum_{\vec{u} \in X_I} \pi_I({\vec{u}})  \sum_{a \in S_i(\vec{u})}  \frac{1}{|S_i(\vec{u}) |}  \expect_{U^{\vec{u}a}}[f^{\vec{u}a}] \\
 & \ge &  -(k+1)\beta+    \sum_{\vec{u} \in X_I} \pi_I({\vec{u}}) \expect_{U^{\vec{u}}}[f^{\vec{u}}] \\
& =& -(k+1)\beta +   \sum_{\vec{u} \in X_I} \pi_I({\vec{u}}) \expect_U[f]   =   -(k+1) \beta  + \expect_U[f]
 \end{eqnarray*}
The penultimate equality holds because the function $f$ is
 independent of the 
coordinates of $I$,   and therefore, for all $\vec{u} \in X_I$,  
$\expect_{U^{\vec{u}}}[f^{\vec{u}}] = \expect_U[f]$.

\end{proof}

\subsection{``Continuity'' for DDWB Processes}\label{continuity-subsec}

\begin{lemma}\label{prob-thing}
Let $\pi$ be a distribution on the product space $\prod_{i=1}^t X_i$ 
given by a DDWB process $\vec{S}$, $\vec{F}$ with covering bound $\kappa$.
Let $ c$ and $d$ be arbitrary.  Let $I_0 \subseteq [t]$
so that $|I_0| = d$.  Let $\vec{u}, \vec{v} \in \prod_{i=1}^t X_i$ be
arbitrary.
If for all $i=1, \ldots t$,
\begin{enumerate}
\item\label{prop-nonzero} $\pi(\vec{u}) >0$ and $\pi(\vec{v}) >0$
\item\label{prop-choicesets-agree} For all $i \in [t] \setminus I_0$,  $S_i(u_1, \ldots u_{i-1}) = S_i(v_1,\ldots v_{i-1})$
\item\label{prop-negblockdiff} For all $i \in [t] \setminus I_0$,  $|F_i(u_1, \ldots u_{i-1}) \oplus F_i(v_1, \ldots v_{i-1})| \le (c/t) |X_i|$
\end{enumerate}
then 
$\pi(\vec{v}) < \kappa^{-d}  e^{c/\kappa} \pi(\vec{u})$.
\end{lemma}

\begin{proof}
Explicit calculation reveals that:
\begin{eqnarray*}
\frac{\pi(\vec{u})}{\pi(\vec{v})}  
& = &  \frac{\prod_{i=1}^t \left( \frac{1}{|S_i(u_1,\ldots u_{i-1}) \setminus F_i(u_1,\ldots u_{i-1}) |} \right) }{ \prod_{i=1}^t \left( \frac{1}{|S_i(v_1,\ldots v_{i-1}) \setminus F_i(v_1,\ldots v_{i-1}) |} \right)}=   \prod_{i=1}^t \frac{|S_i(v_1,\ldots v_{i-1}) \setminus F_i(v_1, \ldots v_{i-1})|}{|S_i(u_1, \ldots,u_{i-1}) \setminus F_i(u_1, \ldots u_{i-1})|} \\
&=  & \prod_{i \in I_0} \frac{|S_i(v_1,\ldots v_{i-1}) \setminus F_i(v_1, \ldots v_{i-1})|}{|S_i(u_1, \ldots,u_{i-1}) \setminus F_i(u_1, \ldots u_{i-1})|} \prod_{ i \in [t] \setminus I_0} \frac{|S_i(v_1,\ldots v_{i-1}) \setminus F_i(v_1, \ldots v_{i-1})|}{|S_i(u_1, \ldots,u_{i-1}) \setminus F_i(u_1, \ldots u_{i-1})|}  \\
& \le  & \prod_{i \in I_0} \frac{|X_i|}{\kappa|X_i|} \prod_{ i \in [t] \setminus I_0} \frac{|S_i(v_1,\ldots v_{i-1}) \setminus F_i(v_1, \ldots v_{i-1})|}{|S_i(u_1, \ldots,u_{i-1}) \setminus F_i(u_1, \ldots u_{i-1})|} \\
& =  & \kappa^{-d} \prod_{ i \in [t] \setminus I_0} \frac{|S_i(v_1,\ldots v_{i-1}) \setminus F_i(v_1, \ldots v_{i-1})|}{|S_i(u_1, \ldots,u_{i-1}) \setminus F_i(u_1, \ldots u_{i-1})|}  =   \kappa^{-d} \prod_{ i \in [t] \setminus I_0} \frac{|S_i(u_1,\ldots u_{i-1}) \setminus F_i(v_1, \ldots v_{i-1})|}{|S_i(u_1, \ldots,u_{i-1}) \setminus F_i(u_1, \ldots u_{i-1})|} \\
& \le  & \kappa^{-d} \prod_{ i \in [t] \setminus I_0} \frac{|S_i(u_1,\ldots u_{i-1}) \setminus F_i(u_1, \ldots u_{i-1})| + |F_i(v_1, \ldots v_{i-1}) \oplus F_i(u_1, \ldots u_{i-1})|}{|S_i(u_1, \ldots,u_{i-1}) \setminus F_i(u_1, \ldots u_{i-1})|} \\
& \le  & \kappa^{-d} \prod_{ i \in [t] \setminus I_0}\left( 1+\frac{(c/t)|X_i|}{\kappa |X_i|} \right) \le \kappa^{-d} e^{(t-d)\frac{c}{t\kappa}} \le \kappa^{-d} e^{\frac{c}{\kappa}}
\end{eqnarray*}
\end{proof}

\section{The Distribution ${\cal{D}}$ is a DDWB Distribution}\label{analyzeDSect}

We give a DDWB process $\vec{S}$, $\vec{F}$ and show that it produces the
distribution ${\cal{D}}$ used to generate reduction layouts used in the
reduction from set-disjointness to the
$FindBadEdge$ search lemma.  This enables us to use the machinery of
DDWB distributions to prove Lemma~\ref{switchable-density} and 
Lemma~\ref{continuity-lemma}.

\begin{definition}\label{ddwb4layoutsDefn}
Let $({\cal{V}}_I,{\cal{V}}_{II})$ be a partition of $MVars_m$.
Let $G$, $N_3(\cdot)$, $N_2(\cdot)$ be as in Definition~\ref{gdefn}.
We define a  DDWB process $\vec{S}$, $\vec{F}$ over the product space
$[m]^{n+1} \times \left([2m+1]^2\right)^n \times \left([2m+1]^3\right)  \times \left([3m]^3\right)^{n+1}$ as follows:

\begin{enumerate}

\item  When choosing $i_k$ given $ i_1, \ldots i_{k-1}$:   
$X_{k} = [m]$,  $S_{k} = G$ and $F_{k}(i_1, \ldots i_{k-1}) = \{i_1, \ldots i_{k-1}\}$.

\item   When choosing $(j_{k,1},j_{k,2})$ given $\vec{\imath},(j_{1,1},j_{1,2}), \ldots (j_{k-1,1},j_{k-1,2})$ (with $k\le n$), 
we have  $X_{n+1+k} = [2m+1]^2$,
$S_{n+1+k}(\vec{\imath},(j_{1,1},j_{1,2}), \ldots (j_{k-1,1},j_{k-1,2}))  =  N_2(i_k)$,  and:   
\begin{eqnarray*}
F_{n+1+k}(\vec{\imath},(j_{1,1},j_{1,2}), \ldots (j_{k-1,1},j_{k-1,2})) &=&  pm_{[2m+1]}\left(\{j_{1,1}, j_{1,2}, \ldots j_{k-1,1}, j_{k-1,2} \}\right) 
\end{eqnarray*}

\item When choosing $(j_{n+1,1}, j_{n+1,2},j_{n+1,3})$ given 
$\vec{\imath},(j_{1,1},j_{1,2}), \ldots (j_{n,1},j_{n,2})$, we have
$X_{2n+2} = [2m+1]^3$, 
$S_{2n+2} \left(\vec{\imath},(j_{1,1},j_{1,2}), \ldots (j_{n,1},j_{n,2})\right) = N_3(i_{n+1})$, and:
\begin{eqnarray*}
F_{2n+2}(\vec{\imath},(j_{1,1},j_{1,2}), \ldots (j_{n,1},j_{n,2})) & = & \{j_{1,1}, j_{1,2}, \ldots j_{n,1}, j_{n,2} \}
\end{eqnarray*}

\item For $k \le n$, when choosing $(u_k,v_k,w_k)$ given
$\vec{\imath},\vec{\jmath},(u_1,v_1,w_1), \ldots (u_{k-1},v_{k-1},w_{k-1})$, 
$X_{2n+2+k}  =  [3m]^3$,
$S_{2n+2+k}(\vec{\imath},\vec{\jmath},(u_1,v_1,w_1), \ldots (u_{k-1},v_{k-1},w_{k-1}))   =   {\cal{K}}_{1,2}\left(E_{i_k}[V_{j_{k,1}} \cap V_{j_{k,2}}] \right)$, and
\begin{eqnarray*}
F_{2n+2+k}(\vec{\imath},\vec{\jmath},(u_1,v_1,w_1), \ldots (u_{k-1},v_{k-1},w_{k-1})) & = &  tm_{[3m]}(\{u_1,v_1,w_1, \ldots u_{k-1},v_{k-1},w_{k-1}\})
\end{eqnarray*}

\item When choosing $(u_{n+1},v_{n+1},w_{n+1})$ given
$\vec{\imath},\vec{\jmath},(u_1,v_1,w_1), \ldots (u_n,v_n,w_n)$, 
$X_{3n+3}  =  [3m]^3$, \newline
$S_{3n+3}(\vec{\imath},\vec{\jmath},(u_1,v_1,w_1), \ldots (u_{k-1},v_{k-1},w_{k-1}))   =   {\cal{K}}_{1,2}\left(E_{i_{n+1}}[V_{j_{n+1,1}} \cap V_{j_{n+1,2}} \cap V_{j_{n+1,3}}] \right)$, and
\begin{eqnarray*}
F_{3n+3}(\vec{\imath},\vec{\jmath},(u_1,v_1,w_1), \ldots (u_n,v_n,w_n)) & = &  tm_{[3m]}(\{u_1,v_1,w_1, \ldots u_n,v_n,w_n\})
\end{eqnarray*}

\end{enumerate}

\end{definition}
\begin{lemma}\label{lemma-DIsNBP}
Let $m \ge 450/\delta^2$.
Let $({\cal{V}}_I,{\cal{V}}_{II})$ be a partition of $MVars_m$.  
Let $\delta = \delta({\cal{V}}_I,{\cal{V}}_{II})$ and
let $\gamma = \frac{n+1}{m}$.
The distribution ${\cal{D}}({\cal{V}}_I,{\cal{V}}_{II})$ is generated 
by the DDWB process $\vec{S}$, $\vec{F}$ over the product space
$[m]^{n+1} \times \left([2m+1]^2\right)^n \times ([2m+1]^3)  \times \left([3m]^3\right)^{n+1}$.  Moreover, this process has blockage bound 
$\le 30\gamma/\delta^2$
and it has  covering bound  
$\ge \min \{\delta^2/10 - 3\gamma, \delta/3-3\gamma,\delta/12-\gamma  \} $.
\end{lemma}
\begin{proof}
That the DDWB process $\vec{S}$, $\vec{F}$ generates the distribution
${\cal{D}}$ follows immediately by comparing the above functions with
the experiment of Definition~\ref{defn-reduction-layout}.
The covering  bounds follow immediately from Lemma~\ref{bountiful-lemma},
and the blockage bounds are implicit in those calculations.
\end{proof}
\begin{corollary}\label{sloppy-cor}
If  $\gamma = \delta^2/60$, then the covering bound of the process
is $\ge \delta^2/20$, ie. $\kappa \ge \delta^2/20$.
\end{corollary}

Now we use Lemma~\ref{prob-thing} to prove the continuity lemma:

\begin{proof}(of the continuity lemma, Lemma~\ref{continuity-lemma})
Let ${L} = (\vec{\imath},\vec{\jmath},\vec{u},\vec{v},\vec{w})$ and
${L}^*=(\vec{\imath}^*,\vec{\jmath}^*,\vec{u}^*,\vec{v}^*,\vec{w}^*)$
be two reduction layouts from 
${\cal{L}}^p$ with $HD({L},{L}^*) \le d$.
Let $\vec{S}$ and $\vec{F}$ be the DDWB process for generating 
the distribution ${\cal{D}}^p$
as described in
Definition~\ref{ddwb4layoutsDefn}.
For the sake of brevity,  in the scope of this proof  we will
write  $S_i({L})$ and $S_i({L}^*)$ instead of with their proper
arguments,  eg. $S_{2n+2+k}({L})$ instead of
$S_{2n+2+k}(\vec{\imath},\vec{\jmath},(u_1,v_1,w_1), \ldots$ $ (u_{k-1},v_{k-1},w_{k-1}))$.  We do the same with the $F_i$'s.
We set $I_0$ to be the set of indices $i$ so that 
$S_i({L}) \neq S_i({L}^*)$.
Checking against the definitions of $\vec{S}$, $\vec{F}$,
it is easily checked by a case-analysis  that $|I_0| \le 2d$.  
We place this argument in the Appendix, Section~\ref{APP-analyzeDSect},
 as Lemma~\ref{differBound}.

We now check that the hypotheses of Lemma~\ref{prob-thing} are met
with the process $\vec{S}$, $\vec{F}$ over $[m]^{n+1} \times \left([2m+1]^2\right)^n \times ([2m+1]^3)   \times \left([3m]^3\right)^{n+1}$,  with $t=3n+3$,
with $\pi = \mu$,
with $I_0$ as above, and with
$\vec{u} = {L^*}$, $\vec{v}={L}$
By Lemma~\ref{lemma-DIsNBP} and Corollary~\ref{sloppy-cor},  the DDWB process generating $\mu$ has 
$\kappa \ge \delta^2/20$
where $\delta=\delta({\cal{V}}_I,{\cal{V}}_{II})$ and 
$\gamma = \frac{n+1}{m} \le \delta^2/60$.

\begin{description}

\item[Property~\ref{prop-nonzero}:] $\mu({L})>0$ and 
$\mu({L}^*)>0$.
This is satisfied because ${L} \in {\cal{L}}$,  and
${L}^* \in {\cal{L}}$.

\item[Property~\ref{prop-choicesets-agree}:]
The set $I_0$ is defined to be the set of $i$
with $S_i({L}) \neq S_i({L}^*)$.

\item[Property~\ref{prop-negblockdiff}:]
In the Appendix, Section~\ref{APP-analyzeDSect}, we show that
for all $i \in [t]$,  $|F_i({L}) \oplus F_i({L}^*)| \le (9d\gamma/(3n+3)) |X_i|$.

\end{description}

By Lemma~\ref{prob-thing}:
\[\mu({L}) \le \kappa^{-2d}e^{9d\gamma/\kappa} \mu({L^*}) \le (\delta^2/20)^{-2d}e^{9d (\delta^2/60)/(\delta^2/20)} \mu({L})  =   (20/\delta^2)^{2d} e^{3d} \mu({L}) = (20 /\delta^2)^{2d}e^{3d} \mu(L) \]

\end{proof}

Now we use Lemma~\ref{lemma-unif2nonblock} to prove the
completeness lemma:

\begin{proof}(of the completeness lemma, Lemma~\ref{switchable-density})
Fix $m$, and let $({\cal{V}}_I,{\cal{V}}_{II})$ be a partition of $MVars_m$ so
with $\delta = \delta({\cal{V}}_I,{\cal{V}}_{II})$.  Let $n$ be given so that
$ n \le \delta^{10}/(2^{10} \cdot 3 \cdot 5^2) m$. Let  $l \in [n]$ be given.
Let $U$ be uniform distribution on 
$[m]^{n+1} \times \left([2m+1]^2\right)^n \times ([2m+1]^3)  \times \left([3m]^3\right)^{n+1}$. Let $\mu$ be  
the mass function for
the distribution ${\cal{D}}$.
Set $\beta$  to be the blockage bound 
for the DDWB process generating ${\cal{D}}$.
Let ${\cal{A}} \subseteq   [m]^{n+1} \times \left([2m+1]^2\right)^n \times ([2m+1]^3)  \times \left([3m]^3\right)^{n+1}$
be the event that $(j_{{n+2},1},j_{l,1},j_{l,2}) \in N_3(i_l)$, $(j_{{n+1},1},j_{{n+1},2},j_{{n+1},3}) \in N_3(i_{n+1})$,
and
$K\left(\{u_{n+1},u_l\},\{v_{n+1},v_l,w_{n+1},w_l\}\right) \subseteq E_{i_l}\left[V_{j_{l,1}}\cap V_{j_{l,2}}  \right] \cap E_{i_{n+1}}\left[ V_{j_{{n+1},1}} \cap V_{j_{{n+1},2}} \cap V_{j_{{n+1},3}} \right]$.
Notice that ${\cal{S}}^{l} = {\cal{L}} \cap {\cal{A}}$,
and that because $\mu({\cal{L}})=1$,  
$\mu({\cal{S}}^{l}) = \mu({\cal{L}} \cap {\cal{A}}) = \mu({\cal{A}})$.

Let $I$ denote the indices $1, \ldots 2n+2$ (so that,
 using our abused notation,
the coordinates of $I$ correspond to $\vec{\imath},\vec{\jmath}$).
Let $A \subseteq [m]^{n+1} \times \left([2m+1]^2\right)^n \times ([2m+1]^3) $ be the event
that $i_l,i_{n+1} \in G$,  $(j_{{n+1},1},j_{{n+1},2},j_{{n+1},3}) \in N_3(i_{n+1})$,
and $(j_{{n+1},2},j_{l,1},j_{l,2}) \in N_3(i_l)$.  Notice that
$A \supseteq {\cal{A}}_I$ and therefore 
$\mu({\cal{A}}) = \sum_{\vec{\imath},\vec{\jmath} \in A} \mu_I(\vec{\imath},\vec{\jmath}) \mu^{\vec{\imath},\vec{\jmath}}({\cal{A}}(\vec{\imath},\vec{\jmath}))$.

For each setting of $\vec{\imath}$ and $\vec{\jmath}$,
the event ${\cal{A}}(\vec{\imath},\vec{\jmath})$
depends only on the values of $(u_{n+1},v_{n+1},w_{n+1})$, and $(u_l,v_l,w_l)$.
Moreover, in the event that ${\cal{A}}$ holds,
we have that $(u_l,v_l,w_l) \in {\cal{K}}_{1,2}(E_{i_l}[V_{j_{l,1}} \cap V_{j_{l,2}}])$
and $(u_{n+1},v_{n+1},w_{n+1}) \in {\cal{K}}_{1,2}(E_{i_{n+1}}[V_{j_{{n+1},1}} \cap V_{j_{{n+1},2}} \cap V_{j_{{n+1},3}}])$.
Therefore we can apply 
Lemma~\ref{lemma-unif2nonblock} and conclude for all $\vec{\imath}$, 
$\vec{\jmath}$:
$\mu^{\vec{\imath},\vec{\jmath}}({\cal{A}}(\vec{\imath},\vec{\jmath}))\ge
U^{\vec{\imath},\vec{\jmath}}({\cal{A}}(\vec{\imath},\vec{\jmath})) - 2 \beta$.

For each $\vec{\imath}$ and
$\vec{\jmath}$   set 
$D(\vec{\imath},\vec{\jmath})= |E_{i_l}\left[V_{j_{l,1}}\cap V_{j_{l,2}}  \right] \cap E_{i_{n+1}}\left[V_{j_{{n+1},1}} \cap V_{j_{{n+1},2}} \cap V_{j_{{n+1},3}} \right] |/{3m \choose 2}$.
Notice that $\delta({\cal{V}}_I,{\cal{V}}_{II})$ is the expectation of
$D$ over the uniform distribution on $[m]^2 \times [2m+1]^5$.
Because the marginal distribution of 
$U^{\vec{\imath}\vec{\jmath}}$  on
$(u_l,v_l,w_l)$ and $(u_{n+1},v_{n+1},w_{n+1})$
is just the uniform distribution  $[3m]^3 \times [3m]^3$,
we can apply Lemma~\ref{supersaturation}:
For each choice of $\vec{\imath},\vec{\jmath}$ we have that $U^{\vec{\imath},\vec{\jmath}} \left({\cal{A}}(\vec{\imath},\vec{\jmath})\right) \ge D(\vec{\imath},\vec{\jmath})^8 -(23/3m)$. Therefore:

\begin{eqnarray*}
\mu({\cal{A}}) & = &   
   \sum_{\vec{\imath} \vec{\jmath} \in A}\mu_I(\vec{\imath},\vec{\jmath}) \mu^{\vec{\imath},\vec{\jmath}}({\cal{A}}(\vec{\imath},\vec{\jmath})) 
 \ge  \sum_{\vec{\imath},\vec{\jmath} \in A} \mu_I(\vec{\imath},\vec{\jmath})\left( U^{\vec{\imath},\vec{\jmath}}\left({\cal{A}}(\vec{\imath},\vec{\jmath})\right) - 2 \beta \right) \ge  - 2\beta + \sum_{\vec{\imath},\vec{\jmath} \in A} \mu_I(\vec{\imath},\vec{\jmath})U^{\vec{\imath},\vec{\jmath}}({\cal{A}}(\vec{\imath},\vec{\jmath})) \\
& \ge & - 2\beta + \sum_{\vec{\imath},\vec{\jmath} \in A} \mu_I(\vec{\imath},\vec{\jmath})(D(\vec{\imath},\vec{\jmath})^8 - (23/3m)) \ge  - 2\beta - (23/3m) + \sum_{\vec{\imath},\vec{\jmath}} \mu_I(\vec{\imath},\vec{\jmath})(   D(\vec{\imath},\vec{\jmath}) \cdot \chi_A(\vec{\imath},\vec{\jmath}))^8 \\ & \ge &  -2\beta - (23/3m) +  \left(\sum_{\vec{\imath},\vec{\jmath}} \mu_I(\vec{\imath},\vec{\jmath})D(\vec{\imath}, \vec{\jmath}) \chi_A(\vec{\imath},\vec{\jmath}) \right)^8 
 =  -2 \beta - (23/3m) +  \left(\expect_{\mu_I}[D \cdot \chi_A] \right)^8 
\end{eqnarray*}

The final task is to get a lower bound for $\expect_{\mu_I}[D \cdot \chi_A]$.
This will follow from an application of Lemma~\ref{lemma-unif2nonblock}.
Let $U$ denote
the uniform distribution over $\vec{\imath},\vec{\jmath}$,  
In the Appendix, Section~\ref{APP-analyzeDSect}, Lemma~\ref{expectedDensity},  it is shown    that:
$\expect_U [ D \cdot \chi_A] \ge \delta({\cal{V}}_I,{\cal{V}}_{II})/2$. 
Notice that  the function $D \cdot \chi_A$ depends only upon $4$ 
coordinates: $i_l$, $i_{n+1}$, the triple $(j_{{n+1},1},j_{{n+1},2},j_{{n+1},3})$ and
the pair $(j_{l,1},j_{l,2})$.  
Moreover,  whenever $D \cdot \chi_A > 0$,
we have that $i_l \in G$, $i_{n+1} \in G$, 
$(j_{{n+1},1},j_{{n+1},2},j_{{n+1},3}) \in N_3(i_{n+1})$, and
$(j_{l,1},j_{l,2}) \in N_2(i_l)$,
so we may apply Lemma~\ref{lemma-unif2nonblock} to conclude that
$\expect_{\mu_I} [\delta \cdot \chi_A] \ge \delta/2 - 4 \beta$.
Therefore: 
\[ \mu({\cal{A}})  \ge -2\beta - (23/3m) + \left(\expect_{\mu_I}[D \cdot \chi_A]\right)^8 \ge -2\beta - (23/3m) + (\delta/2 - 4 \beta)^8 \]
Because $m \ge 31(2/\delta)^8$,  $\frac{23}{3m} \le 0.25(\delta/2)^8$.
By Lemma~\ref{lemma-DIsNBP},
$\beta \le 30 \gamma / \delta^2 \le 30(\delta^{10}/(2^{10} \cdot 3 \cdot 5^2))/\delta^2 = \delta^8/(2^9 \cdot 5)$, therefore:
\begin{eqnarray*}
 \mu({\cal{A}}) &\ge& -  \frac{2\delta^8}{2^9 \cdot 5} -  0.25\left(\frac{\delta}{2}\right)^8 +  \left(\frac{\delta}{2}  -  \frac{4 \delta^8}{2^9 \cdot 5}\right)^8 \\
& > & -  0.2 \left(\frac{\delta}{2}\right)^8 -  0.25\left(\frac{\delta}{2}\right)^8 +  \left(\frac{\delta}{2}\left(1 -  \frac{1}{2^6 \cdot 5}\right)\right)^8  \\
& > &  -  0.45\left(\frac{\delta}{2}\right)^8 +  0.97\left(\frac{\delta}{2}\right)^8 > \frac{\delta^8}{2^9}
\end{eqnarray*}

\end{proof}

\section{Debriefing}\label{debriefing}
After digesting the proof of Theorem~\ref{payoff}, the reader might notice
that there was some  overkill in a few
of the arguments,  and wonder if a tighter argument could
improve the constants of Theorem~\ref{payoff}.
This seems  likely,  however, it was decided that 
optimizing between different values of ``astronomical'' was not worth the
added length.

There are two points in the argument particularly worthy of mention.
The first is that Definition~\ref{defn-lswitch} is bit stronger than is 
needed to prove Lemma~\ref{lemma-switcheroo}, 
and it may be possible with a more careful definition to reduce the
exponent of $8$ (which comes  from trying to randomly find
a $K_{2,4}$ in a graph of edge density $\alpha$) to something smaller, like
$4$ or $6$. This would clearly improve the bound in
 Lemma~\ref{switchable-density}. Furthermore, it might also allow a
slackening of the
definition of partition density, Definition~\ref{densityDefn},
 so that a larger value is
guaranteed by an analog to Lemma~\ref{part-lemma}.
Furthermore, the DDWB machinery introduces a fair a amount of slop
because the blockage bounds (coverage bounds) are taken as a 
maximum (minimum) over all coordinates, whereas a more careful coordinate-wise
analysis of the particular transformation of Lemma~\ref{lemma-switcheroo}
would improve the constants seen in  Lemma~\ref{switchable-density} and
Lemma~\ref{continuity-lemma}.  Of course, this would likely be a more lengthy
analysis.

\bibliography{OBDDtl}

\begin{appendix}

\section{Proofs and Calculations for Section~\ref{notationSect}}\label{APP-notationSect}

\begin{proof}(of Lemma~\ref{convexity-lemma})
A standard application of the convexity of the function 
$x \mapsto x^k$.   For each $x \in X$, let 
$d_x = | \{ i \in [n] \mid x \in Y_i \}|$.  Set 
$\bar{d_x} = \frac{1}{|X|} \sum_{x \in X} d_x$.  We  have that
$\bar{d_x} = \frac{1}{|X|} \sum_{x \in X} d_x = \frac{1}{|X|} \sum_{i=1}^n |Y_i| = \alpha n$,
and therefore by Jensen's Inequality:
\[
\frac{1}{n^k} \sum_{\vec{\imath} \in [n]^k} | \bigcap_{l=1}^k Y_{i_l}| =
\frac{1}{n^k} \sum_{x \in X} d_x^k 
 \ge  \frac{1}{n^k} |X| \left(  \bar{d_x} \right)^k  \ge  \frac{1}{n^k} |X| \left(  \alpha n \right)^k = \alpha^k |X|
\]
\end{proof}

\begin{proof}(of Lemma~\ref{supersaturation})

\begin{enumerate}
\item   Conditioned on the choice of $u_1$, the probability
that $\{u_1,u_2\} \in E$ and $\{u_1,u_3\} \in E$ is 
$\left( \frac{d_{u_1}}{N} \right)^2$.  
Because $\frac{1}{N} \sum_u d_u = \frac{1}{N} 2 \alpha {N \choose 2}=  \alpha (N-1)$, convexity shows that
the probability that
$\{u_1,u_2\} \in E$ and $\{u_1,u_3\} \in E$ is
at least $N^{-3} \cdot N(\alpha (N-1))^2 = \alpha^2( 1 - 2/N + 1/N^3)$.
We now subtract out the probability
that $u_1, u_2, u_3$ are not all distinct, which is clearly
no more than $3/N$, and we obtain the stated bound.

\item For each $u_1$ and $u_2$, let $D(u_1,u_2)$ be the number of common
neighbors of $u_1$ and $u_2$.  Because the average degree of $u \in V$ is $\alpha(N-1)$,
Lemma~\ref{convexity-lemma} shows that
$\frac{1}{N^2} \sum_{\vec{u} \in V^2} D(u_1,u_2) \ge \alpha^2 ((N-1)/N)^2 (N-1) \ge \alpha^2(1-2/N)$.
Conditioned on the choice of $u_1,u_2$,  the probability that
all edges are present is clearly $\left( D(u_1,u_2)  / N \right)^4$.
Apply Jensen's Inequality  and we have that the probability that all edges are 
present is at least $\left( \alpha^2 (1-2/N) \right)^4 = \alpha^8(1-2/N)^4 \ge \alpha^8 (1-8/N)$.
We now subtract out the probability
that $u_1, u_2, u_3,u_4,u_5,u_6$ are not all distinct, which is clearly
no more than ${6 \choose 2}/N = 15/N$, and we obtain the stated bound.
\end{enumerate}

\end{proof}

\section{Proofs and Calculations for Section~\ref{layoutSect}}\label{APP-layoutSect}
\begin{proof}(of Lemma~\ref{lemma-gdense})
Let $\delta = \delta({\cal{V}}_I,{\cal{V}}_{II})$. 
Notice that when  $m \ge 3\delta \ge ((6/\delta) -1)/2$,
we have that
$3/(2m+1) \le \delta/2$.
 By Definition~\ref{densityDefn} , we have that
\[ \frac{1}{m^2(2m+1)^5} \sum_{\vec{\imath}\in [m]^2} \sum_{\vec{\jmath} \in [2m+1]^5} |\bigcap_{k=1}^5 \left( E_{i_1}[V_{j_k}] \cap E_{i_2}[V_{j_k}]\right) | = \delta {3m \choose 2} \]

And therefore
$\frac{1}{m(2m+1)^3} \sum_{i \in [m]} \sum_{\vec{\jmath} \in [2m+1]^3} | E_{i}[V_{j_1} \cap V_{j_2} \cap V_{j_3}] | \ge \delta {3m \choose 2}$.
Because the number of terms with $j_1=j_2$, $j_2=j_3$ or
$j_1=j_3$ is at most $3m(2m+1)^2$,  such terms can contribute
at most $\frac{1}{m(2m+1)^3}3m(2m+1)^2 {3m\choose 2}=\frac{3}{2m+1}{3m \choose 2}$
to this sum, so we have:
\[ \frac{1}{m(2m+1)^3} \sum_{i \in [m]} \sum_{\vec{\jmath} \in [2m+1]^3 \atop{\mbox{\tiny{$\vec{\jmath}$ distinct}}}} | E_{i}[V_{j_1} \cap V_{j_2} \cap V_{j_3}] | \ge (\delta - 3/(2m+1) ){3m \choose 2} \ge (\delta/2){3m \choose 2} \]

Combining this with the fact that for
each $i \in [m]$, $|E_i| \le {3m \choose 2}$,  by averaging,
we have that with probability at least $\delta/6$  over  the choice of
$i$,$j_1$,$j_2$,$j_3$, with $j_1,j_2,j_3$ all distinct, that
$|E_{i}[V_{j_1} \cap V
_{j_2} \cap V_{j_3}] | \ge (\delta/3) {3m \choose 2}$.
Therefore,  with probability at least $\delta/12$ over choices of $i$,
there are at least $(\delta/12)[2m+1]^3$ many triples $j_1,j_2,j_3$
that are distinct and have
$|E_{i}[V_{j_1} \cap V_{j_2} \cap V_{j_3}] | \ge (\delta/3) {3m \choose 2}$.
Therefore, $|G|  \ge (\delta/12)m$.

\end{proof}

\begin{proof}(of Lemma~\ref{layout-plant-lemma})
Let ${L}=\left(\vec{\imath},\vec{\jmath},\vec{u},\vec{v},\vec{w}\right)$ be a reduction layout, and let $X_1, \ldots X_n, Y_1, \ldots Y_n$ be a
set intersection instance.  
Let $e$ be a bad edge of $A_{{L},\vec{X},\vec{Y}}$.
We recall two useful definitions for the proof of this lemma:
From Definition~\ref{defn-reduction-layout},
the planted edge under $\vec{X}, \vec{Y}, {{L}}$ is defined as
$pe\left(\vec{X},\vec{Y},{{L}}\right) = \{u_{n+1}, w_{n+1} \}$.
From Definition~\ref{defn-assignment},
the assignment $A_{L,\vec{X},\vec{Y}}$ is defined as follows:
We set $I = \{i_1, \ldots i_{n+1} \}$, set $J = \{ j_{1,1}, j_{1,2}, \ldots j_{n,1}, j_{n,2},j_{n+1,1}, j_{n+1,2}, j_{n+1,3} \}$, and set
$V = \{u_1, \ldots u_{n+1}, v_1, \ldots v_{n+1}, w_1, \ldots w_{n+1} \}$.
We set $\beta$, $\beta\left(  {{L}} \right)$ to
be the lexicographically first assignment to the variables
$\{ x^i_e \mid i \in [m] - I, \ e \in [3m-V]^2 \} \cup \{y^j_u \mid j \in [2m+1] - J, \ u \in [3m]-V \}$,
so that $\beta$ defines a matching of size $m-n-1$ and an independent set of
size $2(m-n-1)$.

\begin{eqnarray*}
A_{{L}, \vec{X}, \vec{Y}}(x^i_e) & = & \left\{ \begin{array}{cc}
\beta(x^i_e) & {\mbox{ if $i \in [m]-I$ and $e \in \left([3m] -V\right)^2$}} \\
X_k     & {\mbox{ if $i=i_k$ and $e = \{u_k,v_k\}$ for some $k \in [n]$ }}\\
\neg X_k & {\mbox{ if $i=i_k$ and $e = \{u_k,w_k\}$ for some $k \in [n]$ }}\\
1 & {\mbox{ if $i= i_{n+1}$ and $e = \{u_{n+1},w_{n+1}\}$}} \\
0  &  {\mbox{ otherwise}}
\end{array}\right.
\end{eqnarray*}

\begin{eqnarray*}
A_{{L}, \vec{X}, \vec{Y}}(y^j_x) & = & \left\{ \begin{array}{cc}
\beta(y^j_x) & {\mbox{ if $j \in [2m+1]-j$ and $u \in [3m] -V$}} \\
1    & {\mbox{if  $j = j_{k,1}$ and $x = v_k$ for some $k \in [n] $}} \\
 Y_k  & {\mbox{if  $j = j_{k,2}$ and $x= u_k$ for some $k \in [n]$}} \\
\neg Y_k  & {\mbox{if  $j = j_{k,2}$ and $x= w_k$ for some $k \in [n]$}} \\
1  & {\mbox{if  $j = j_{n+1,1}$ and $x= u_{n+1}$}} \\
1  & {\mbox{if  $j = j_{n+1,2}$ and $x= v_{n+1}$}} \\
1  & {\mbox{if  $j = j_{n+1,3}$ and $x= w_{n+1}$}} \\
0  &  {\mbox{ otherwise}}
\end{array}\right.
\end{eqnarray*}

Let $e$ be a bad edge for the assignment $A_{L,\vec{X},\vec{Y}}$.
First of all, because $\beta$ sets no bad edges,  $e \cap V \neq \emptyset$.
Furthermore, for all  $e$ with $|e \cap V | = 1$ have $A_{{L},\vec{X},\vec{Y}}(x^i_e)=0$ for all $i$,  so $e \subseteq V$.
Finally, for $e \subseteq V$,  with some 
$A_{{L},\vec{X},\vec{Y}}(x^i_e)=1$,  we have that for some
$k \in [n]$, $e = \{u_k,v_k\}$ or $e=\{u_k,w_k\}$.
Choose $k$ so that $e = \{u_k,v_k\}$ or $e=\{u_k,w_k\}$.
If $k=n+1$ then we must have that $e = \{u_{n+1},w_{n+1}\}$, and
$e$ the bad edge, so consider the case when $k \le n$.

Notice that for all $i' \neq i_k$, 
$A_{{L},\vec{X},\vec{Y}}(x^{i'}_e)=0$. On the other hand,
$e$ is a bad edge, so
there is some $x^i_e$ that gets set to $1$, therefore
$A_{{L},\vec{X},\vec{Y}}(x^{i_k}_e)=1$.

We now rule out the case that $e=\{u_k,w_k\}$.  Because 
$A_{{L},\vec{X},\vec{Y}}(x^{i_k}_e)=1$,  we have by construction that
$X_k = 0$.  Because $e$ is bad,  for some $j,j'$,
$A_{{L},\vec{X},\vec{Y}}(y^j_{u_k})=1$ and
$A_{{L},\vec{X},\vec{Y}}(y^{j'}_{w_k})=1$. However, 
$y^j_{u_k}$ and $y^{j'}_{w_k}$ cannot both be set to $1$.

Suppose that $e=\{u_k,v_k\}$.  
Because $A_{{L},\vec{X},\vec{Y}}(x^{i_l}_e)=1$, we have by construction
that $X_l = 1$.
If $(X_l,Y_l)=(1,1)$, then the lemma holds.  Otherwise,
$Y_l=0$.  But in this case, we have that for all $j$, 
$A_{{L},\vec{X},\vec{Y}}(y_{u_l}^{j})=0$, contradiction to
 $e$ being a bad edge.

\end{proof}

\section{Proofs and Calculations for Section~\ref{distributionSect}}\label{APP-distributionSect}

\begin{proof}(of Lemma~\ref{bountiful-lemma})
For each $k=1, \ldots n$, as we choose $(j_{k,1},j_{k,2})$ (and
$(j_{n+1,1},j_{{n+1},2},j_{{n+1},3})$),    $|J| \le 2n < 2(n+1)=2\gamma m$  and
as we choose each $(u_k,v_k,w_k)$,  $|V^*| \le 3n < 3(n+1) = 3\gamma m$.  
\begin{enumerate}
\item By Lemma~\ref{lemma-gdense}, $|G| \ge (\delta/12)m$.  On the other hand,
$|\{i_1, \ldots i_{k-1}\}| \le n < \gamma m$. Therefore,
$|G \setminus \{i_1, \ldots i_{k-1}\}| > ((\delta/12)-\gamma)m$.

\item 
Because $|J| \le 2n$, we have that $pm_{[2m+1]}(J) \le 2n(2m+1)+(2m+1)2n < 2( 2 \gamma m )(2m+1)= 2(2 \gamma m)(2 m +1)  = 2 \gamma (2m)(2m+1) < 2\gamma (2m+1)^2$.
Combining this with the fact that $i_k \in G$ and therefore $|N_2(i_k)| \ge |N_3(i_k)| \ge (\delta/3)(2m+1)^2$ we have that
$|N_2(i_k) \setminus pm_{[2m+1]}(J)| \ge ((\delta/3)  - 2\gamma )(2m+1)^2 $.

\item Because $|J| \le 2n$ we have that
$tm_{[2m+1]}(J) \le 3(2n)(2m+1)^2 < 3(2\gamma m )(2m+1)^2=3\gamma(2 m)(2m+1)^2< 3\gamma(2m+1)^3 $.
Because  $i_p \in G$, 
$|N_3(i_p)| \ge (\delta/3)(2m+1)^3$.  Therefore:
$ |N_3(i_p) \setminus tm_{[2m+1]}(J)| \ge ((\delta/3)- 3\gamma)(2m+1)^3 $.

\item Because $|V^*| \le 3n$,  $|tm(V^*)| \le 3(3n)(3m)^2 < 3(3\gamma m )(3m)^2 = 3\gamma (3m)^3$.
We now get a lower bound on the size of ${\cal{K}}_{1,2}\left( E_{i_k}[V_{j_{k,1}} \cap V_{j_{k,2}}]\right)$:
First, because $(j_{k,1},j_{k,2}) \in N_2(i_k)$, there exists some $j'$
with
$|E_i[V_{j'} \cap V_{j_{k,1}} \cap V_{j_{k,2}}]| \ge(\delta/3){3m \choose 2}$,   so we have that
$|E_{i_k} \left[ V_{j_{k,1}} \cap V_{j_{k,2}} \right] \ge (\delta/3){3m \choose 2}$.
Feeding this lowerbound on the edge density
into Lemma~\ref{supersaturation}, we have that:
\[|{\cal{K}}_{1,2}(E_{i_k} \left[ V_{j_{k,1}} \cap V_{j_{k,2}} \right])| \ge  \left( \delta^2/9 - (5/m) \right) \cdot (3m)^3 \]

Combining the upper bound on $|tm(V^*)|$ and with the preceding lower bound:
\[|{\cal{K}}_{1,2}(E_{i_k} \left[ V_{j_{k,1}} \cap V_{j_{k,2}} \right]) \setminus tm(V^*)| \le \left( (\delta^2/9) - (5/m) - 3\gamma  \right) (3m)^3\]
Because $m \ge 450/\delta^2$, we have that $5/m \le \delta^2/90$ and therefore
the above quantity is 
$\ge (\delta^2/9 - \delta^2/90 -3\gamma)(3m)^3 = (\delta^2/10) - 3 \gamma)(3m)^3$.
\item This derivation is  identical to the previous, except that it
uses the lower bound of
$|E_i [V_{j_{p,3}} \cap V_{j_{p,1}} \cap V_{j_{p,2}}]| \ge (\delta/3){3m \choose 2}$ that
holds because $(j_{p,1},j_{p,2},j_{p,3}) \in N_3(i_p)$.
\end{enumerate}

\end{proof}

\begin{proof}(details for Lemma~\ref{lemma-switcheroo})

{\bf{The proof that $f$ is an involution.}}  
Let ${L}= (\vec{\imath},\vec{\jmath},\vec{u},\vec{v},\vec{w})$
be a reduction layout,  and let
$\left(\vec{\imath}^*,\vec{\jmath}^*,\vec{u}^*,\vec{v}^*,\vec{w}^* \right) = f({L})$, and let
$\left(\vec{\imath}^{**},\vec{\jmath}^{**},\vec{u}^{**},\vec{v}^{**},\vec{w}^{**} \right) =f(f({L}))$.
Applying the definitions shows that:

\[n
\begin{array}{ccccccc}
i^{**}_k & = & \left\{ \begin{array}{cc}
i_{n+1}^*=i_l & {\mbox{ if $k=l$}} \\
i_l^*=i_{n+1} & {\mbox{ if $k={n+1}$}} \\
i_k^* =i_k& {\mbox{ otherwise}}
\end{array} \right.  & & u^{**}_k &= & \left\{ \begin{array}{cc}
u_l^* =u_{n+1}& {\mbox{ if $k={n+1}$}} \\
u_{n+1}^* =u_l& {\mbox{ if $k=l$}} \\
u_k^* = u_k & {\mbox{ otherwise}}
\end{array} \right. \\
j^{**}_{k,1} & = & \left\{ \begin{array}{cc}
j^*_{{n+1},3}=j_{l,1} & {\mbox{ if $k=l$}} \\
j^*_{l,2} = j_{n+1} & {\mbox{ if $k=n+1$}} \\
j^*_{k,1} = j_{k,1} & {\mbox{ otherwise}}
\end{array}  \right. & &
v^{**}_k &= & \left\{ \begin{array}{cc}
w^*_{n+1} =v_l & {\mbox{ if $k=l$}} \\
v^*_k = v_k & {\mbox{ otherwise}}
\end{array} \right. \\
j^{**}_{k,2} & = & \left\{ \begin{array}{cc}
j^*_{{n+1},1} = j_{l,2} & {\mbox{ if $k=l$}} \\
j^*_{k,2} = j_{k,2} & {\mbox{ otherwise}}
\end{array} \right. &&
w^{**}_k &= & \left\{ \begin{array}{cc}
v_l^* = w_{n+1} & {\mbox{ if $i={n+1}$}} \\
w_k^* = w_k& {\mbox{ otherwise}} \\
\end{array} \right. \\
j^{**}_{{n+1},3} & =&  j^*_{l,1} = j_{{n+1},3} & && &
\end{array}
\]

{\bf{The proof that $A_{f({L}),\vec{X},\vec{Y}} = A_{{L},\vec{X},\vec{Y}}$}}
We expand the definitions of $A_{{L},\vec{X},\vec{Y}}$ and 
$A_{f({L}),\vec{X},\vec{Y}}$,  per definition \ref{defn-assignment}
Notice that $\{i_1, \ldots i_{n+1} \} = \{i^*_1, \ldots i^*_{n+1} \}$,
$\{ j_{1,1}, j_{1,2}, \ldots j_{n,1}, j_{n,2}, j_{n+1,1}, j_{n+1,2}, j_{{n+1},3}, \} = \{ j_{1,1}^*, j_{1,2}^*, \ldots j_{n,1}^*, j_{n,2}^*, j_{n+1,1}^*, j_{n+1,2}^*, j_{n+1,3}^*, \}$,
and $\{u_1, \ldots u_{n+1}, v_1, \ldots v_{n+1}, w_1, \ldots w_{n+1} \}
= \{u_1^*, \ldots u_{n+1}^*,$ $v_1^*, \ldots v_{n+1}^*, w_1^*, \ldots w_{n+1}^* \}$.  Let $I$, $J$,  and $V$ respectively
denote these three sets.
Because $\beta({L})$ and $\beta({L}^*)$ are both
the lexicographically first assignment to the variables
\[ \{ x^i_e \mid i \in [m] - I, \ e \in [3m-V]^2 \} \cup \{y^j_u \mid j \in [2m+1] - J, \ u \in [3m]-V \}\]
so that $\beta$ defines a matching of size $m-n-1$ and an independent set of
size $2(m-n-1)$,
we have that $\beta({L}) = \beta(L^*)$. Write $\beta$ for this
assignment.
We compare $A_{{L},\vec{X},\vec{Y}}$ and $A_{f({L}),\vec{X},\vec{Y}}$ directly:

\begin{eqnarray*}
A_{{L}, \vec{X}, \vec{Y}}(x^i_e) & = & \left\{ \begin{array}{cc}
\beta(x^i_e) & {\mbox{ if $i \in [m]-I$ and $e \in \left([3m] -V\right)^2$}} \\
X_k     & {\mbox{ if $i=i_k$ and $e = \{u_k,v_k\}$ for some $k \in [n] \setminus \{l \}$ }}\\
\neg X_k & {\mbox{ if $i=i_k$ and $e = \{u_k,w_k\}$ for some $k \in [n] \setminus \{l \}$ }}\\
1 (=X_l)     & {\mbox{ if $i=i_l$ and $e = \{u_l,v_l\}$ }}\\
0 (=\neg X_l) & {\mbox{ if $i=i_l$ and $e = \{u_l,w_l\}$ }}\\
1 & {\mbox{ if $i= i_{n+1}$ and $e = \{u_{n+1},w_{n+1}\}$}} \\
0  &  {\mbox{ otherwise}}
\end{array}\right.
\end{eqnarray*}

\begin{eqnarray*}
A_{f(L), \vec{X}, \vec{Y}}(x^i_e) & = & \left\{ \begin{array}{cc}
\beta(x^i_e) & {\mbox{ if $i \in [m]-I$ and $e \in \left([3m] -V\right)^2$}} \\
X_k     & {\mbox{ if $i=i_k$ and $e = \{u_k,v_k\}$ for some $k \in [n] \setminus \{l \}$ }}\\
\neg X_k & {\mbox{ if $i=i_k$ and $e = \{u_k,w_k\}$ for some $k \in [n] \setminus \{l \}$ }}\\
1      & {\mbox{ if $i=i_l(=i_{n+1}^*)$ and $e = \{u_l,v_l\} (= \{u_{n+1}^*,w_{n+1}^*\})$ }} \\
0  & {\mbox{ if $i=i_l(=i_{n+1}^*)$ and $e = \{u_l,w_l\} (= \{u^*_{n+1}, w_l^*\} )$}}\\
1 (=X_l) & {\mbox{ if $i= i_{n+1}(=i_l^*)$ and $e = \{u_{n+1},w_{n+1}\} (= \{u^*_l,v^*_l \} )$}} \\
0  &  {\mbox{ otherwise}}
\end{array}\right.
\end{eqnarray*}

\begin{eqnarray*}
A_{{L}, \vec{X}, \vec{Y}}(y^j_x) & = & \left\{ \begin{array}{cc}
\beta(y^j_x) & {\mbox{ if $j \in [2m+1]-J$ and $u \in [3m] -V$}} \\
1    & {\mbox{if  $j = j_{k,1}$ and $x = v_k$ for some $k \in [n] $}} \\
 Y_k  & {\mbox{if  $j = j_{k,2}$ and $x= u_k$ for some $k \in [n] \setminus \{l\}$}} \\
\neg Y_k  & {\mbox{if  $j = j_{k,2}$ and $x= w_k$ for some $k \in [n] \setminus \{l \}$}} \\
 1 (=Y_l)  & {\mbox{if  $j = j_{l,2}$ and $x= u_l$}} \\
 0 (=\neg Y_l)  & {\mbox{if  $j = j_{l,2}$ and $x= w_l$}} \\
1  & {\mbox{if  $j = j_{n+1,1}$ and $x= u_{n+1}$}} \\
1  & {\mbox{if  $j = j_{n+1,2}$ and $x= v_{n+1}$}} \\
1  & {\mbox{if  $j = j_{n+1,3}$ and $x= w_{n+1}$}} \\
0  &  {\mbox{ otherwise}}
\end{array}\right.
\end{eqnarray*}

\begin{eqnarray*}
A_{f(L), \vec{X}, \vec{Y}}(y^j_x) & = & \left\{ \begin{array}{cc}
\beta(y^j_x) & {\mbox{ if $j \in [2m+1]-J$ and $u \in [3m] -V$}} \\
1    & {\mbox{if  $j = j_{k,1}$ and $x = v_k$ for some $k \in [n] $}} \\
 Y_k  & {\mbox{if  $j = j_{k,2}$ and $x= u_k$ for some $k \in [n] \setminus \{l\}$}} \\
\neg Y_k  & {\mbox{if  $j = j_{k,2}$ and $x= w_k$ for some $k \in [n] \setminus \{l \}$}} \\
 1   & {\mbox{if  $j = j_{l,2} (= j_{n+1,1}^*)$ and $x= u_l (=u_{n+1}^*)$}} \\
 0   & {\mbox{if  $j = j_{l,2} (=j_{n+1,1}^*$ and $x= w_l(=w^*_l)$}} \\
1 (=X_l)  & {\mbox{if  $j = j_{n+1,1} (=j_{l,2}^*)$ and $x= u_{n+1}=u^*_l$}} \\
1  & {\mbox{if  $j = j_{n+1,2} (=j_{n+1,2}^*)$ and $x= v_{n+1}=v^*_{n+1}$}} \\
1  & {\mbox{if  $j = j_{n+1,3}(=j_{l,1}^*)$ and $x= w_{n+1}=v_l^*$}} \\
0  &  {\mbox{ otherwise}}
\end{array}\right.
\end{eqnarray*}

\end{proof}

\section{Proofs and Calculations for Section~\ref{analyzeDSect}}\label{APP-analyzeDSect}

\begin{lemma}\label{differBound}
If ${L}$ and ${L}^*$ are reduction layouts with 
$HD({L},{L}^*) \le d$,  then there are
at most $2d$ positions $i$ with $S_i({L}) \neq S_i ({L}^*)$.
\end{lemma}
\begin{proof}
Let ${L} =(\vec{\imath},\vec{\jmath},\vec{u},\vec{v},\vec{w})$
and let ${L}^* =(\vec{\imath}^*,\vec{\jmath}^*,\vec{u}^*,\vec{v}^*,\vec{w}^*)$.  We  consider each position
where ${L}$ and ${L}^*$ might differ and see how 
each affects the functions
$\vec{S}$ given in Definition~\ref{ddwb4layoutsDefn}.
\begin{enumerate}

\item If $i_k \neq i^*_k$, with $k \le n$, then we might have that
$S_{n+1+k}({L})=N_2(i_k) \neq N_2(i^*_k) =S_{n+1+k}({L}^*)$,
or that 
$S_{2n+2+k}({L})={\cal{K}}_{1,2}(E_{i_k}[V_{j_{k,1}} \cap V_{j_{k,2}}]) \neq {\cal{K}}_{1,2}(E_{i_k^*}[V_{j_{k,1}^*} \cap V_{j_{k,2}^*}])S_{2n+2+k}({L}^*)$.

\item If $i_{n+1} \neq i^*_{n+1}$,  then we might have that
$S_{2n+2}({L})=N_3(i_{n+1}) \neq N_3(i^*_{n+1}) =S_{2n+2}({L}^*)$,
or that 
$S_{3n+3}({L})={\cal{K}}_{1,2}(E_{i_{n+1}}[V_{j_{{n+1},1}} \cap V_{j_{{n+1},2}} \cap V_{j_{{n+1},3}} ]) \neq {\cal{K}}_{1,2}(E_{i_{n+1}^*}[V_{j_{{n+1},1}^*} \cap V_{j_{{n+1},2}^*} \cap V_{j_{{n+1},3}^*} ])= S_{3n+3}({L}^*)$.

\item If, for some $k \le n$, $(j_{k,1},j_{k,2}) \neq (j_{k,1}^*,j_{k,2}^*)$ then we might
have that $S_{2n+2+k}({L})={\cal{K}}_{1,2}(E_{i_k}[V_{j_{k,1}} \cap V_{j_{k,2}}]) \neq {\cal{K}}_{1,2}(E_{i_k^*}[V_{j^*_{k,1}} \cap V_{j^*_{k,2}}]) =S_{2n+2+k}({L}^*)$.

\item If $(j_{{n+1},1},j_{{n+1},2},j_{{n+1},3}) \neq (j_{{n+1},1}^*,j_{{n+1},2}^*,j_{{n+1},3}^*)$
then we might have that
\[S_{3n+3}({L}) ={\cal{K}}_{1,2}(E_{i_{n+1}}[V_{j_{{n+1},1}} \cap V_{j_{{n+1},2}} \cap V_{j_{{n+1},3}}]) \neq {\cal{K}}_{1,2}(E_{i_{n+1}^*}[V_{j_{{n+1},1}^*} \cap V_{j_{{n+1},2}^*} \cap V_{j_{{n+1},3}^*}]) =  S_{3n+3}({L}^*)\]

\item Differences between $(u_k,v_k,w_k)$ and $(u_k^*,v_k^*,w_k^*)$ do
not affect any of the $S_i$'s.
\end{enumerate}
\end{proof}

\begin{proof}(The calculations ensuring Property~\ref{prop-negblockdiff},
of Lemma~\ref{prob-thing} as applied in the proof of Lemma~\ref{continuity-lemma} .)

\begin{enumerate}

\item Coordinates $1, \ldots n+1$:
 $F_{k}(i_1, \ldots i_{k-1}) = \{i_1, \ldots i_{k-1}\}$ and $X_k=[m]$,
therefore:
\begin{eqnarray*}
|F_k({L}) \oplus F_k({L}^*)|  &= & |\{i_1, \ldots i_{k-1}\}  \oplus \{i_1^*, \ldots i_{k-1}^* \}|  \le  d \\
& = & \frac{d}{3n+3} \frac{3n+3}{m}m =\frac{d}{3n+3} \frac{3 \gamma m}{m}m = \frac{3d\gamma}{3n+3}|X_k|
\end{eqnarray*}

\item   For coordinates $n+2, \ldots 2n+1$, $X_{n+1+k} = [2m+1]^2$ and
\begin{eqnarray*}
F_{n+1+k}(\vec{\imath},(j_{1,1},j_{1,2}), \ldots (j_{k-1,1},j_{k-1,2})) &=& pm_{[2m+1]}\left(  \{j_{1,1}, j_{1,2}, \ldots j_{k-1,1}, j_{k-1,2} \} \right) \\
F_{n+1+k}(\vec{\imath}^*,(j_{1,1}^*,j_{1,2}^*), \ldots (j_{k-1,1}^*,j_{k-1,2}^*)) &=& pm_{[2m+1]}\left(  \{j_{1,1}^*, j_{1,2}^*, \ldots j_{k-1,1}^*, j_{k-1,2}^* \} \right) 
\end{eqnarray*}
Notice that for any $X,Y$, $pm_{[2m+1]}(X) \oplus pm_{[2m+1]}(Y) \subseteq pm_{[2m+1]}(X \oplus Y)$.
On the other hand, $HD({L},{L}^*) \le d$, so
$|\{j_{1,1}, j_{1,2}, \ldots j_{k-1,1}, j_{k-1,2}\} \oplus
\{j_{1,1}^*, j_{1,2}^*, \ldots j_{k-1,1}^*, j_{k-1,2}^*\}| \le 2d$,
and therefore
\begin{eqnarray*}
|F_{n+1+k}({L}) \oplus F_{n+1+k}({L}^*)| &\le& 2\cdot  2d \cdot (2m+1)= \frac{4d}{(3n+3)(2m+1)}(3n+3)(2m+1)^2 \\
&=& \frac{4d}{3n+3}\frac{3\gamma m}{2m+1}|X_{n+1+k}| <\frac{4d}{3n+3}\frac{3\gamma m}{2m}|X_{n+1+k}| = \frac{6d\gamma}{3n+3}|X_{n+1+k}|
\end{eqnarray*}

\item   At coordinate $2n+2$,   $X_{2n+2} = [2m+1]^3$ and
\begin{eqnarray*}
F_{2n+2}(\vec{\imath},(j_{1,1},j_{1,2}), \ldots (j_{n,1},j_{n,2})) &=& tm_{[2m+1]}\left(  \{j_{1,1}, j_{1,2}, \ldots j_{n,1}, j_{n,2} \} \right) \\
F_{2n+2}(\vec{\imath}^*,(j_{1,1}^*,j_{1,2}^*), \ldots (j_{n,1}^*,j_{n,2}^*)) &=& tm_{[2m+1]}\left(  \{j_{1,1}^*, j_{1,2}^*, \ldots j_{n,1}^*, j_{n,2}^* \} \right) 
\end{eqnarray*}
Notice that for any $X,Y$, $tm_{[2m+1]}(X) \oplus tm_{[2m+1]}(Y) \subseteq tm_{[2m+1]}(X \oplus Y)$.
On the other hand, $HD({L},{L}^*) \le d$, so
$|\{j_{1,1}, j_{1,2}, \ldots j_{n,1}, j_{n,2}\} \oplus
\{j_{1,1}^*, j_{1,2}^*, \ldots j_{n,1}^*, j_{n,2}^*\}| \le 2d$,
and therefore
\begin{eqnarray*}
|F_{2n+2}({L}) \oplus F_{2n+2}({L}^*)| &\le& 3 \cdot  2d \cdot (2m+1)^2 = \frac{6d}{(3n+3)(2m+1)}(3n+3)(2m+1)^3 \\
&=& \frac{6d}{3n+3}\frac{3n+3}{2m+1}(2m+1)^3  =\frac{6d}{3n+3}\frac{3  \gamma m}{2m+1}(2m+1)^3  \\
& < & \frac{6d}{3n+3}\frac{3  \gamma m}{2m}(2m+1)^3  = \frac{9d \gamma }{3n+3}(2m+1)^3 
\end{eqnarray*}

\item For  coordinates $2n+3, \dots 3n+3$, $X_{2n+2+k} = [3m]^3$ and
\begin{eqnarray*}
F_{2n+2+k}(\vec{\imath},\vec{\jmath},(u_1,v_1,w_1), \ldots (u_{k-1},v_{k-1},w_{k-1})) & = &  tm(\{u_1,v_1,w_1, \ldots u_{k-1},v_{k-1},w_{k-1}\})
\end{eqnarray*}
Notice that for any finite sets $X$ and $Y$: $tm_{[3m]}(X) \oplus tm_{[3m]}(Y) \subseteq tm_{[3m]}(X \oplus Y)$.
\begin{eqnarray*}
\lefteqn{|F_{2n+2+k}({{L}}) \oplus F_{2n+2+k}({L}^*)|} \\
& = & |tm_{[3m]}(\{u_1,v_1,w_1, \ldots u_{k-1},v_{k-1},w_{k-1}\}) \oplus tm_{[3m]}(\{u_1^*,v_1^*,w_1^*, \ldots u_{k-1}^*,v_{k-1}^*,w_{k-1}^*\}) | \\
& \le &  |tm_{[3m]}(\{u_1,v_1,w_1, \ldots u_{k-1},v_{k-1},w_{k-1}\} \oplus\{u_1^*,v_1^*,w_1^*, \ldots u_{k-1}^*,v_{k-1}^*,w_{k-1}^*\})| \\
& \le & 3 \cdot 3d \cdot (3m)^2  = \frac{9d}{3n+3}\frac{3n+3}{3m}(3m)^3 = \frac{9d}{3n+3}\frac{3\gamma m}{3m}(3m)^3 =\frac{9d\gamma}{3n+3}(3m)^3
\end{eqnarray*}

Therefore,  for every $i=1, \ldots 3n$, 
$|F_i({L}) \oplus F_i({L}^*)| \le \frac{9d\gamma}{3n+3} |X_i|$.
\end{enumerate}
\end{proof}

\begin{lemma}\label{expectedDensity}
Let $\delta > 0$ be given,  and let $m$ be an integer
$\ge 36/\delta$.
Let $({\cal{V}}_I,{\cal{V}}_{II})$ be  a partition of $MVars_m$,
with $ \delta({\cal{V}}_I,{\cal{V}}_{II}) \ge \delta$.  
Let $D$ be as in the Proof of Lemma~\ref{switchable-density}.
Let $G$, $N_2$ and $N_3$ be as in Definition~\ref{gdefn}.
Let $U$ be the uniform distribution over $i_{n+1},i_l \in [m]$,
$(j_{{n+1},1},j_{{n+1},2},j_{{n+1},3}) \in [2m+1]^3$,
and $j_{l,1},j_{l,2}) \in [2m+1]^2$.
Let $A$ be the event that $i_{n+1} \in G$, $i_l \in G$, 
$(j_{{n+1},1},j_{{n+1},2},j_{{n+1},3}) \in N_3(i_p)$,  
and $(j_{{n+1},2},j_{l,1},j_{l,2}) \in N_3(i_l)$.
\[ \expect_U [ D \cdot \chi_A ] \ge \delta({\cal{V}}_I,{\cal{V}}_{II})/2 \]

\end{lemma}

\begin{proof}
Let $B_0$ be the event that either  $j_{l,1}=j_{l,2}$, $j_{{n+1},1}=j_{{n+1},2}$,
$j_{{n+1},2}=j_{{n+1},3}$, $j_{{n+1},3}=j_{{n+1},1}$, $j_{l,1}=j_{{n+1},2}$, or
$j_{l,1}=j_{{n+1},2}$. 
Let $B_1$  be the event that $i_{n+1} \not\in G$,  let $B_2$ be the event that
$i_l \not\in G$,  let $B_3$ be the event that 
$(j_{{n+1},1},j_{{n+1},2},j_{{n+1},3}) \not\in N_3(i_{n+1})$, 
and let $B_4$ be the event that
$(j_{{n+1},2},j_{l,1},j_{l,2}) \not\in N_3(i_l)$.
For each $i=0, \ldots 4$, let
$B^*_i = B_i \cap \bigcap_{j=0}^{i-1} B_j^c$. Because the $B^*$'s partition
$A^c$ we  have that:
\[ \expect_U [D] = \expect_U[ D \cdot \chi_A] + 
\sum_{i=0}^4 \expect_U [D \cdot \chi_{B_i^*} ] \]

Set $\delta^* = \delta({\cal{V}}_I,{\cal{V}}_{II})$.
The  calculations below show that $U(B^*_0) \le 6/(2m+1)$ and
for each $i=1,\ldots 4$, 
$E_U[D \cdot \chi_{B^*_i}] \le (5 \delta^*/12) U(B^*_i)$.  
Modulo those calculations, we have the lemma:

\begin{eqnarray*}
 \expect_U[ D\cdot \chi_A] & = & \expect_U [D] - \sum_{i=0}^4 \expect_U [D \cdot \chi_{B_i^*} ] \ge   \delta^* - 6/(2m+1) - \sum_{i=1}^4 (5\delta^*/12)U(B^*_i) \\
 &\ge &  \delta^* - 5\delta^*/12 - 6/(2m+1) \ge 7\delta/12 - 6/(2(36/\delta)) = 7\delta^*/12 - \delta/12  \ge \delta^*/2
\end{eqnarray*}

For each of the six pairs $j_{l,1}$ and $j_{l,2}$,  $j_{{n+1},1}$ and
$j_{{n+1},2}$,  $j_{{n+1},2}$ and $j_{{n+1},3}$, $j_{{n+1},3}$ and $j_{{n+1},1}$,
$j_{{n+1},2}$ and $j_{j_{l,1}}$,  and $j_{{n+1},2}$ and $j_{l,2}$,  
there is a collision with probability $1/(2m+1)$. Therefore by the union
bound, $U(B^*_0)=U(B_0) \le  6/(2m+1)$.
We now bound the expectation over the  pieces $B^*_1$, $B^*_2$,
$B^*_3$, and $B^*_4$.  Because these events are contained in $B_0^c$,
for elements drawn from these sets,
the tuples $(j_{l,1},j_{l,2})$, $(j_{{n+1},1},j_{{n+1},2},j_{{n+1},3})$,
and $(j_{n+1,2},j_{l,1},j_{l,2})$ each contain
distinct elements.  To denote this,  we will use
$Z$ to denote the set of pairs tuples $(\vec{\jmath_{n+1}},\vec{\jmath_l})$
with $j_{l,1} \neq j_{l,2}$,  $j_{{n+1},1} \neq j_{{n+1},2}$,  $j_{{n+1},2} \neq j_{{n+1},3}$,
$j_{{n+1},3} \neq j_{{n+1},1}$,  $j_{{n+1},1} \neq j_{j_{l,1}}$,  and
$j_{{n+1},1} \neq j_{l,2}$,  
let $[2m+1]_2$ denote all ordered pairs from $[2m+1]$
with two distinct values and let $[2m+1]_3$ denote all ordered triples from
$[2m+1]$ with three distinct values.  Finally,
set $M = m^2(2m+1)^5$,

\begin{eqnarray*}
\expect_U [D \cdot \chi_{B^*_1}] & = & \frac{1}{M} \sum_{i_{n+1} \not\in G} \sum_{i_l \in [m]} \sum_{(\vec{\jmath_{n+1}}, \vec{\jmath_l}) \in Z}  D(i_{n+1},i_l,\vec{\jmath_{n+1}},\vec{\jmath_l}) \\
& \le &
\frac{1}{M} \sum_{i_{n+1} \not\in G} \sum_{i_l \in [m]} \sum_{\vec{\jmath_{n+1}} \in  [2m+1]_3 \atop \vec{\jmath_l} \in [2m+1]_2}  D(i_{n+1},i_l,\vec{\jmath_{n+1}},\vec{\jmath_l}) \\
& = & \frac{1}{M} \sum_{i_{n+1} \not\in G} \sum_{\vec{\jmath_{n+1}} \in [2m+1]_3} \sum_{i_l \in [m] \atop \vec{\jmath_l} \in [2m+1]_2} D(i_{n+1},i_l,\vec{\jmath_{n+1}},\vec{\jmath_l})  \\
& = & \frac{1}{M} \sum_{i_{n+1} \not\in G} \sum_{\vec{\jmath_{n+1}} \in N_3(i_{n+1})} \sum_{i_l \in [m] \atop \vec{\jmath_l} \in [2m+1]_2} D(i_{n+1},i_l,\vec{\jmath_{n+1}},\vec{\jmath_l}) \\
&& + \frac{1}{M} \sum_{i_{n+1} \not\in G} \sum_{\vec{\jmath_{n+1}} \in [2m+1]_3 \setminus N_3(i_{n+1})} \sum_{i_l \in [m] \atop \vec{\jmath_l} \in [2m+1]_2} D(i_{n+1},i_l,\vec{\jmath_{n+1}},\vec{\jmath_l}) \\
& \le & \frac{1}{M} \sum_{i_{n+1} \not\in G} \sum_{\vec{\jmath_{n+1}} \in N_3(i_{n+1})} \sum_{i_l \in [m] \atop \vec{\jmath_l} \in [2m+1]_2} 1  + \frac{1}{M} \sum_{i_{n+1} \not\in G} \sum_{\vec{\jmath_{n+1}} \in [2m+1]_3 \setminus N_3(i_{n+1})} \sum_{i_l \in [m] \atop \vec{\jmath_l} \in [2m+1]_2} (\delta/3) \\
& = & \frac{1}{M} \sum_{i_{n+1} \not\in G}  \sum_{\vec{\jmath_{n+1}} \in N_3(i_{n+1})} m(2m+1)^2 + \frac{1}{M} \sum_{i_{n+1} \not\in G} \sum_{\vec{\jmath_{n+1}} \in [2m+1]_3 \setminus N_3(i_{n+1})} (\delta/3)m(2m+1)^2 \\
& \le & \frac{1}{M} \sum_{i_{n+1} \not\in G}  (\delta/12)(2m+1)^3 m(2m+1)^2 + \frac{1}{M} \sum_{i_{n+1} \not\in G} (2m+1)^3 (\delta/3)m(2m+1)^2 \\
& = & \frac{1}{M} \sum_{i_{n+1} \not\in G}  (\delta/12 + \delta/3) (2m+1)^3 m(2m+1)^2 = (5\delta/12)U(B^*_1)
\end{eqnarray*}

To bound $E_U[D \cdot \chi_{B^*_2}]$ 
we need first show that for all $i_{n+1}, i_l \in [m]$, all
$\vec{\jmath_{n+1}} \in [2m+1]_3$, and all 
$\vec{\jmath_l} \in [2m+1]_2 \setminus  N_2(i_l)$
$D(i_{n+1},i_l,\vec{\jmath_{n+1}},\vec{\jmath_l}) \le \delta/3$. To see this
choose $j^* \in \{j_{{n+1},1},j_{{n+1},2},j_{{n+1},3}\} \setminus \{j_{l,1},j_{l,2}\}$
 and calculate:
\begin{eqnarray*}
D(i_{n+1},i_l,\vec{\imath_{n+1}},\vec{\jmath_l}) & = & \frac{|E_{i_{n+1}}[V_{j_{{n+1},1}} \cap V_{j_{{n+1},2}} \cap V_{j_{{n+1},3}}] \cap E_{i_l}[V_{j_{l,1}} \cap V_{j_{l,2}}]|}{{3m \choose 2}} \\
 &\le & \frac{|E_{i_l}[V_{j_{l,1}} \cap V_{j_{l,2}} \cap V_{j^*}]|}{{3m \choose 2}} \le \delta/3 \end{eqnarray*}

\begin{eqnarray*}
\expect_U [D \cdot \chi_{B^*_2}] & = &
\frac{1}{M}  \sum_{i_{n+1} \in G }\sum_{i_l \not\in G} \sum_{(\vec{\jmath_{n+1}},\vec{\jmath_l}) \in Z }  D(i_{n+1},i_l,\vec{\jmath_{n+1}},\vec{\jmath_l}) \\
& \le & \frac{1}{M}\sum_{i_{n+1} \in G } \sum_{i_l \not\in G} \sum_{\vec{\jmath_l} \in N_2(i_l)} \sum_{\vec{\jmath_{n+1}} \in [2m+1]_3} D(i_{n+1},i_l,\vec{\jmath_{n+1}},\vec{\jmath_l})   \\
&& +\frac{1}{M} \sum_{i_{n+1} \in G } \sum_{i_l \not\in G} \sum_{\vec{\jmath_l} \in[2m+1]_2 \setminus  N_2(i_l)} \sum_{\vec{\jmath_{n+1}} \in [2m+1]_3} D(i_{n+1},i_l,\vec{\jmath_{n+1}},\vec{\jmath_l})  \\
& \le & \frac{1}{M} \sum_{i_{n+1} \in G } \sum_{i_l \not\in G} \sum_{\vec{\jmath_l} \in N_2(i_l)} \sum_{\vec{\jmath_{n+1}} \in [2m+1]_3} 1    +\frac{1}{M} \sum_{i_{n+1} \in G }  \sum_{i_l \not\in G} \sum_{\vec{\jmath_l} \in [2m+1]_2 \setminus N_2(i_l)} \sum_{\vec{\jmath_{n+1}} \in [2m+1]_3} (\delta/3)   \\
& \le & \frac{1}{M} \sum_{i_{n+1} \in G } \sum_{i_l \not\in G} (\delta/12)(2m+1)^5    +\frac{1}{M} \sum_{i_l \not\in G} (\delta/3)(2m+1)^5 = (5\delta/12)U(B^*_2(i_{n+1}))  
\end{eqnarray*}

To bound $\expect_U [ D \cdot \chi_{B^*_3}]$,
note that for all $(i_{n+1},i_l,\vec{\jmath_{n+1}},\vec{\jmath_l})\in B^*_3$,
because $(j_{{n+1},1},j_{{n+1},2},j_{{n+1},3}) \in [2m+1]_3 \setminus N_3(i_{n+1})$:
\begin{eqnarray*}
D(i_{n+1},i_l,\vec{\jmath_{n+1}},\vec{\jmath_l})  &=&  \frac{|E_{i_l}[V_{j_{l,1}} \cap V_{j_{l,2}}] \cap E_{i_{n+1}} [V_{j_{{n+1},1}} \cap V_{j_{{n+1},2}} \cap V_{j_{{n+1},3}}]|}{{3m \choose 2}} \\
 &\le&  \frac{|E_{i_{n+1}}[V_{j_{{n+1},1}} \cap V_{j_{{n+1},2}} \cap V_{j_{{n+1},3}}]|}{{3m \choose 2}}   \le \delta/3 \end{eqnarray*}
Therefore 
$\expect_U [ D \cdot \chi_{B^*_3}] \le (\delta/3) U(B^*_3)$.

Similarly, to bound $\expect_U [ D \cdot \chi_{B^*_4}]$, observe that 
for all $(i_{n+1},i_l, \vec{\jmath_{n+1}},\vec{\jmath_l}) \in B^*_4$,
because $(j_{{n+1},2},j_{l,1},j_{l,2}) \in [2m+1]_3 \setminus N_3(i_l)$:
\begin{eqnarray*}
D(i_{n+1},i_l,\vec{\jmath_{n+1}},\vec{\jmath_l})  & = & \frac{|E_{i_l}[V_{j_{l,1}} \cap V_{j_{l,2}}] \cap E_{i_{n+1}} [V_{j_{{n+1},1}} \cap V_{j_{{n+1},2}} \cap V_{j_{{n+1},3}}]|}{{3m \choose 2}} \\
& \le & \frac{|E_{i_l}[V_{j_{{n+1},2}} \cap V_{j_{l,1}} \cap V_{j_{l,2}}]|}{{3m \choose 2}}   \le \delta/3 
\end{eqnarray*}
Therefore
$\expect_U [ D \cdot \chi_{B^*_4}] \le (\delta/3) U(B^*_4)$.

\end{proof}

\end{appendix}

\end{document}